\documentclass[a4paper,11pt]{article}
\pdfoutput=1 
\usepackage{jheppub} 
\usepackage[T1]{fontenc}
\usepackage{xcolor}
\usepackage{epigraph}
\usepackage{comment}
\usepackage{amssymb}
\usepackage{amsmath}
\usepackage{graphicx}
\usepackage{framed}
\usepackage{xltabular}
\usepackage{float}
\usepackage{caption}
\usepackage{subcaption}
\usepackage{changepage}
\usepackage{multirow}

\usepackage{amsfonts}
\usepackage{amsthm}
\usepackage{amsbsy}
\usepackage{mathtools}
\usepackage{bm}
\usepackage{hyperref}
\providecommand{\eprint}[1]{}\renewcommand{\eprint}[1]{\href{https://arxiv.org/abs/#1}{arXiv:#1}}

\usepackage{lmodern}
\usepackage{booktabs,array,enumitem,graphicx}
\usepackage{longtable}
\usepackage[expansion=false]{microtype}
\usepackage{tikz}
\usetikzlibrary{arrows.meta,positioning,calc,decorations.pathmorphing}
\usepackage[framemethod=TikZ]{mdframed}

\newcommand{\kk}{\kappa}
\newcommand{\la}{\lambda}
\newcommand{\lah}{\hat\la}
\newcommand{\Bhalf}{B_{1/2}}
\newcommand{\BhalfD}{\tilde{B}_{1/2}}
\newcommand{\Bsix}{B^{\varphi}_{1/6}}
\newcommand{\Bsixth}{B^{\theta}_{1/6}}
\newcommand{\Eis}[1]{E_{#1}}
\newcommand{\thh}[1]{\vartheta_{#1}}
\newcommand{\hmod}{\hat m}

\newcommand{\vev}[1]{\left\langle #1\right\rangle}
\newcommand{\dk}{\partial_\kk}

\newcommand{\Res}{\mathop{\mathrm{Res}}}
\newcommand{\im}{\operatorname{Im}}
\newcommand{\re}{\operatorname{Re}}
\newcommand{\Dsc}{\Delta}
\newcommand{\PhB}{\Phi_B}
\newcommand{\Lcl}{\Lambda}

\newcommand{\Cop}{\mathsf{C}}

\newcommand{\OmB}{\Omega_{B}}

\DeclareMathOperator{\Tr}{Tr}

\newcommand{\Sh}{\mathcal S}
\newcommand{\Ccor}{\mathcal C}
\newcommand{\Iinst}{\mathcal I}
\makeatletter
\newcommand{\labelalias}[2]{\AtBeginDocument{\@ifundefined{r@#2}{}{\global\expandafter\let\csname r@#1\expandafter\endcsname\csname r@#2\endcsname}}}
\makeatother

\title{Radiation from periods: Bremsstrahlung functions, wound loops, latitudes and integral constraints in ABJM}

\emailAdd{michelangelo.preti$\bullet$unito.it}

\author{Michelangelo Preti}

\affiliation{
Dipartimento di Fisica, Universit\`a di Torino and INFN - Sezione di
Torino, Via P. Giuria 1, Torino 10125, Italy
}

\abstract{
The Bremsstrahlung functions of the $\tfrac12$BPS and $\tfrac16$BPS Wilson lines of ABJM theory measure their response to small
deformations. We show that a single geometric object, the spectral curve of the localisation matrix model, captures both of them
in closed form, together with the planar wound and latitude Wilson loops at any complex winding and framing. The fermionic
function turns out to be a period of the second kind: a single hypergeometric function, which solves a second-order differential
equation and is quasimodular. The bosonic one is not a period but a ratio of periods. At every coupling it is the smaller of the two, so radiation grows along the defect flow from the bosonic to the fermionic line. The key role is played by a new observable, the winding velocity, which turns out to be the 't~Hooft
coupling shifted by four times the fermionic function. It is at the same time the winding derivative of the wound loop and the
curvature of the latitude, and the curve captures it in both incarnations through its logarithmic periods and through its periods with
two charges. It is also a datum of the defect conformal field theory on the $\tfrac12$BPS line, since it fixes a new integrated
four-point function of the tilt operators. All results extend to ABJ theory when the ranks of the gauge groups are not equal.
}

\begin{document} 
\maketitle
\flushbottom

\section{Introduction and summary of the results}
\label{sec:intro}

Among gauge theories with a holographic dual, ABJM theory \cite{Aharony:2008ug} certainly occupies a prominent role. It is the three-dimensional $\mathcal N=6$ Chern-Simons-matter theory
with gauge group $U(N)_k\times U(N)_{-k}$ with $k$ the Chern-Simons level. It describes stacks of M2-branes in M-theory on $AdS_4\times S^7/\mathbb Z_k$ and, in the regime of large $k$, it is dual to type IIA string theory on $AdS_4\times\mathbb{CP}^3$. This makes ABJM a unique window on M-theory: for instance the celebrated $N^{3/2}$ growth of the number of degrees of freedom of
$N$ M2-branes, predicted by 11D supergravity, was derived from the field theory by supersymmetric localisation \cite{Kapustin:2009kz,Drukker:2010nc}. The same localisation captured the partition function at \emph{finite} $N$ and $k$, as an Airy function up to exponentially small
corrections \cite{Fuji:2011km,Marino:2011eh} organised by the topological string on a local Calabi-Yau \cite{Hatsuda:2015gca,Hatsuda:2013oxa}. On a second front, ABJM is considered to be integrable in the planar limit
\cite{Minahan:2008hf,Gromov:2008qe,Cavaglia:2014exa}. In other words, the spectrum of single-trace operators is governed by the Quantum Spectral Curve (QSC) similarly to $\mathcal N=4$ super Yang-Mills (SYM) in 4D, but with one important difference: the coupling the integrability techniques see is not the usual 't~Hooft coupling $\lambda=N/k$, but a non-trivial function of it, the \emph{interpolating function} $h(\la)$. This is not fixed by symmetry and its closed form has been conjectured in \cite{Gromov:2014eha,Cavaglia:2016ide}. On a third front, the ABJM
matrix model solution \cite{Marino:2009jd,Drukker:2010nc} lives on the mirror curve of local $\mathbb P^1\times\mathbb P^1$ giving access to the rich structure of topological strings on this geometry. 
Thus, ABJM provides a rare setting in which three powerful exact approaches such as localisation, integrability, and holography coexist, making the quantities that connect these approaches particularly valuable.

Supersymmetric Wilson loops are the foremost such observables. In $\mathcal N=4$ SYM, the circular $\tfrac12$BPS loop, computed exactly by a Gaussian matrix model \cite{Erickson:2000af,Drukker:2000rr,Pestun:2007rz}, was among the first observables to be matched with string theory at strong coupling and remains one of the cleanest probes of the AdS/CFT correspondence. In ABJM, supersymmetric Wilson loops can be constructed from two distinct gauge connections (see \cite{Drukker:2019bev} for a review): a bosonic connection coupled to a scalar bilinear \cite{Drukker:2008zx,Chen:2008bp,Rey:2008bh,Berenstein:2008dc}, and a fermionic superconnection mixing the two gauge-group factors \cite{Drukker:2009hy}. If supported on a circle or a straight line, they preserve respectively $\tfrac16$ and $\tfrac12$ of the supersymmetries. Both are captured by the localisation matrix model of \cite{Kapustin:2009kz}, with the fermionic loop related to a combination of bosonic loops by a cohomological equivalence.

A supersymmetric Wilson line also defines a one-dimensional superconformal defect, and a first non-trivial datum of a line
defect is its response to deformations, both geometric and internal. This is encoded in the \emph{Bremsstrahlung function},
which measures the leading response to a small deformation away from the BPS configuration. It is the coefficient of the cusp
anomalous dimension at small angles, it controls the energy radiated by an accelerated probe and the normalisation of the
two-point function of the displacement operator \cite{Correa:2012at}, and it fixes the one-point function of the stress tensor
in the presence of the line \cite{Lewkowycz:2013laa,Bachas:2025brp}. For the $\tfrac12$BPS fermionic line, supersymmetry
relates the geometric $\varphi$ and internal $\theta$ angular responses, giving a single function $\Bhalf$. For the
$\tfrac16$BPS bosonic line they are independent, giving two functions, $\Bsix$ and $\Bsixth$, and the internal one is half of
the geometric one \cite{Bianchi:2018scb}. The $\tfrac13$BPS line of \cite{Drukker:2022txy}, built from two $\tfrac12$BPS
superconnections that share eight supercharges, has four Bremsstrahlung functions, all of which are fixed by $\Bhalf$. Through supersymmetry and defect conformal symmetry, the Bremsstrahlung functions can be written
as derivatives of supersymmetric Wilson loops with respect to deformations that preserve some of the supercharges, and a
combination of localisation, perturbative calculations and defect-CFT arguments relates them to the matrix model
\cite{Griguolo:2012iq,Lewkowycz:2013laa,Bianchi:2014laa,Aguilera-Damia:2014qgy,Bianchi:2017svd,Bianchi:2017ozk,Bianchi:2018scb,Bianchi:2018bke,Bianchi:2020hsz}.
In analogy with $\mathcal{N}=4$ SYM, their connection to integrability is expected to be especially direct.
Therefore, the Bremsstrahlung functions are simultaneously intrinsic defect-CFT data, probes of geometric and internal deformations, and
a natural interface between localisation, integrability and holography.

The status of these functions is uneven, and the results are scattered over different methods. The fermionic
function at planar level is known exactly from \cite{Bianchi:2018scb}. There the framing phase of the bosonic circle is evaluated on the
planar solution of the matrix model and it gives $\Bhalf$ as a combination of complete elliptic integrals of the three kinds. The formula leaves its own structure unexplained. For instance, it does not say why elliptic integrals appear, why the perturbative expansion at strong coupling stops after two terms and  what fixes the constant term of that expansion. The bosonic function is known much less. Its two exact
prescriptions need a quantity that the planar matrix model did not provide: the derivative of the multiply wound loop in the
winding number \cite{Lewkowycz:2013laa}, whose closed form was known only at integer winding \cite{Klemm:2012ii}, or the
derivative of the phase of the bosonic latitude in the latitude parameter \cite{Bianchi:2018scb}. Away from the great circle
the latitude matrix model \cite{Bianchi:2018bke} deforms the measure of the eigenvalues, and its planar limit has not been
solved: the latitudes are known through three loops at weak coupling and, at strong coupling, through the Airy function of the
Fermi gas, which misses the worldsheet instantons. As a result, the bosonic function is known as a weak-coupling series
\cite{Lewkowycz:2013laa,Bianchi:2017afp} and through its first orders at strong coupling \cite{Aguilera-Damia:2014qgy,Aguilera-Damia:2014bqa}.
Finally, in $\mathcal N=4$ SYM the Bremsstrahlung function was computed also by integrability
\cite{Correa:2012hh,Drukker:2012de,Gromov:2012eu} and the agreement with localisation was extended to its excited states \cite{Gromov:2012eu, Bonini:2015fng, Giombi:2018qox}. In ABJM
the same comparison would test the interpolating function $h(\la)$ itself, but no integrability computation of the
Bremsstrahlung functions exists, and it is not known in which form their result should appear as a function of $h$.

These gaps concern the nature of the Bremsstrahlung functions more than their values, and they lead to the questions that this paper answers:
\begin{enumerate}
\item \emph{What kind of function is $\Bhalf$?} In $\mathcal N=4$ SYM the Gaussian matrix model has one cut and its curve
is a sphere, while in ABJM the curve is a torus. The coupling and the circular Wilson loops are periods of the curve
\cite{Marino:2009jd,Drukker:2010nc,Klemm:2012ii}, and the topological-string free energies are quasimodular
\cite{Aganagic:2006wq,Haghighat:2008gw}. The Bremsstrahlung function instead is fixed by the phase of a loop, so is it a period too? Of which kind? What are its modular properties?
\item \emph{Do the wound loops have a closed form that can be continued to non-integer winding?} The bosonic Bremsstrahlung function needs this continuation near $n=1$. In $\mathcal N=4$ SYM it is immediate
since the winding only rescales the coupling, while in ABJM the derivatives in the winding and in the coupling do not
coincide.
\item \emph{Does $\Bsix$ have a
closed form at every coupling, and is it a function of the same kind?} The two lines are the ends of a defect
renormalisation-group flow \cite{Castiglioni:2022yes,Castiglioni:2023uus}, and the answer also says which of them radiates
more.
\item \emph{Can the latitudes be computed in the planar limit at every value of the deformation and of the coupling?}
They are the one-parameter family from which both Bremsstrahlung functions descend.
\item \emph{What do these quantities say
about the defect conformal field theory on the $\tfrac12$BPS line beyond the two-point function that defines $\Bhalf$?} In
$\mathcal N=4$ SYM the derivatives of the latitude are integrated correlators of the defect, and they are exact input for its
bootstrap \cite{Cavaglia:2022qpg,Cavaglia:2022yvv}.
\item \emph{What should an integrability computation of the Bremsstrahlung functions return
as a function of $h$?}
\end{enumerate}

We answer these questions with a single geometric object, the spectral curve of the planar matrix model. The eigenvalues of
the two nodes condense on two cuts, and the resolvent lives on a genus-one curve, the mirror curve of local
$\mathbb P^1\times\mathbb P^1$ on the slice selected by ABJM \cite{Aganagic:2002wv,Marino:2009jd}, whose modulus is related to
the 't~Hooft coupling by the mirror map. The planar loops wound
$n$ times are periods of the curve \cite{Marino:2009jd,Klemm:2012ii}. Our tool is the Gauss-Manin connection of these periods,
a linear system with coefficients rational in the modulus, together with the Picard-Fuchs equations that follow from it. The first step is to merge the extended special geometry language of \cite{Klemm:2012ii} with the framing relation
\cite{Bianchi:2014laa,Bianchi:2017svd} leading to our first results. The winding
derivative needs the extension of the connection to logarithmic periods and the continuation of the wound loops to complex
winding. The latitudes need the Fermi gas, in which they are loops with two charges
\cite{Bianchi:2018bke}, and we obtain their planar limit from a Ward identity of the gas.

\subsection*{Summary of results}

All the quantities above are periods of one curve, or elementary functions of its periods, and this is what answers the six
questions. The curve $\Sigma_\kk$ and the mirror map $\la(\kk)$ are
\begin{equation}
\Sigma_\kk:\quad u^2=\big(Z^2-i\kk Z+1\big)^2-4Z^2,\qquad
\la(\kk)=\frac{\kk}{8\pi}\;{}_3F_2\left(\genfrac{}{}{0pt}{}{\frac12,\ \frac12,\ \frac12}{1,\ \frac32}\,\Biggl|\,-\frac{\kk^2}{16}\right),
\label{eq:curveintro}
\end{equation}
with $\dk\la$ the period of the holomorphic differential $dZ/u$ on the cycle $A$ around one of the two cuts.

1. \emph{$\Bhalf$ is a period of the second kind}.
\begin{equation}
{\ \Bhalf=\oint_A\OmB=-\frac14\,(16+\kk^2)\,\partial_\kk^2\la
=\frac{\kk}{64\pi}\;{}_2F_1\left(\genfrac{}{}{0pt}{}{\frac12,\ \frac12}{2}\,\Biggl|\,-\frac{\kk^2}{16}\right)\ }
\label{eq:mainperiod}
\end{equation}
The differential $\OmB$ has poles without residues (second-kind) and with the
holomorphic differential it spans the cohomology of the torus: the Bremsstrahlung function fills the place that the coupling
leaves free. Its properties follow. As a period, it can be written with just complete elliptic integral $E$ and $K$, so the integral of the third kind of
\cite{Bianchi:2018scb} drops out. Its Picard-Fuchs equation is singular only where the curve degenerates. In the modular
parameter of the torus it is quasimodular of weight one and depth one. This fixes its expansion at strong coupling because the mirror map is quadratic in $\log\kk$ at strong coupling and the
operator in \eqref{eq:mainperiod} lowers the degree by one. The constant appearing in the expansion is the anomaly of the quasimodular form $\Eis2$ under S-duality, and the worldsheet instantons are integral $q$-series, whose prefactors are predictions for the string. The $B$-period
of $\OmB$ is a second, dual, function, and the duality that exchanges the two cusps maps the strong-coupling expansion into
the weak-coupling ones at the coupling $16/\kk$. Since $B=2h_w$ for superconformal lines in three dimensions
\cite{Lewkowycz:2013laa,Bachas:2025brp}, the one-point function of the stress tensor in the presence of the $\tfrac12$BPS line
is half a period of the second kind.

2. \emph{The wound loops in closed form}. The wound loops are periods at every winding, and
their closed form depends on $n$ only through the parameters of two hypergeometric functions as follows
\begin{equation}
\begin{split}
\vev{W^{1/6}_n}=\;&\frac{\kk}{8\pi\la}\,{}_4F_3\left(\genfrac{}{}{0pt}{}{\frac{1-n}2,\frac{1-n}2,\frac{1+n}2,\frac{1+n}2}{\frac12,\ 1,\ \frac32}\,\Biggl|\,-\frac{\kk^2}{16}\right)\\
&\qquad+\frac{i\,n^2\kk^2}{64\pi\la}\,{}_4F_3\left(\genfrac{}{}{0pt}{}{\frac{2-n}2,\frac{2-n}2,\frac{2+n}2,\frac{2+n}2}{\frac32,\ \frac32,\ 2}\,\Biggl|\,-\frac{\kk^2}{16}\right).
\end{split}
\label{eq:towerintro}
\end{equation}
This is the continuation that the matrix model selects and it is unique. At integer winding one of the
two series terminates, and the fermionic loop is a polynomial of degree $n$ in $\kk$ divided by $8\pi n\la$. Its expansion at
strong coupling is finite, and its leading term $e^{n\pi\sqrt{2\lah}}/8\pi n\la$ is the string wound $n$ times. At $n=1$ the
derivative is purely imaginary, $\partial_n\vev{W^{1/6}_n}|_{n=1}=4\pi i\,\Lcl\,\vev{W^{1/2}}$, and it defines the
\emph{winding velocity} $\Lcl$, which is the running mean of the coupling.

\begin{figure}[t]
\centering
\includegraphics[width=0.6\textwidth]{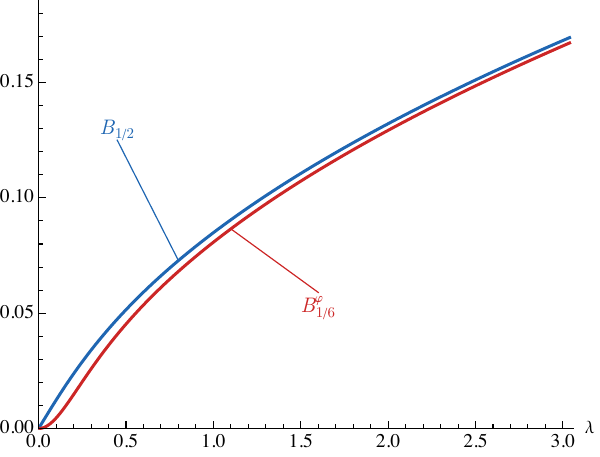}
\caption{The Bremsstrahlung functions of the fermionic $\tfrac12$BPS line and of the bosonic $\tfrac16$BPS line
\eqref{eq:mainperiod} and \eqref{eq:B16intro}, against the 't~Hooft coupling. The bosonic line radiates less than the
fermionic one at every coupling, and the two functions approach each other at strong coupling.}
\label{fig:Bintro}
\end{figure}

3. \emph{$\Bsix$ is a projection}. With the winding velocity, the bosonic function follows in
closed form,
\begin{equation}
{\ \Bsix=\frac{8\,\Bhalf\,\Lcl}{1+64\pi^2\Bhalf^2},\qquad
\Lcl=\la-4\Bhalf=\frac{\kk}{16\pi}\;{}_3F_2\left(\genfrac{}{}{0pt}{}{\frac12,\ \frac12,\ \frac12}{\frac32,\ 2}\,\Biggl|\,-\frac{\kk^2}{16}\right).\ }
\label{eq:B16intro}
\end{equation}
The bosonic function is a rational function of a period and of the flat coordinate, a projection rather than a period: it
obeys a nonlinear differential equation of third order instead of the linear second-order equation of a period, it is not
quasimodular, and its expansion at strong coupling does not stop, although the closed form resums it. The bosonic line
radiates less than the fermionic one at every coupling, $0<\Bsix<\Bhalf$ (figure~\ref{fig:Bintro}), by a relative amount that the drop of the defect
entropy along the flow bounds from both sides. The derivation extends to ABJ theory where as a byproduct we can extend also $B_{1/2}$ fixing the way in which the cusp anomalous dimension exponentiates for different ranks of the gauge groups.

4. \emph{The latitudes in closed form}. In the Fermi gas the latitude is a loop with two charges
\cite{Bianchi:2018bke}. A Ward identity of the gas, together with the wound loops, shows that in the planar limit these loops
are again periods of $\Sigma_\kk$, with the second coordinate of the curve in the integrand. The bosonic and the fermionic
latitudes follow at every $\nu$
\begin{equation}\begin{split}
\frac{\vev{W_B(\nu)}_\nu}{\vev{W^{1/2}}}=1+i\tan\PhB(\nu)&,\qquad
\frac{\vev{W_F(\nu)}_\nu}{\vev{W^{1/2}}}=1-\cot\frac{\pi\nu}{2}\,\tan\PhB(\nu),\\
\tan\PhB(\nu)=\frac{\nu\kk}{8}\;&{}_3F_2\left(\genfrac{}{}{0pt}{}{\frac12,\ 1-\frac\nu2,\ 1+\frac\nu2}{\frac32,\ 2}\,\Biggl|\,-\frac{\kk^2}{16}\right) .
\label{eq:latintro}
\end{split}\end{equation}
The deformation enters only
through the parameters, and not through a rescaling of the coupling as in $\mathcal N=4$ SYM \cite{Drukker:2006ga}. At
$\nu=1$, $\tan\PhB=8\pi\Bhalf$ and its derivative is $2\pi\Lcl$, the two data of \eqref{eq:B16intro}. At weak coupling
\eqref{eq:latintro} reproduces the three-loop latitude of \cite{Bianchi:2018bke}. At strong coupling the fermionic latitude is
exactly the sum of the two saddles of the string: the first is the Airy form of \cite{Bianchi:2018bke}, while the second carries the worldsheet instantons.

5. \emph{Integrated correlators of the defect}. The logarithmic derivatives of the fermionic latitude
$\Ccor_n=\partial_\nu^n\log\vev{W_F(\nu)}_\nu|_{\nu=1}$ are fixed by the integrated connected $2n$ points
correlators of the tilt operators on the line. The first is the Bremsstrahlung function
$\Ccor_1=4\pi^2\Bhalf$. The second is the integrated four-point function which the closed forms fix to be 
\begin{equation}
\Ccor_2=2\pi^2\Lcl-16\pi^4\Bhalf^2 .
\label{eq:C2intro}
\end{equation}
The winding velocity is therefore a datum of the defect conformal field theory, and \eqref{eq:B16intro} and \eqref{eq:latintro}
are exact constraints for the bootstrap of the $\tfrac12$BPS line at every coupling.

6. \emph{What integrability should compute}. Setting $\kk=4\sinh2\pi h$, a M\"obius map which changes the coordinates on $\Sigma_\kk$, the curve branch points are mapped to the ends of the cut $[-2h,2h]$ of the quantum spectral curve \cite{Gromov:2014eha}. In this frame every planar quantity above is an integral over the cut: $\Bhalf$ is the area
under the hyperbolic semicircle $\sqrt{\sinh^22\pi h-\sinh^2\pi u}$ in units of the $\tfrac12$BPS circle, the winding velocity
is its first moment in the rapidity. Then $\Bhalf$ is a single combination of two Legendre functions of $\cosh4\pi h$, where $\mathcal N=4$ SYM gives a ratio of Bessel functions. In the same coordinate the Picard-Fuchs equation of $\Bhalf$ becomes the zero-energy P\"oschl-Teller problem of the hyperbolic Calogero-Sutherland model.

The manuscript is organised as follows. 
We begin in section~\ref{sec:setup} by reviewing the two lines, the localisation matrix model with its spectral curve, and the
exact prescriptions for the Bremsstrahlung functions. In section~\ref{sec:deriv} we show that $\Bhalf$ is a period of the second
kind of this curve, whose Picard-Fuchs equation and modular properties explain why its strong-coupling expansion terminates. The
bosonic function needs one more ingredient, the winding velocity. Section~\ref{sec:logperiods} obtains it from the logarithmic
periods, as the derivative at unit winding of the wound loops, which we write in closed form at every complex winding. With it,
section~\ref{sec:bosonic} gives $\Bsix$ in closed form, compares the two lines along the defect flow and at strong coupling, and
extends the results to ABJ theory. Section~\ref{sec:latexact} turns to the latitudes, which are loops with two charges of the Fermi
gas, and reads their derivatives at the great circle as integrated correlators of the defect conformal field theory. We close with the discussion of section~\ref{sec:outlook} with what an integrability computation should reproduce and with the extension to finite $N$ and $k$.
The appendices collect the technical steps.

\section{Wilson loops and Bremsstrahlung functions in ABJM}
\label{sec:setup}

\subsection{The bosonic and the fermionic BPS Wilson loops}
\label{sec:sota:lines}

ABJM theory \cite{Aharony:2008ug} is the three-dimensional $\mathcal N=6$ superconformal
Chern-Simons-matter theory with gauge group $U(N)_k\times U(N)_{-k}$. It has two gauge
fields $A_\mu$ and $\hat A_\mu$ with Chern-Simons levels $k$ and $-k$, and four complex
scalars  and fermions $C_I$ and $\bar\psi^I$ ($I=1,\dots,4$) in the bifundamental
representation of the gauge group, transforming in the fundamental and in the antifundamental of the
$SU(4)_R$ R-symmetry respectively. Their conjugates $\bar C^I$ and $\psi_I$ are antibifundamental.
The superconformal group is $OSp(6|4)$ which contains 24
supercharges in addition to the bosonic subgroup $SU(4)_R\times Sp(4)$. The only parameters are the rank $N$ and the
integer level $k$ whose inverse plays the role of the coupling. In the planar limit $N,k\to\infty$ and quantities reorganise themselves in terms of the 't~Hooft coupling $\la=N/k$ with $\la$ fixed. With unequal ranks, the theory is known as ABJ \cite{Aharony:2008gk}, with couplings $\la_1=N_1/k$ and $\la_2=N_2/k$.

The theory supports several supersymmetric Wilson loop operators from which the Bremsstrahlung functions originate. One can divide them into two families: bosonic and fermionic ones. Writing them along a contour $\Gamma$ parametrised by $x^\mu(\tau)$ in Euclidean signature, we have:
\begin{itemize}
\item\emph{The bosonic loop.} The natural generalisation of the Maldacena-Wilson loop
\cite{Maldacena:1998im} couples the gauge field of one node to a bilinear in the scalars
\cite{Drukker:2008zx,Chen:2008bp,Rey:2008bh,Berenstein:2008dc},
\begin{equation}
W^{1/6}=\frac1N\Tr\,\mathcal P\exp\Big[-i\oint_\Gamma d\tau\Big(A_\mu\dot x^\mu
-\frac{2\pi i}{k}\,|\dot x|\,M_J{}^I\,C_I\bar C^J\Big)\Big],
\label{eq:W16def}
\end{equation}
with $\hat W^{1/6}$ the analogous loop for the second node, built from $\hat A_\mu$ and
$\hat M_J{}^I\bar C^JC_I$ with $\hat M=M$. 
On the straight line $M=\mathrm{diag}(-1,-1,1,1)$ and it
preserves four of the 24 supercharges, hence the name $\tfrac16$BPS. From a string theory point of view it has been interpreted as a string smeared over a $\mathbb{CP}^1\subset\mathbb{CP}^3$ \cite{Drukker:2008zx,Rey:2008bh}.
\item\emph{The fermionic loop.} The fundamental string in $AdS_4\times\mathbb{CP}^3$ ending on a line
preserves half of the supercharges. Its field-theory dual is the holonomy of a $U(N|N)$ superconnection that contains also fermions \cite{Drukker:2009hy} as follows
\begin{equation}
\begin{gathered}
W^{1/2}=\frac{1}{\mathrm{STr}\mathcal{T}}\mathrm{STr}\Big[\mathcal P\exp\Big(-i\oint_\Gamma d\tau\,\mathcal L\Big)\,
\mathcal T\Big],\\
\mathcal L=\begin{pmatrix} A_\mu\dot x^\mu-\frac{2\pi i}{k}|\dot x|M_J{}^IC_I\bar C^J &
-i\sqrt{\frac{2\pi}{k}}\,|\dot x|\,\eta_I\bar\psi^I\\[1mm]
-i\sqrt{\frac{2\pi}{k}}\,|\dot x|\,\psi_I\bar\eta^I &
\hat A_\mu\dot x^\mu-\frac{2\pi i}{k}|\dot x|\hat M_J{}^I\bar C^JC_I\end{pmatrix},
\end{gathered}
\label{eq:W12def}
\end{equation}
If $\Gamma$ is a straight line, $M=\hat M=\mathrm{diag}(-1,1,1,1)$ and 
commuting spinor couplings $\eta_I\propto\delta_I^1$. The
superconnection is invariant under half of the supercharges only up to a supergauge transformation. The constant twist supermatrix $\mathcal T$ is
what makes the traced holonomy invariant under such transformations. On the line, it turns the supertrace into a trace with $\mathrm{STr}\mathcal{T}=2N$. More general BPS loops exist: on arbitrary contours
\cite{Cardinali:2012ru}, one-parameter families of fermionic $\tfrac16$BPS loops \cite{Ouyang:2015iza,Ouyang:2015bmy,Mauri:2018fsf} and $\tfrac13$BPS loops \cite{Drukker:2022txy}. The last one deserves a special mention. Indeed, it settled the long-standing question of whether ABJM admits $\tfrac13$BPS Wilson loops as predicted from M-theory \cite{Drukker:2008jm}. It is constructed with a larger block supermatrix with multiple copies of the superconnection $\mathcal{L}$ in two different R-symmetry directions I. The string dual of $W^{1/3}$ is not known, but it has been suggested to be a superposition of two strings at different points of $\mathbb{CP}^3$ \cite{Drukker:2022txy}.
\end{itemize}

In general, the fermionic loop shares the supercharges of the bosonic one. As a consequence, the two are in the same cohomology class and when supported on a straight line or a circle we have
\begin{equation}
W^{1/2}=\tfrac12\big(W^{1/6}+\hat W^{1/6}\big)+\mathcal Q V ,
\label{eq:cohom}
\end{equation}
for a combination $\mathcal Q$ of the shared supercharges and some functional $V$ of the fields. 
The factor $\tfrac12$ reflects the normalisations $1/N$ and $1/2N$ of \eqref{eq:W16def} and \eqref{eq:W12def}. 
This relation is what makes the
fermionic loop computable by supersymmetric localisation, but in general it requires the \emph{framing} regularisation to hold. 
Wilson loops in a Chern-Simons theory are defined by
point splitting, and the result depends on the linking number $f$ of the contour with its displaced
copy. In pure Chern-Simons theory the dependence is a phase \cite{Witten:1988hf,Guadagnini:1989am}. Localisation on $S^3$
computes the loops at framing $f=1$, because the point splitting compatible with supersymmetry
displaces the circle along a Hopf fibre \cite{Kapustin:2009kz}. This is also the framing within which \eqref{eq:cohom} holds for the expectation values. 
In ABJM the framing dependence of the bosonic loop is still a phase
\begin{equation}
\vev{W^{1/6}}_f=e^{i\PhB(f)}\,|\vev{W^{1/6}}|
\label{eq:PhBdef}
\end{equation}
but it is no longer a one-loop effect, and $\Phi_B$ is a non-trivial function of $f$ and the 't~Hooft coupling.
For the fermionic loop the framing factor is argued to be $e^{i\pi f(N_1-N_2)/k}$ that is the one of the Chern-Simons theory with gauge supergroup $U(N_1|N_2)$. In ABJM, it becomes trivial since $N_1=N_2=N$ \cite{Drukker:2010nc}. 

A Wilson loop can be wrapped around multiple times defining new operators $W_n$ where $n$ is the winding number. Since the fermionic couplings are antiperiodic, the cohomological equivalence of the wound loops carries a relative sign $(-1)^n$ and it is given by
\begin{equation}
\vev{W^{1/2}_n}=\tfrac12\,i^{1-n}\left(\vev{W^{1/6}_n}-(-1)^n\vev{\hat W^{1/6}_n}\right),\qquad
\vev{\hat W^{1/6}_n}=\overline{\vev{W^{1/6}_n}}\,.
\label{eq:node}
\end{equation}
where the second relation holds at equal ranks, where the two nodes are exchanged by $k\to-k$.
With this overall normalisation, the wound $\tfrac12$BPS loop is real and for $n=1$ it reduces to
\begin{equation}\label{eq:W12reW16}
\vev{W^{1/2}}=\re\vev{W^{1/6}}
\end{equation}
where we used the shorthand notation $W=W_1$.

The phase in \eqref{eq:PhBdef} is entirely due to the framing, since the great circle is real at framing zero. This follows
from parity. The reflection through the plane of the circle reverses the sign of the Chern-Simons levels so, combined with the
exchange of the two gauge groups, it is a symmetry of ABJM \cite{Aharony:2008ug}. It maps the circle to itself, the loop
$W^{1/6}$ of the first node to the loop $\hat W^{1/6}$ of the second, and the framing $f$ to $-f$. At framing zero it gives
$\vev{W^{1/6}}_0=\vev{\hat W^{1/6}}_0$ and, since at equal ranks the two are complex conjugate as in \eqref{eq:node}, the
great circle at framing zero is real.

\subsection{The matrix model and its loops}
\label{sec:sota:mm}

Supersymmetric localisation on $S^3$ \cite{Kapustin:2009kz} reduces the ABJM partition function to the matrix integral
\begin{equation}
Z=\frac{1}{(N!)^2}\int\prod_{i}\frac{d\mu_i}{2\pi}\prod_j\frac{d\hat\mu_j}{2\pi}\;
\frac{\prod_{i<j}\big[2\sinh\tfrac{\mu_{ij}}{2}\big]^2\big[2\sinh\tfrac{\hat\mu_{ij}}{2}\big]^2}
{\prod_{i,j}\big[2\cosh\tfrac{\mu_i-\hat\mu_j}{2}\big]^2}\;
e^{\frac{ik}{4\pi}\left(\sum_i\mu_i^2-\sum_j\hat\mu_j^2\right)} ,
\label{eq:MM}
\end{equation}
with $\mu_{ij}=\mu_i-\mu_j$. The bosonic loops at framing one are given by the eigenvalue averages
\begin{equation}
\vev{W^{1/6}_n}=\Big\langle\frac1N\sum_ie^{n\mu_i}\Big\rangle,\qquad
\vev{\hat W^{1/6}_n}=\Big\langle\frac1N\sum_je^{n\hat\mu_j}\Big\rangle .
\label{eq:MMloops}
\end{equation}
while the fermionic one is given by \eqref{eq:node}. The model \eqref{eq:MM} is an analytic continuation
of the Chern-Simons matrix model on the lens space $L(2,1)$ \cite{Marino:2002fk,Aganagic:2002wv}, and it
was solved at planar level in \cite{Marino:2009jd}. The eigenvalues condense on two cuts (one for each node) and the
planar resolvent lives on a genus-one spectral curve which will play a central role throughout this manuscript. The solutions are organised around the modulus $\kk$ of that curve, related to $\la$ by the mirror map 
\begin{equation}
\la(\kk)=\frac{\kk}{8\pi}\;{}_3F_2\left(\genfrac{}{}{0pt}{}{\frac12,\ \frac12,\ \frac12}{1,\ \frac32}\,\Biggl|\,-\frac{\kk^2}{16}\right).
\label{eq:lamkappaintro}
\end{equation}
For the singly wound bosonic loop, the authors of \cite{Marino:2009jd} found the following combination that can be entirely expressed in terms of complete elliptic integrals
\begin{equation}
\dk\big(\la\,\vev{W^{1/6}}\big)=-\,\frac{\sqrt{ab}\,\big[\,b\,K(\tilde k)-(a+b)\,\Pi(\tilde n|\tilde k)\big]}{2\pi^2\,(1+ab)} ,
\label{eq:MP443}
\end{equation}
which is the ABJM slice of \cite{Drukker:2010nc} with
\begin{equation}
\tilde k^2=\frac{(a^2-1)(b^2-1)}{(1+ab)^2},\qquad \tilde n=\frac{1-a^2}{1+ab},
\end{equation}
\begin{equation}
a(\kk)=\frac12\left[(2+i\kk)-\sqrt{(2+i\kk)^2-4}\right],\qquad
b(\kk)=\frac12\left[(2-i\kk)-\sqrt{(2-i\kk)^2-4}\right].
\label{eq:defab}
\end{equation}
The quantities $a(\kk)$ and $b(\kk)$ \footnote{In our notation they are the inverse of the $a$, $b$ of \cite{Marino:2009jd}.} fix the endpoints of the two cuts, while $K$ and $\Pi$ are the complete
elliptic integrals of the first and third kind, with modulus $\tilde k$ and characteristic $\tilde n$.
The same combination for the fermionic loop follows from the cohomological equivalence
\eqref{eq:W12reW16}. The real part of \eqref{eq:MP443} is a constant, and gives
\begin{equation}
\dk\big(\la\vev{W^{1/2}}\big)=1/8\pi\,,
\label{eq:dW12MP}
\end{equation}
while the imaginary part, which carries the framing phase of the bosonic loop, is not.

Both lines admit a supersymmetric deformation along a latitude of a two-sphere in $S^3$ at an azimuthal angle $\theta_0$.
The matter couplings can also be rotated by an internal angle $\alpha$. The resulting loops \cite{Cardinali:2012ru} depend on the two angles only through
\begin{equation}
\nu=\sin2\alpha\,\cos\theta_0\;\in\,[0,1].
\label{eq:nudef}
\end{equation}
The bosonic latitude $W_B(\nu)$ is \eqref{eq:W16def} with scalar coupling depending on $\nu$. In general it preserves two
supercharges so it is $\tfrac1{12}$BPS. The fermionic latitude $W_F(\nu)$ is the deformation of \eqref{eq:W12def} with $\nu$-dependent matter couplings and twist. It preserves four supercharges so it is $\tfrac16$BPS. At $\nu=1$ they reduce to $W^{1/6}$ and $W^{1/2}$ on the great circle. 

In \cite{Bianchi:2014laa}, on the basis of their two-loop result, it was conjectured that the cohomological equivalence survives the deformation provided that the loops are evaluated at the non-integer framing $f=\nu$ as follows
\begin{equation}
\vev{W_F(\nu)}_\nu=\frac{e^{-\frac{i\pi\nu}2}\vev{W_B(\nu)}_\nu-e^{\frac{i\pi\nu}2}\vev{\hat W_B(\nu)}_\nu}
{e^{-\frac{i\pi\nu}2}-e^{\frac{i\pi\nu}2}} ,
\label{eq:cohomnu}
\end{equation}
where $\vev{W_B(\nu)}_\nu=e^{i\Phi_B(\nu)}\,|\vev{W_B(\nu)}|$. 
Notice that at $\nu=1$ relation \eqref{eq:cohomnu} becomes \eqref{eq:cohom}.
A matrix model for the latitude loops was proposed in
\cite{Bianchi:2018bke} on the basis of perturbative data and internal consistency. It computes the latitudes at framing $\nu$, as \eqref{eq:cohomnu} requires.
Its expansion agrees with a three-loop computation at generic framing, winding and representation \cite{Bianchi:2018bke}
and with a five-loop test \cite{Bianchi:2020cfn}. In \cite{Griguolo:2021rke} it was later derived from supersymmetric
localisation, with the framing $f=\nu$ as an input.

\subsection{The spectral curve and its periods}
\label{sec:curve}

In the planar limit, the eigenvalue integral \eqref{eq:MM} is dominated by a saddle-point configuration in which the eigenvalues condense onto a collection of cuts. The planar resolvent encodes the moments of the eigenvalue density and develops square-root branch points at the endpoints of these cuts. Therefore, it naturally defines a function on a two-sheeted cover of the complex plane, branched at the cut endpoints: the \emph{spectral curve}. For a generic configuration with $n$ cuts, the corresponding curve has genus $n-1$. In the case of ABJM \cite{Marino:2009jd,Drukker:2010nc}, there are two cuts, so the spectral curve is elliptic, or equivalently a torus.

Introducing the exponentiated eigenvalues $Z=e^{\mu}$ for the first node and $Z=-e^{\hat\mu}$ for the second, the curve takes the form
\begin{equation}
\Sigma_\kk:\qquad
u^2=P(Z;\kk)=\big(Z^2-i\kk Z+1\big)^2-4Z^2 ,
\label{eq:curve}
\end{equation}
which factorises as
\begin{equation}
P=Z^4-2i\kk Z^3-(\kk^2+2)Z^2-2i\kk Z+1
=\big(Z^2-(2+i\kk)Z+1\big)
\big(Z^2+(2-i\kk)Z+1\big).
\label{eq:curveexp}
\end{equation}
The four branch points occur in reciprocal pairs, $a,a^{-1},-b,-b^{-1}$, satisfying
\begin{equation}
a+a^{-1}=2+i\kk,\qquad
b+b^{-1}=2-i\kk.
\end{equation}
The parameter $\kk$ is related to the 't~Hooft coupling $\la$ through the map \eqref{eq:lamkappaintro}. Geometrically, $\kk$ is an algebraic coordinate on the moduli space of the curve, whereas $\la$ is the ``flat'' coordinate natural to the physical theory. More precisely, $\la$ is a period of the spectral curve.

\begin{figure}[!t]
    \centering   
    \includegraphics[width=\columnwidth]{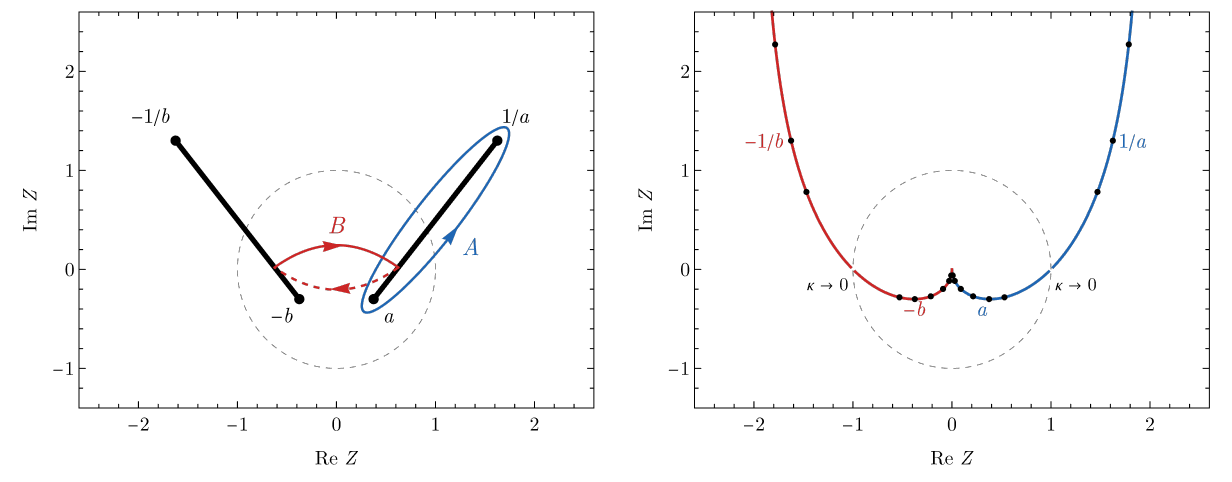}
    \caption{The spectral curve $\Sigma_\kk$. Left: branch points, cuts and the cycles A and B at $\kk=1$. The cycle B is drawn solid on the first sheet and dashed on the second. Right: motion of the branch points on the physical slice $0 < \kk < \infty$. The dots mark $\kk=1/2, 1, 2, 4, 8, 16$. At weak coupling the cuts shrink to $Z=\pm1$ ($A$ vanishes), while at strong coupling $a$ and $-b$ collide at $Z = 0$ ($B$ vanishes).}
    \label{fig:cuts}
\end{figure}

To make this statement precise, recall that a period associated with a meromorphic differential $\omega$ on $\Sigma_\kk$ is an integral
\begin{equation}
\Pi_\gamma(\kk)=\oint_\gamma\omega
\end{equation}
over a closed cycle $\gamma$ on the curve. As functions of $\kk$, periods satisfy linear differential equations with rational coefficients. They are multivalued only through the monodromy of the cycles around the points where the curve degenerates. A torus has two
independent cycles. On $\Sigma_\kk$ we take $A$, which encircles the cut $[a^{-1},a]$, and $B$, which runs from the first cut to the second on one sheet and back on the other (figure~\ref{fig:cuts}, left).
Integrals over them are the $A$- and the $B$-periods.

This cycle structure is the geometric origin of \emph{special geometry}. In a matrix model with several cuts, an $A$-period of the resolvent counts the eigenvalues supported on the corresponding cut and therefore determines the associated 't~Hooft coupling. The dual $B$-period gives the derivative of the planar free energy with respect to that coupling. In other words, it measures the cost of moving an eigenvalue from one cut to another. Thus the $A$-periods provide flat coordinates on the moduli space, while the $B$-periods are their conjugate coordinates, obtained as derivatives of a prepotential. This is the same special-geometric structure that underlies Seiberg-Witten theory \cite{Seiberg:1994rs}. In particular, the curve \eqref{eq:curve} exhibits the characteristic period and degeneration structure of the Seiberg-Witten curve of pure $SU(2)$.

The degenerating points of this geometry are identifiable from the discriminant of $P$ which is proportional to
\begin{equation}
\Dsc(\kk)=\kk^2\,(16+\kk^2)\,.
\label{eq:discdef}
\end{equation}
Then the singular points occur at $\kk=0$ and $\kk=\pm4i$ in addition to a degeneration in the strong-coupling limit at $\kk=\infty$. The corresponding limits have different geometric interpretations:
\begin{itemize}
\item \emph{Weak coupling}, $\kk=0$. The curve degenerates to
$P=(Z-1)^2(Z+1)^2$, so both cuts shrink to points, at $Z=1$ and $Z=-1$. The $A$-cycle collapses. Consequently, the period of a meromorphic differential around this cycle is governed by the residue at the collision point, while the dual $B$-period develops logarithmic behaviour.

\item \emph{Strong coupling}, $\kk=\infty$. The endpoints of the two cuts collide pairwise: $a$ and $-b$ approach $Z=0$, while $a^{-1}$ and $-b^{-1}$ approach $Z=\infty$. The two cuts merge in this limit, and the $B$-cycle shrinks.

\item \emph{Conifold points}, $\kk=\pm4i$. The curve becomes
$P(Z;\pm4i)=(Z\pm1)^2\big(Z^2\pm6Z+1\big)$.
At $\kk=4i$, the branch points $a$ and $a^{-1}$ collide at $Z=-1$. However, the vanishing cycle is not simply $A$. When the curve is analytically continued from the physical slice through the upper half of the $\kk$-plane, the branch point $-b$ crosses the segment $[a^{-1},a]$ and the vanishing cycle is therefore $2B-A$. At $\kk=-4i$, it is the second cut that collapses, at $Z=1$, and continuation through the lower half-plane gives the vanishing cycle $2B+A$.
\end{itemize}
The physical slice $\kk\in(0,\infty)$ does not encounter any of these degenerations at finite coupling since $\Dsc(\kk)>0$ there (figure~\ref{fig:cuts}, right). The conifold points are reached only by analytically continuing the coupling into the complex $\kk$-plane.

Since periods depend on the choice of cycle and on the branch of the differential, we now specify our conventions. For real $\kk>0$, we choose the branch-points to be the ones appearing in \eqref{eq:defab} where the square root is the principal one, so that $|a|<1$. The branch of $b$ is chosen similarly, giving $b=\bar a$. We take
\begin{equation}
u(Z)=
\sqrt{(Z-a)(Z-a^{-1})}\,
\sqrt{(Z+b)(Z+b^{-1})},
\label{eq:ubranch}
\end{equation}
with principal square roots for the individual factors. This choice is continuous along the straight segment from $a^{-1}$ to $a$. When $\kk$ is analytically continued away from the physical slice, the branches and cycles are continued along the chosen path. The overall sign of periods depends also on the orientation of the cycle on which they are computed. For the $A$ cycle, we fix this convention physically by requiring that the 't~Hooft coupling increase with $\kk$, namely $d\la/d\kk>0$ on the physical slice.

We can now make the relation between the coupling and the spectral curve explicit. Let $R_{\pm}^2=Z^2\pm(2\mp i\kk)Z+1$
with branches chosen consistently with \eqref{eq:ubranch}, so that $u=R_+R_-$. The planar resolvent can then be written as
\begin{equation}
\varrho_0(Z):=2\log\big[R_+(Z)-R_-(Z)\big].
\label{eq:resolvent}
\end{equation}
Its integral along the $A$-cycle gives the 't~Hooft coupling $\la$ through the following differential
\begin{equation}
\la=\frac1{8\pi^2}\oint_A\varrho_0\frac{dZ}{Z}.
\label{eq:lambdafed}
\end{equation}
This is the precise sense in which the coupling is a period of the curve. The corresponding quantity associated with the $B$-cycle is the dual coordinate $\la_D$ of \cite{Drukker:2010nc}, which determines the derivative of the planar free energy according to $\partial_\la F_0=4\pi^3\la_D$. Similarly to the coupling \eqref{eq:lamkappaintro}, it has a closed hypergeometric form \cite{Drukker:2010nc} that reads
\begin{equation}
\la_D=\frac{\log\kk}{2\pi}+\frac{1}{\pi\kk^2}\;{}_4F_3\left(\genfrac{}{}{0pt}{}{1,\ 1,\ \frac32,\ \frac32}{2,\ 2,\ 2}\,\Biggl|\,-\frac{16}{\kk^2}\right),
\label{eq:laD}
\end{equation}
in which the variable is $-16/\kk^2$ instead of $-\kk^2/16$. As in the $A$-cycle case, we chose the orientation of the contour in order to have $d\lambda/d\kk>0$. The two closed forms are expansions around opposite ends of the moduli space. The coupling is a power series in $-\kk^2/16$ analytic at weak coupling, where the $A$-cycle shrinks. The dual coupling $\la_D$ is, up to the logarithm, a power series in $-16/\kk^2$ around strong coupling, where the $B$-cycle shrinks. The series in \eqref{eq:laD} converges for $\kk>4$, but ${}_4F_3$ is analytic along the negative real axis, so that \eqref{eq:laD} holds on the whole physical slice, and $\la_D$ vanishes at $\kk=0$.

Although the resolvent $\varrho_0$ is logarithmic, its derivative with respect to the parameter $\kk$ is particularly simple
\begin{equation}
\dk\,\varrho_0(Z)=\frac{iZ}{u(Z)}.
\label{eq:loginsertion}
\end{equation}
The logarithmic dependence is thereby replaced by an algebraic differential. Moreover, the factor of $Z$ precisely cancels the $1/Z$ in \eqref{eq:lambdafed}, leaving the holomorphic differential $dZ/u$. Its two periods are
\begin{equation}
\oint_A\frac{dZ}{u}=-8\pi^2i\,\dk\la=-2i\,K\left(-\tfrac{\kk^2}{16}\right),\qquad
\oint_B\frac{dZ}{u}=4\pi^2\,\dk\la_D=\frac4\kk\,K\left(-\tfrac{16}{\kk^2}\right),
\label{eq:ABperiods}
\end{equation}
where $K(m)$ is the complete elliptic integral of the first kind. Thus $\la$ and $\la_D$ are primitives with respect to $\kk$ of the $A$- and $B$-periods of a single holomorphic differential, as the derivatives of \eqref{eq:lamkappaintro} and \eqref{eq:laD} confirm. This is precisely the structure expected from special geometry.

Let's focus on the $A$-period. We introduce the integrals
\begin{equation}
J_n(\kk):=\int_{a^{-1}}^{a}\frac{Z^n\,dZ}{u}=\frac{1}{2}\oint_A\frac{Z^n\,dZ}{u}
\label{eq:Jdef}
\end{equation}
where it was used that the integral around the cycle is twice the one along the cut. From the definition and \eqref{eq:ABperiods}, this implies
\begin{equation}
J_0=-4\pi^2 i\,\dk\la \,,
\label{eq:J0im}
\end{equation}
which is purely imaginary since we chose $\partial_{\kk}\lambda>0$.
This relation makes the distinction between the two notions of period transparent: $\la$ is the period of the logarithmic differential in \eqref{eq:lambdafed}, whereas $d\la/d\kk$ is proportional to the $A$-period of the holomorphic differential $dZ/u$.

We now derive the differential equations obeyed by the integrals $J_n$. The key observation is that, although the periods depend non-trivially on the parameter $\kk$, their derivatives can be expressed in terms of a finite basis of differentials, modulo exact ones. 
Indeed, the cycle can be held fixed while $\kk$ varies, so the derivative of a period is the period of the derivative of
the differential. Reducing the resulting differential to a basis modulo exact forms defines the \emph{Gauss-Manin connection}. In the present case, this reduction can be performed algebraically using $P$ and its derivatives, and it closes on the three differentials
$\left\{\frac{dZ}{u},\frac{Z\,dZ}{u},\frac{Z^2\,dZ}{u}\right\}$.
Equivalently, the polynomial reduction known as the \emph{Griffiths-Dwork method} (see appendix~\ref{sec:GM}) expresses the derivative of each representative as a rational linear combination of these three differentials, up to an exact differential.

The same reduction also explains why higher powers of $Z$ introduce no new independent periods. For every integer $\ell$, the function $Z^\ell u$ is regular along the cut, which does not contain $Z=0$, and vanishes at its endpoints, so that $\int_{a^{-1}}^ad(Z^\ell u)=0$. Since $d(Z^\ell u)=\big(2\ell Z^{\ell-1}P+Z^\ell P'\big)\,dZ/2u$, writing $P=\sum_{j=0}^4p_jZ^j$ as in \eqref{eq:curveexp}, with $p_0=p_4=1$, $p_1=p_3=-2i\kk$ and $p_2=-(\kk^2+2)$, this gives the five-term relation
\begin{equation}
\sum_{j=0}^{4}p_j(2\ell+j)\,J_{\ell-1+j}=0,
\qquad \ell\in\mathbb Z.
\label{eq:contig}
\end{equation}
This is a recurrence in the moment index, rather than a differential equation in $\kk$. Its origin is nevertheless the same reduction modulo exact differentials that underlies the Gauss-Manin connection. In the variable $x=Z+Z^{-1}$ the corresponding relation among the moments of $y\,dx$ is the recursion of Suyama \cite{Suyama:2025ydf}. In \cite{Suyama:2026nzo} it is extended to ABJ theory, where it is again a five-term relation, and used to fix the dependence of the local solutions on the winding.

The case $\ell=0$ is particularly useful, since the coefficient of $J_{-1}$ vanishes and the relation gives
\begin{equation}
J_3=\frac{i\kk}{2}J_0
+\frac{\kk^2+2}{2}J_1
+\frac{3i\kk}{2}J_2.
\end{equation}
For $\ell\geq1$, the coefficient $2\ell+4$ of the highest moment is non-zero. Hence, by induction, every $J_m$ with $m\geq0$ is a polynomial linear combination of $J_0,J_1,J_2$, with coefficients depending on $\kk$, for instance
\begin{equation}
\begin{split}
J_4&=-\frac{5 \kappa ^2+2}{6}J_0+\frac{i \kappa(5 \kappa ^2+16)}{6}    J_1+\frac{8-11 \kappa ^2}{6}  J_2\,,\\
J_5&=-\frac{i \kappa (13 \kappa ^2-2) }{12} J_0+\frac{12-\kappa ^2 (13 \kappa ^2+38)}{12} J_1-\frac{5 i \kappa  (5 \kappa ^2-14)}{12}  J_2\,,
\end{split}
\end{equation}
and so on. This is the finite-dimensionality underlying both constructions: the differentials $Z^m\,dZ/u$, $m\geq0$, span the algebraic de Rham cohomology of the elliptic curve with its two points at infinity removed. Its dimension is $2g+1$ that for genus $g=1$ is 3 \cite{Grothendieck:1966}.

Applying this reduction to \eqref{eq:Jdef}, we obtain
\begin{align}
\dk J_0
&=-\frac{\kk(\kk^2+4)}{\Dsc}J_0
-\frac{4i\kk^2}{\Dsc}J_1
+\frac{4\kk}{\Dsc}J_2,
\label{eq:GM0}\\
\dk J_1
&={\frac{3i\kk^2}{\Dsc}J_0
+\frac{\kk^3}{\Dsc}J_1
+\frac{i\kk^2}{\Dsc}J_2},
\label{eq:GM1}\\
\dk J_2
&={-\frac{\kk(\kk^2+4)}{\Dsc}J_0
+\frac{i\kk^2(\kk^2+12)}{\Dsc}J_1
+\frac{4\kk}{\Dsc}J_2},
\label{eq:GM2}
\end{align}
where $\Dsc$ is the discriminant \eqref{eq:discdef}. These equations exhibit a key feature of periods: although the individual $J_n$ are transcendental functions of $\kk$, their derivatives are governed by a first-order linear system with rational coefficients. The transcendental information is contained in the solutions, while the differential connection itself is algebraic. A similar reduction, organised in a different way, appears in \cite{Klemm:2012ii}.

Finally, it is useful to distinguish the three standard types of meromorphic differentials, which are classified by their poles. A differential of the \emph{first kind} is holomorphic, one of the \emph{second kind} has poles but zero residue at every pole, and one of the \emph{third kind} has poles with nonzero residues. Two facts follow from the definition. Adding a multiple of $dZ/u$ to a differential of the second kind gives another one
with the same poles, so a second-kind differential is fixed by its poles only up to a multiple of $dZ/u$. Exact
differentials $df$ are also of the second kind, and all their periods vanish. Up to exact differentials, the first and the
second kind together form the two-dimensional de Rham cohomology of the curve.

In our basis, $dZ/u$ is holomorphic and therefore of the first kind, while $Z\,dZ/u$ has simple poles at the two points at infinity, with residues $\mp1$, and is of the third kind. The differential $Z^2dZ/u$ has double poles at infinity, but also residues $\mp i\kk$, so that it is a mixture of the second and the third kind. Subtracting $i\kk$ times $Z\,dZ/u$ removes the residues and leaves
\begin{equation}
\omega=\left(Z^2-i\kk Z\right)\frac{dZ}{u},
\label{eq:defsecondkind}
\end{equation}
which has the same double poles at infinity and no residues, making it of the second kind. This is not the only possible choice, since $(Z^2-i\kk Z)\,dZ/u+c\,dZ/u$ is of the second kind for every $c$. The Bremsstrahlung differential $\OmB$ of section \ref{sec:differential} is, up to an overall factor and an exact differential, the member of this family with $c=-(\kk^2+4)/4$.

The Gauss-Manin system \eqref{eq:GM0}-\eqref{eq:GM2} already expresses the derivatives of the two planar loops quoted in section \ref{sec:sota:mm} through a single period, before any period is evaluated. Indeed, one can express the framed Wilson loop in terms of the period data. The planar one-wound bosonic loop is the first moment of the eigenvalue density \eqref{eq:MMloops} and, since on the curve the
density is the discontinuity of the resolvent across the cut, one can write \cite{Marino:2009jd,Drukker:2010nc}
\begin{equation}
\vev{W^{1/6}}=\frac{J_2-J_0}{4\pi^2\la}.
\label{eq:W1}
\end{equation}
The combination $J_2-J_0$ has a simple origin. The coupling \eqref{eq:lambdafed} is the zeroth moment of the
eigenvalue density of the first node. The loop averages over
the same eigenvalues, so $\la\vev{W^{1/6}}$ is the first moment, the same period with one more factor of $Z$. The resolvent is a logarithm, but after an integration by parts only its derivative
enters, and that derivative is algebraic giving rise to the combination of periods.

 The derivative of \eqref{eq:W1} in $\kk$ follows from the Gauss-Manin system alone. In the difference of
\eqref{eq:GM2} and \eqref{eq:GM0} the coefficients of $J_0$ and $J_2$ cancel, while that of $J_1$ does not. Hence, with \eqref{eq:W12reW16} for the fermionic loop,
\begin{equation}
\dk\big(\la\vev{W^{1/6}}\big)=\frac{i}{4\pi^2}\,J_1,\qquad
\dk\big(\la\vev{W^{1/2}}\big)=-\frac{1}{4\pi^2}\,\im J_1 .
\label{eq:dWJ1}
\end{equation}
The $\kk$-derivative of the framed loop is therefore a single period, that of the third-kind differential $Z\,dZ/u$.

\subsection{The three basic periods $J_0$, $J_1$, $J_2$}
\label{sec:threeperiods}

We now solve the Gauss-Manin system \eqref{eq:GM0}-\eqref{eq:GM2} in terms of the physical quantities. One of the three is already known by \eqref{eq:J0im}: $J_0$ is the holomorphic period, fixed directly by the derivative of the
coupling. This section shows that the other two are then fixed as well.

A linear system such as \eqref{eq:GM0}-\eqref{eq:GM2} generally has no constant solutions. When it does, it is usually a special
feature of the geometry, and the constant is a topological invariant. This quantity is the following linear combination
\begin{equation}
\mathcal Q(\kk):=J_1-\frac{i\kk}{4}\,J_0\,.
\label{eq:defQ}
\end{equation}
Differentiating with respect to $\kk$ and using the Gauss-Manin connection \eqref{eq:GM0}-\eqref{eq:GM1} it is straightforward to prove that $\mathcal Q'=0$, so that it does not depend on $\kk$. Notice that all three coefficients of $J_i$ in $\mathcal Q'$ vanish separately, which is a stronger statement than simply the vanishing derivative: the combination is annihilated by the connection identically, not by a cancellation
between the periods. Geometrically, $\mathcal Q$ is the residue pairing of the third-kind differential
$Z\,dZ/u$ with the cycle $A$, and residues do not move as the modulus varies. 

Since $\mathcal Q$ does not depend on $\kk$, one may compute it at the most convenient
coupling, for instance $\kk\to0$. There the branch points $a$ and $a^{-1}$ merge at $Z=1$, the cut shrinks to a
point, and the cycle $A$ can be taken to be a fixed small circle around $Z=1$, on which the curve
becomes simply $u=\pm(Z^2-1)$. The term proportional to $\kk J_0$ in \eqref{eq:defQ} vanishes in this limit because $J_0$
stays finite, while $J_1$ becomes a residue. Indeed, in this limit it reduces to the integral of $Z/(Z^2-1)$ that has a simple pole at $Z=1$ with residue $\tfrac12$, so one turn around it gives $J_1=- i\pi/2$ for $\kappa\rightarrow 0$. The sign is fixed by the branch \eqref{eq:ubranch} and the orientation of $A$. Hence $\mathcal Q=-i\pi/2$ at every coupling. This also explains why a bare $\pi$ appears although every coefficient of the connection is
rational: the flat combination is normalised by a residue, and residues come with $2\pi i$. Therefore, given the residue that fixes the flat invariant $\mathcal{Q}$, we can conclude that 
\begin{equation}
{\ J_1=\frac{i\kk}{4}J_0-\frac{i\pi}{2}.\ }
\label{eq:J1}
\end{equation}

With $J_0$ and $J_1$ in hand, the connection gives $J_2$ without a further integration. Indeed, solving \eqref{eq:GM1} for $J_2$ and substituting $\partial_{\kk}J_1$ with the derivative of \eqref{eq:J1}, we get
\begin{equation}
{\ J_2=J_0+\frac{\kk(16+\kk^2)}{4}\,\dk J_0+\frac{\pi\kk}{2}.\ }
\label{eq:J2J0}
\end{equation}
A similar result was obtained in different notation in \cite{Klemm:2012ii}.

Finally, since $J_0$ is given in terms of the derivative of the 't~Hooft coupling by \eqref{eq:J0im}, one can  express the three basic periods through $\{1,\dk\la,\partial_\kk^2\la\}$ with coefficients rational in $\kk$ as follows
\begin{align}
J_0&=-4\pi^2i\,\dk\la,\\
J_1&=\pi^2\kk\,\dk\la-\frac{i\pi}2,\label{eq:J1lam}\\
J_2&=-4\pi^2i\,\dk\la-i\pi^2\kk(16+\kk^2)\,\partial_\kk^2\la+\frac{\pi\kk}2 .
\end{align}
By \eqref{eq:contig}, the same holds for every $J_m$ with $m\ge0$
with coefficients that are polynomials in $\kk$. This representation is unique, because the three functions are linearly independent.

\subsection{The planar loops in closed form}
\label{sec:loops}

Once the Gauss-Manin system is solved, the derivatives \eqref{eq:dWJ1} can finally be written in terms of the coupling. Indeed, using \eqref{eq:J1}, we have 
\begin{equation}
\dk\big(\la\,\vev{W^{1/6}}\big)=\frac1{8\pi}+\frac{i\kk}{4}\,\dk\la ,\qquad
\dk\big(\la\vev{W^{1/2}}\big)=\frac{1}{8\pi}\,,
\label{eq:dW16}
\end{equation}
where we used that $\dk \lambda>0$, so it is purely real. These are the two results quoted in section \ref{sec:sota:mm}, namely \eqref{eq:MP443} and \eqref{eq:dW12MP}. Notice that, since the combination of $K(\tilde k)$ and $\Pi(\tilde n|\tilde k)$
found in \cite{Marino:2009jd} equals to \eqref{eq:dW16}, the complete integral of the third
kind is reducible! Indeed, in the derivative with respect to the coupling only the elliptic integral of the first kind contributes, as follows
\begin{equation}
\dk\big(\la\,\vev{W^{1/6}}\big)=
\frac{1}{8 \pi }+\frac{i \kappa}{16 \pi ^2}  K\left(-\frac{\kappa ^2}{16}\right)
\end{equation}
Neither derivative \eqref{eq:dW16} needs the explicit form of $J_2$: only the difference of two rows of the connection and the flat invariant enter.

However, since we computed the period $J_2$ in terms of the coupling, we can exploit \eqref{eq:W1} directly to compute the exact closed form of the loops instead of their derivative. Indeed, substituting \eqref{eq:J2J0} into the bosonic loop definition and using \eqref{eq:W12reW16}, we have 
\begin{equation}
\vev{W^{1/6}}=\frac{\kappa}{8 \pi \lambda}-i\frac{\kappa}{4\lambda}(16+\kappa^2)\partial^2_\kappa\lambda\qquad
\text{and}\quad\vev{W^{1/2}}=\text{Re}\vev{W^{1/6}}=\frac{\kappa}{8 \pi \lambda}
\label{eq:Wexact}
\end{equation}
where the conifold factor $16+\kappa^2$ of the discriminant naturally appears. The same quantities can be derived integrating \eqref{eq:dW16} directly. Considering that $\lambda=0$ for $\kk=0$ while the Wilson loops are normalized to 1, the combination $\lambda\vev{W}$ vanishes at $\kk=0$ and then the integration is well defined with no unknown constants. For the fermionic Wilson loop we obtain the same relation, while for the bosonic one this procedure gives the following integral representation\footnote{The same formula was derived in \cite{Klemm:2012ii}, but the printed version does not agree with our \eqref{eq:W16planar}. In their normalisation, where the planar term is $g_sN\vev{W^{1/6}}$ with $g_s=2\pi i/k$, it should read $\tfrac{i\kk}{4}-\tfrac\pi2\int_0^\kk\kk'\,\partial_{\kk'}\la\,d\kk'$, while the printed right-hand side is real for real $\kk$ and grows like $\kk^2\log^2\kk$, against their (2.60).}
\begin{equation}
\vev{W^{1/6}}=\frac{\kk}{8\pi\la}+\frac{i}{4\la}\int_0^{\kk}\kk'\,\partial_{\kk'}\la(\kk')\,d\kk'\ .
\label{eq:W16planar}
\end{equation}
This fact has an interesting consequence. We recall the definition of the phase of the framed loop defined in section \ref{sec:sota:lines}. Localisation computes the loop at framing $f=1$, so in \eqref{eq:PhBdef} $\vev{W^{1/6}}_{f=1}~\!\!=~\vev{W^{1/6}}$ and $\PhB\equiv\PhB(1)$. The cohomological equivalence
\eqref{eq:W12reW16} identifies the $\tfrac12$BPS loop with the real part $\vev{W^{1/2}}=|\vev{W^{1/6}}|\cos\PhB$, where we wrote the exponential in polar coordinates. Since $\vev{W^{1/2}}>0$ on the physical slice from \eqref{eq:Wexact}, $\cos\PhB\neq0$, so
the modulus can be traded for the $\tfrac12$BPS loop such that
\begin{equation}
\vev{W^{1/6}}=\vev{W^{1/2}}\,\big(1+i\tan\PhB\big).
\label{eq:W16W12tanphi}
\end{equation}
Thus the imaginary part of the loop measures the framing phase.
Applied to \eqref{eq:W16planar} and \eqref{eq:J2J0}, this gives the phase in two closed forms
\begin{equation}
\tan\PhB=\frac{2\pi}{\kk}\int_0^\kk\kk'\,\partial_{\kk'}\la\,d\kk'=-2\pi\,(16+\kk^2)\,\partial_\kk^2\la .
\label{eq:tanPhB}
\end{equation}
Their equality is a differential equation for the coupling, integrated once, and it is worth making this equation explicit, since it governs much of what follows. The Gauss-Manin system \eqref{eq:GM0}-\eqref{eq:GM2} is first order in three unknowns. The flat invariant \eqref{eq:defQ} has removed one of them, and \eqref{eq:J2J0} expresses $J_2$ through $J_0$ and its first derivative. Inserting \eqref{eq:J2J0} into the last row \eqref{eq:GM2}, with $J_1$ from \eqref{eq:J1}, it closes the system on $J_0$ alone. The constants $\pi\kk/2$ and $-i\pi/2$ cancel identically, and what is left for the holomorphic period $g:=\dk\la=iJ_0/4\pi^2$ is the second-order equation
\begin{equation}
\kk(16+\kk^2)\,g''+(16+3\kk^2)\,g'+\kk\,g=0 .
\label{eq:PFg}
\end{equation}
This is the \emph{Picard-Fuchs equation} of the curve: the scalar linear differential equation in the modulus satisfied by its periods, obtained from the Gauss-Manin system by eliminating all but one unknown, and whose order equals the rank of the cohomology, two for a torus. Its coefficients are polynomials in $\kk$, and the discriminant \eqref{eq:discdef} stands in front of the highest derivative: the equation is singular exactly where the curve degenerates, at $\kk=0$, at the conifold points $\kk=\pm4i$ and at $\kk=\infty$, and nowhere else. It is the equation of the ABJM slice found in \cite{Marino:2009jd,Drukker:2010nc,Suyama:2025ydf}, and we have obtained it from the algebraic connection rather than from a special-function representation. The two are of course consistent: in the variable $x=-\kk^2/16$, \eqref{eq:PFg} is the hypergeometric equation $x(1-x)\,\partial_x^2g+(1-2x)\,\partial_xg-\tfrac14\,g=0$, whose solution regular at $x=0$ is ${}_2F_1\left(\genfrac{}{}{0pt}{}{\frac12,\ \frac12}{1}\,\Biggl|\,x\right)=\tfrac2\pi K(x)$, as \eqref{eq:ABperiods} requires. The second solution carries $\log\kk$: it is the $B$-period of $dZ/u$, which grows logarithmically when both cuts shrink. 

\subsection{The Bremsstrahlung functions}
\label{sec:sota:brem}

Two rays meeting at a point with Euclidean external angle $\varphi$ define a cusped line, whose expectation value has a logarithmic divergence governed by the cusp anomalous
dimension $\vev{W_{\rm cusp}}\sim e^{-\Gamma_{\rm cusp}(\varphi)\log(L/\epsilon)}$, with $L$ and $\epsilon$
infrared and ultraviolet cutoffs \cite{Polyakov:1980ca,Korchemsky:1987wg}. For loops that couple to
matter one can also rotate the couplings on one ray by an internal angle $\theta$, which defines the
generalised cusp $\Gamma_{\rm cusp}(\varphi,\theta)$ \cite{Drukker:1999zq,Drukker:2011za}. In a conformal
theory the small angle behaviour of the cusp anomalous dimension defines the Bremsstrahlung function $B$, which has three physical interpretations \cite{Correa:2012at}: 
\begin{itemize}
    \item it is the energy radiated by a slowly moving heavy probe  $\Delta E=2\pi B\int dt\,\dot v^2$;
\item  it normalises the two-point function of the displacement operator on the straight line
$\vev{\!\vev{\mathbb D_i(t)\,\mathbb D_j(0)}\!}=C_D\,\delta_{ij}/t^4$, so that $C_D=12B$;
\item it is the first-order response of
a loop to a deformation of its contour or internal couplings.
\end{itemize}
In $\mathcal N=4$ SYM, the authors of \cite{Correa:2012at} showed that
\begin{equation}
B_{\mathcal N=4}=\frac{1}{2\pi^2}\,\la\partial_\la\log\vev{W},\qquad\text{with}\qquad
\vev{W}=\frac{2}{\sqrt\la}I_1(\sqrt\la) .
\label{eq:BN4}
\end{equation}
where the expectation value of the circular Wilson loop was computed in \cite{Erickson:2000af,Drukker:2000rr,Pestun:2007rz}.
The same function follows from the near-BPS limit of the boundary TBA of the cusp
\cite{Correa:2012hh,Drukker:2012de,Gromov:2012eu}, and the comparison of the two confirms that in
$\mathcal N=4$ the coupling of integrability is the 't~Hooft coupling itself. For $\mathcal N=2$
superconformal theories see \cite{Fiol:2015spa,Bianchi:2018zpb}, and for the defect CFT on the $\mathcal N=4$ line \cite{Cooke:2017qgm,Giombi:2017cqn}.

In ABJM both bosonic and fermionic lines define a generalised cusp \cite{Griguolo:2012iq,Aguilera-Damia:2014qgy}.
For the fermionic cusp made by two $\tfrac12$BPS lines the configurations with $\theta=\pm\varphi$ are still BPS, so a single function
controls both angles, while for the bosonic one there is no BPS configuration and two functions are needed as follows
\begin{equation}
\Gamma_{1/2}(\varphi,\theta)\simeq\big(\theta^2-\varphi^2\big)\,\Bhalf,\qquad
\Gamma_{1/6}(\varphi,\theta)\simeq\theta^2\,\Bsixth-\varphi^2\,\Bsix .
\label{eq:Bdefs}
\end{equation}
The cusps themselves are known at two loops \cite{Griguolo:2012iq,Castiglioni:2023tci}, in the ladder
limit \cite{Bonini:2016fnc}, at three loops at $\varphi=0$
\cite{Bianchi:2017svd} (see also \cite{Preti:2017fjb}), and, for small angles, at one loop at strong
coupling \cite{Forini:2012bb,Aguilera-Damia:2014bqa}. On the integrability side, an open spin chain with all-loop
reflection matrices describes insertions into the $\tfrac12$BPS line, and a $Y$-system for the cusp
reproduces its one-loop value \cite{Correa:2023lsm} (see also \cite{Correa:2026akb}). The analogue of
the $\mathcal N=4$ derivation of $B$ from the cusp TBA, which would test $h(\la)$ directly, is not yet available.
It is interesting to mention also the cusp made by $\tfrac13$BPS rays since it has a richer structure. Indeed, it breaks three
inequivalent sets of R-symmetry generators leading to
\begin{equation}
\Gamma_{1/3}(\varphi,\{\theta,\theta',\theta''\})\simeq\theta^2B^\theta_{1/3}+\theta'^2B^{\theta'}_{1/3}+\theta''^2B^{\theta''}_{1/3}
-\varphi^2B^\varphi_{1/3},
\label{eq:B13defs}
\end{equation}
As for the other lines, each function
is the normalisation of a two-point function on the line.

Each of the three functions appearing in \eqref{eq:Bdefs} has been related to a BPS loop, while the four functions of the $\tfrac13$BPS loop are fixed by $\Bhalf$. In particular:
\begin{itemize}[leftmargin=1.6em]
\item The radiated energy fixes the coefficient $h_w$ of the one-point function of the stress tensor in the presence of the
line $\Bsix=2h_w$. Conjectured in
\cite{Lewkowycz:2013laa}, this is equivalent to $C_D=24h_w$, the 3D case of the relation between the norm of the displacement
operator and the stress-tensor one-point function \cite{Bianchi:2015liz,Bianchi:2016xvf}. It has been recently proved for every 3D superconformal line in \cite{Bachas:2025brp}. Since $h_w$ was originally obtained by localisation on the squashed
sphere which at first order in the squashing is the winding derivative of the wound loop, we have
\begin{equation}
\Bsix=\frac1{4\pi^2}\,\partial_n\log\big|\vev{W^{1/6}_n}\big|\,\Big|_{n=1},
\label{eq:LM}
\end{equation}
The proposal was tested at weak and strong coupling and agrees with the two-loop cusp \cite{Griguolo:2012iq,Bianchi:2014laa} and with the classical and one-loop
string \cite{Aguilera-Damia:2014qgy,Aguilera-Damia:2014bqa}.
\item By analogy with $\mathcal N=4$ SYM, the $\tfrac12$BPS function is the latitude derivative of the fermionic latitude \cite{Bianchi:2014laa}
\begin{equation}
\Bhalf=\frac1{4\pi^2}\,\partial_\nu\log\vev{W_F(\nu)}\,\Big|_{\nu=1},
\label{eq:B12lat}
\end{equation}
checked against the two- and three-loop cusp \cite{Griguolo:2012iq,Bianchi:2017svd} at weak coupling and at leading order at strong coupling \cite{Aguilera-Damia:2014qgy}. An exact derivation of this formula was provided in \cite{Bianchi:2017ozk} treating the $\tfrac12$BPS line as a superconformal defect with $su(1,1|3)$ symmetry.
\item For the internal angle of the bosonic line the analogous statement
\begin{equation}
\Bsixth=\frac1{4\pi^2}\,\partial_\nu\log\big|\vev{W_B(\nu)}\big|\,\Big|_{\nu=1},
\label{eq:B16lat}
\end{equation}
was derived in \cite{Aguilera-Damia:2014qgy} and tested at two loops in \cite{Bianchi:2014laa}. The modulus, which removes
framing-independent imaginary terms that appear starting from three loops, was introduced in \cite{Bianchi:2018bke,Bianchi:2018scb}.
From a four-loop computation of $\Bsixth$ \cite{Bianchi:2017afp,Bianchi:2017ujp}, it was conjectured that
\begin{equation}
\Bsixth=\tfrac12\,\Bsix ,
\label{eq:BthetaBphi}
\end{equation}
and later proved in \cite{Bianchi:2018scb}. Combined with \eqref{eq:LM}, it identifies the latitude with the winding expressing the winding number as a function of the latitude angle $n(\nu)$ \cite{Bianchi:2018scb}.
\item For the $\tfrac13$BPS loop no new prescription is needed \cite{Drukker:2022txy}. Indeed, analysing the $su(1,1|2)$ Ward identities, one obtains that all the Bremsstrahlung functions are proportional to $\Bhalf$
\begin{equation}
B^\varphi_{1/3}=B^\theta_{1/3}=B^{\theta'}_{1/3}+B^{\theta''}_{1/3}=\Bhalf,\qquad
B^{\theta'}_{1/3}=\frac{n_1}{n_1+n_4}\,\Bhalf,\qquad B^{\theta''}_{1/3}=\frac{n_4}{n_1+n_4}\,\Bhalf ,
\label{eq:B13rel}
\end{equation}
where $n_{1,4}$ are the numbers of times the superconnection $\mathcal{L}$ defined in \eqref{eq:W12def} appears in the $\tfrac13$BPS block superconnection in the directions $I=1$ and $I=4$.
\end{itemize}
A direct consequence of these relations is that, among the large family of Bremsstrahlung functions in ABJM, only $\Bhalf$ and $\Bsix$ seem to be independent. For this reason, from now on we concentrate just on these two quantities. From the definition \eqref{eq:B12lat}, using the cohomological equivalence \eqref{eq:cohomnu}, $\Bhalf$ splits into
two terms \cite{Bianchi:2014laa}
\begin{equation}
\Bhalf=\frac1{4\pi^2}\Big[\partial_\nu\log\big(\vev{W_B(\nu)}_\nu+\vev{\hat W_B(\nu)}_\nu\big)\Big|_{\nu=1}
+\frac\pi2\tan\PhB\Big].
\label{eq:B12split}
\end{equation}
The first term vanishes. This was conjectured in \cite{Bianchi:2014laa} by trading the latitude for the
winding and later proved as an exact property of the
latitude matrix model for any $\nu$ \cite{Bianchi:2018bke}. 
The second one is the tangent of the phase of the expectation value of the bosonic latitude on the maximal circle ($\nu=1$)
as computed by localisation ($f=1$). Thus, we have
\begin{equation}
\Bhalf=\frac{1}{8 \pi}\tan\Phi_B=-\frac{i}{8\pi}\,\frac{\vev{W^{1/6}}-\vev{\hat W^{1/6}}}{\vev{W^{1/6}}+\vev{\hat W^{1/6}}}=\frac{1}{8 \pi}\frac{\im \vev{W^{1/6}}}{\re \vev{W^{1/6}}}
\label{eq:framing}
\end{equation}
In other words, the modulus of the framed bosonic loop carries no information about $\Bhalf$, while its phase
is precisely $\Bhalf$. Using this, the authors of \cite{Bianchi:2018scb} obtained the planar $\Bhalf$ in closed form, as a combination of complete elliptic integrals of the
first, second and third kind with algebraic coefficients in $\kk$.

Written for the modulus and the phase, the
vanishing of the first term of \eqref{eq:B12split} gives the following relation \cite{Bianchi:2018scb}
\begin{equation}
\Bsix=\frac4\pi\,\Bhalf\,\partial_\nu\Phi_B(\nu)\big|_{\nu=1}
=\frac{1}{2\pi^2}\,\tan\Phi_B(\nu)\,\partial_\nu\Phi_B(\nu)\big|_{\nu=1},
\label{eq:BPVrel}
\end{equation}
In conclusion, the two Bremsstrahlung functions are not truly independent. Both are determined by the phase of the bosonic latitude at framing $\nu$. However, while its value on the great circle $\Phi_B=\Phi_B(1)$, is fixed by $\Bhalf$, its dependence on $\nu$ was known only perturbatively at weak coupling, up to three loops \cite{Bianchi:2018scb}, and at strong coupling through the Airy form of the fermionic latitude \cite{Bianchi:2018bke}. Consequently, no closed form for $\Bsix$ was available. In the rest of the manuscript we will show how to compute the exact closed form for both the Bremsstrahlung functions and we will show also how to obtain the exact bosonic phase reducing both quantities to a single exact function.

\section{The fermionic Bremsstrahlung functions from the curve}
\label{sec:deriv}

\subsection{\texorpdfstring{$\Bhalf$}{B1/2} is the second-kind period of the curve}
\label{sec:mainthm}

The framing relation \eqref{eq:framing} trades the Bremsstrahlung function for the phase of the planar bosonic
loop at framing one. This phase is not an independent piece of data. In section~\ref{sec:loops} it was obtained
directly from the periods of the spectral curve, using only the Gauss-Manin connection
\eqref{eq:GM0}-\eqref{eq:GM2}, its flat invariant \eqref{eq:defQ} with the residue value \eqref{eq:J1}, and the
dictionary \eqref{eq:W1} between the framed loop and the periods. The two closed forms \eqref{eq:tanPhB} of
$\tan\PhB$ can therefore be inserted directly into \eqref{eq:framing}, and they give the Bremsstrahlung function
exactly in terms of a double derivative of the coupling as follows
\begin{equation}
\Bhalf(\kk)=-\frac14\,(16+\kk^2)\,\partial_\kk^2\la(\kk)
\label{eq:main}
\end{equation}
or equivalently, in terms of the integral 
\begin{equation}
\Bhalf(\kk)=\frac{1}{4\kk}\int_0^\kk \kk'\,\partial_{\kk'}\la(\kk')\,d\kk' .
\label{eq:Bint}
\end{equation}
where their equivalence is the sign of the underlying Picard-Fuchs equation \eqref{eq:PFg} integrated once, as shown at the end of section~\ref{sec:loops}.

It is instructive to see how the Bremsstrahlung function sits inside the periods themselves. By \eqref{eq:W1} the
framed loop is the period combination $J_2-J_0$, its real part is the $\tfrac12$BPS loop \eqref{eq:Wexact}, and by
\eqref{eq:framing} its imaginary part measured in units of the real one is $8\pi\Bhalf$. Therefore, the two loops are related by \eqref{eq:W16W12tanphi}, which in terms of the periods reads
\begin{equation}
J_2-J_0=\frac{\pi\kk}{2}\,\big(1+8\pi i\,\Bhalf\big),\qquad\Rightarrow\qquad
\Bhalf=\frac{1}{4\pi^2\kk}\,\im\big(J_2-J_0\big).
\label{eq:BJ}
\end{equation}
In this form the content of \eqref{eq:main} is transparent. The identity \eqref{eq:J2J0} writes $J_2-J_0$ as the
sum of two terms of a quite different nature: the residue constant $\pi\kk/2$ of the flat invariant and $\tfrac14\kk(16+\kk^2)\,\dk J_0$, the derivative of the holomorphic period with respect to
the modulus. Since $J_0=-4\pi^2 i\,\dk\la$ is purely imaginary on the physical slice, the two terms live in
different components. The real one, which is a residue and therefore elementary, is the $\tfrac12$BPS loop, while
the imaginary one, the genuinely transcendental part of the period, is the Bremsstrahlung function. This also
explains the factor $16+\kk^2$ in \eqref{eq:main}. The Gauss-Manin connection \eqref{eq:GM1} has the conifold factor $16+\kk^2$ of the discriminant
\eqref{eq:discdef} in its denominator, and solving it for $J_2$ moves this factor in front of $\dk J_0$ in
\eqref{eq:J2J0}. It vanishes at the conifold points $\kk=\pm4i$, where a cycle of the curve shrinks.

\begin{figure}[t]
\centering
\includegraphics[width=\textwidth]{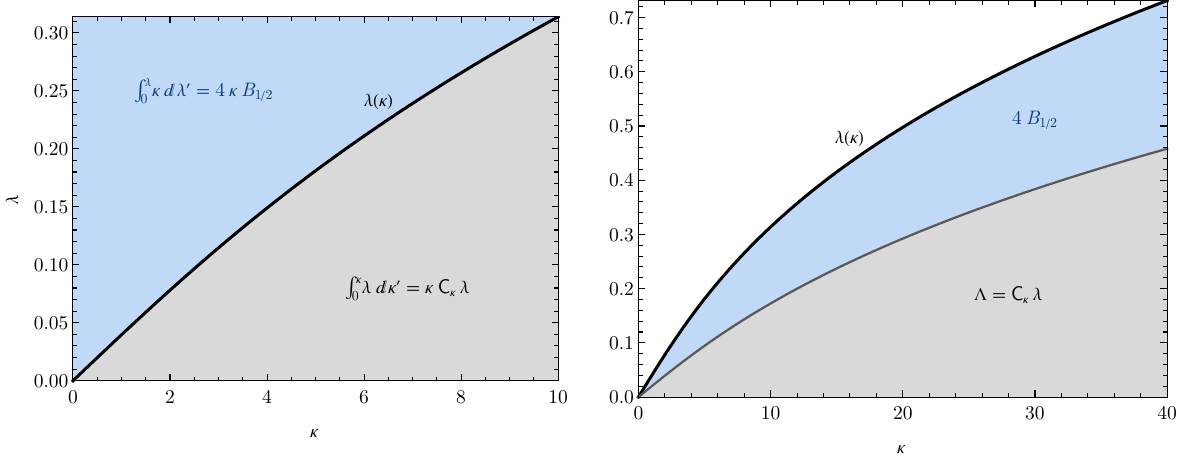}
\caption{The Bremsstrahlung function as an area. Left: at $\kk=10$, the mirror map $\la(\kk)$ divides the rectangle of
sides $\kk$ and $\la$ into the region below the curve, of area $\kk\,\Cop_\kk\la$ (grey), and the region to its left, of
area $4\kk\Bhalf$ (blue), as in \eqref{eq:meanbody} and \eqref{eq:BW12}. Right: the same splitting at every coupling,
divided by $\kk$. The coupling is the sum of its running mean $\Cop_\kk\la$, which is the winding velocity $\Lcl$ of
section~\ref{sec:B12wind} (grey), and of $4\Bhalf$ (blue).}
\label{fig:area}
\end{figure}

Integrating \eqref{eq:Bint} by parts, since the
boundary term at $\kk'=0$ vanishes because $\la=\kk/8\pi+O(\kk^3)$, one obtains
\begin{equation}
\Bhalf=\frac{\la}{4}-\frac{1}{4\kk}\int_0^\kk\la(t)\,dt=\frac14\,(1-\Cop_\kk)\,\la ,
\label{eq:meanbody}
\end{equation}
where $\Cop_x$ denotes the running mean of a function over $[0,x]$ defined as $(\Cop f)(x):=\tfrac1x\int_0^{x}f(t)\,dt$.
The Bremsstrahlung function is a quarter of the amount by which the 't~Hooft coupling exceeds its own running mean (figure~\ref{fig:area}).
In particular $\Bhalf>0$ at every coupling, since $\la$ grows with $\kk$ and therefore exceeds its mean. Through \eqref{eq:main}, the same positivity says that the mirror map is concave $\partial_\kk^2\la<0$. The splitting
$\la=4\Bhalf+\Cop_\kk\la$ is more than a rewriting: the running mean of the coupling is itself an observable, the winding
velocity $\Lcl$ of section~\ref{sec:B12wind}, which enters the bosonic Bremsstrahlung function and the integrated
four-point function of the defect.

The integral representation of $B_{1/2}$ can also be written as an integral over $\lambda$. Indeed, using $d\lambda'=\partial_{\kk'}\lambda(\kk')d\kk'$ and inverse map $\kk'=\kk(\lambda')$, equation \eqref{eq:Bint} becomes
\begin{equation}
\Bhalf=\frac{1}{4\kk}\int_0^\la\kk(\la')\,d\la'=
\frac{1}{4\la\vev{W^{1/2}}}\int_0^\la \la'\,\vev{W^{1/2}}\,d\la' ,
\label{eq:BW12}
\end{equation}
where we used $\kk=8\pi\la\vev{W^{1/2}}$ \eqref{eq:Wexact}.
The planar Bremsstrahlung function of the $\tfrac12$BPS line is thus determined by the $\tfrac12$BPS circle as a
function of the 't~Hooft coupling. This is the ABJM counterpart of the relation between the Bremsstrahlung function and the circular loop of $\mathcal N=4$ super Yang-Mills theory \cite{Correa:2012at}:
\begin{equation}
\text{$\mathcal{N}=4$ SYM}:\;B=\frac{1}{2\pi^2}\frac{\la\partial_\la\vev{W}}{\vev{W}},\qquad\qquad
\text{ABJM}:\;B_{1/2}=\frac{1}{4}\frac{\Cop_\la[\la\vev{W^{1/2}}]}{\vev{W^{1/2}}} \,.
\end{equation}
In $\mathcal N=4$ the relation holds at finite $N$, while in ABJM we can only establish it in the planar limit. Confronting \eqref{eq:meanbody} with \eqref{eq:BW12}, one recognises that the two running means are Legendre transforms of each other, so $\kk$ and $\la$ are conjugate variables, $\partial_\kk(\kk\,\Cop_\kk\la)=\la$ and $\partial_\la(\la\,\Cop_\la\kk)=\kk$.

Equation \eqref{eq:main} says that the Bremsstrahlung function is no more complicated than the derivative of the
coupling. Since $\dk\la$ is the holomorphic period \eqref{eq:ABperiods}, it can be written as a complete elliptic integral that is a Gauss hypergeometric in disguise. Then, \eqref{eq:main} gives $\Bhalf$ in closed form,
\begin{equation}
\Bhalf
=\frac{(16+\kk^2)\,K\left(-\tfrac{\kk^2}{16}\right)-16\,E\left(-\tfrac{\kk^2}{16}\right)}{16\pi^2\kk}
=\frac{\kk}{64\pi}\;{}_2F_1\left(\genfrac{}{}{0pt}{}{\frac12,\ \frac12}{2}\,\Biggl|\,-\frac{\kk^2}{16}\right)\,.
\label{eq:BPVreduced}
\end{equation}
The first exact expression of the planar $\Bhalf$ was obtained in \cite{Bianchi:2018scb} from the same framing
relation, evaluated with the generating function of \cite{Klemm:2012ii} for the moments of the eigenvalue density. It looks
quite different from \eqref{eq:BPVreduced}, but it is the same function, as we show below. 

In terms of the branch points $a$
and $b$ of section~\ref{sec:curve} it reads
\begin{equation}
\Bhalf=\frac{i}{8\pi}\Big[\,1+c_1\,K(m_*)-c_2\,E(m_*)-c_3\,\Pi(n_*,m_*)\,\Big],
\label{eq:BPV616}
\end{equation}
with modulus and characteristic
\begin{equation}
m_*=\frac{(1-a^2)(1-b^2)}{(1+ab)^2}=1-\nu^4,\qquad
n_*=\frac{(a^2-1)\,b}{a\,(1+ab)},\qquad
\nu:=\frac{2\sqrt{ab}}{1+ab}=\frac{\sqrt{16+\kk^2}-\kk}{4},
\label{eq:BPVmn}
\end{equation}
and coefficients\footnote{The $a$ and $b$ of \cite{Bianchi:2018scb} are our $a^{-1}$ and $b^{-1}$. After this
substitution we have simplified with $a+a^{-1}=2+i\kk$ and $b+b^{-1}=2-i\kk$, which give $(a+b)(1+ab)=4ab$ and
$\kk\nu=2(1-\nu^2)$; before simplification, for instance,
$c_1=4a\,(3ab^2+a^2b+b-a)/\pi\sqrt{ab}\,(a-b)(a^2b^2-1)$. For $\kk>0$ one has $b=\bar a$ and $\sqrt{ab}=|a|$.}
\begin{equation}
c_1=\frac{2\nu}{\pi}\Big(\frac1b+\frac{i\kk}{4}\Big)-\frac{2i\,(1+\nu^4)}{\pi\kk\nu},\qquad
c_2=-\frac{4i}{\pi\kk\nu},\qquad
c_3=\frac{8\nu}{\pi(1+ab)} .
\label{eq:BPVc}
\end{equation}
The modulus $m_*$ has a simple geometric meaning. A M\"obius transformation of $Z$ that sends three of the four branch
points to $0$, $1$ and $\infty$ brings $\Sigma_\kk$ to the Legendre form $y^2=x(x-1)(x-m)$: the fourth branch point lands at
$x=m$, which is therefore a cross-ratio of the four branch points, and the periods become complete elliptic integrals of
modulus $m$. Grouping the branch points by cut, and using $(a+b)(1+ab)=4ab$, the cross-ratio is
\begin{equation}
\frac{(a+b)(a^{-1}+b^{-1})}{(a+b^{-1})(a^{-1}+b)}=\frac{(a+b)^2}{(1+ab)^2}
=\nu^4=\left(\frac{\sqrt{16+\kk^2}-\kk}{\sqrt{16+\kk^2}+\kk}\right)^{2},
\label{eq:crossratio}
\end{equation}
so that $m_*=1-\nu^4$ is a Legendre modulus of $\Sigma_\kk$ itself (relabelling the branch points maps $m$ to $1-m$,
$1/m$ and their combinations). The formula of \cite{Bianchi:2018scb} is therefore written with the elliptic integrals of the
curve, while the modulus $-\kk^2/16$ of \eqref{eq:BPVreduced} belongs to a curve related to $\Sigma_\kk$ by the quadratic
transformation used below (section~\ref{sec:modular} relates both to the modular parameter).

At first sight \eqref{eq:BPV616} has little in common with \eqref{eq:BPVreduced}. Its coefficients are irrational
functions of $\kk$, its reality is not manifest, and it contains the complete elliptic integral of the third kind.
However, the curve explains it term by term: \eqref{eq:BPV616} is the framing relation in which the flat invariant has not
been used. Indeed, \eqref{eq:GM0} gives $J_2-J_0=\tfrac14\kk(16+\kk^2)\,\dk J_0+i\kk\,\mathcal Q$, with
$\mathcal Q=J_1-\tfrac{i\kk}{4}J_0$ the flat combination \eqref{eq:defQ}, and inserted in the framing relation
\eqref{eq:BJ} this gives
\begin{equation}
\Bhalf=\frac{i}{8\pi}\Big[\,1-\frac{16+\kk^2}{2\pi}\,\dk J_0-\frac{2i}{\pi}\,\mathcal Q\,\Big].
\label{eq:Bframed}
\end{equation}
Now, writing the periods in Legendre normal form as follows
\begin{equation}
J_0=-i\nu\,K(m_*),\qquad
J_1=i\nu\Big[\frac{K(m_*)}{b}-\frac{4\,\Pi(n_*,m_*)}{1+ab}\Big],
\label{eq:JLegendre}
\end{equation}
it is evident that the third-kind integral enters only through the flat combination
\begin{equation}
-\frac{2i}{\pi}\,\mathcal Q=\frac{2\nu}{\pi}\Big(\frac1b+\frac{i\kk}{4}\Big)K(m_*)-c_3\,\Pi(n_*,m_*),
\end{equation}
and the second-kind integral only through the derivative of the holomorphic period
\begin{equation}
-\frac{16+\kk^2}{2\pi}\,\dk J_0=-\frac{2i}{\pi\kk\nu}\Big[(1+\nu^4)\,K(m_*)-2E(m_*)\Big].
\end{equation}
The two terms add up to $c_1K(m_*)-c_2E(m_*)-c_3\Pi(n_*,m_*)$ with the coefficients \eqref{eq:BPVc}, so that
\eqref{eq:Bframed} is \eqref{eq:BPV616}. Therefore, the formula of \cite{Bianchi:2018scb} differs from \eqref{eq:main} by the term
$\tfrac{i}{8\pi}(1-\tfrac{2i}{\pi}\mathcal Q)$ which vanishes upon using that $\mathcal{Q}=-\tfrac{i\pi}{2}$ from section \ref{sec:threeperiods}. This is equivalent to rewriting the third-kind period as
\begin{equation}
c_3\,\Pi(n_*,m_*)=1+\frac{2\nu}{\pi}\Big(\frac1b+\frac{i\kk}{4}\Big)K(m_*) ,
\label{eq:Pired}
\end{equation}
making \eqref{eq:BPV616} collapse to
\begin{equation}
\Bhalf=\frac{(1+\nu^4)\,K(m_*)-2E(m_*)}{4\pi^2\kk\,\nu} ,
\label{eq:Bmstar}
\end{equation}
which is manifestly real. Finally, the quadratic transformation
\begin{equation*}
K\left(-\frac{\kk^2}{16}\right)=\nu\,K(m_*),\qquad
E\left(-\frac{\kk^2}{16}\right)=\frac{E(m_*)+\nu^2K(m_*)}{2\nu}
\end{equation*}
turns \eqref{eq:Bmstar} into \eqref{eq:BPVreduced}.

\subsection{The Bremsstrahlung differential}
\label{sec:differential}

The derivative in \eqref{eq:main} has a direct geometric meaning. With $J_0=-4\pi^2i\,\dk\la$ from
\eqref{eq:J0im}, the Bremsstrahlung function is the derivative of the holomorphic period with respect to the modulus up to the conifold factor $16+\kk^2$.
Since the cycle can be held fixed while $\kk$ varies, the derivative of the period is the period of the derivative so that the Bremsstrahlung function is an $A$-period
\begin{equation}
\Bhalf=\oint_A\OmB,\qquad\text{with}\qquad
\OmB:={-\frac{i(16+\kk^2)}{32\pi^2}}\,\dk\!\left(\frac{dZ}{u}\right),
\label{eq:OmBdef}
\end{equation}
where performing the derivative we obtain
\begin{equation}
\OmB=
{\frac{16+\kk^2}{32\pi^2}}\,\frac{Z(Z^2-i\kk Z+1)dZ}{u^{3}} .
\label{eq:OmBclosed}
\end{equation}
To check that $\OmB$ is of the second kind, it has to have poles but no residues. Since the numerator is a polynomial, at finite $Z$ \eqref{eq:OmBclosed} can be singular only where $u$ vanishes, i.e. at the four branch points
$a,a^{-1},-b,-b^{-1}$. Take, for instance, the branch point $a$ with $Z=a+t^2$. The differential becomes
\begin{equation}
\OmB={\frac{16+\kk^2}{32\pi^2}}\,\frac{a^2}{2\,(a^2-1)^{3/2}}\,\frac{dt}{t^{2}}+\dots.
\end{equation}
The expansion is even in $t$ so no term $t^{-1}dt$ can appear. In conclusion, the
pole at $Z=a$ is double with zero residue. The argument applies unchanged at the other branch points. At infinity it is regular. Then, the first derivative of the coupling is a period of the first-kind differential $dZ/u$, and
the Bremsstrahlung function, which involves one more derivative, is a period of a second-kind one.

The differential $\OmB$ can be written in the basis of section \ref{sec:curve} up to an exact
differential. The Gauss-Manin system gives the decomposition explicitly that reads
\begin{equation}\begin{split}
\OmB&=-\frac{i}{8\pi^2\kk}\,\big(Z^2-i\kk Z\big)\frac{dZ}{u}+\frac{i(\kk^2+4)}{32\pi^2\kk}\,\frac{dZ}{u}+d\tilde{\omega},
\label{eq:OmBZbasis}
\end{split}\end{equation}
with $\tilde{\omega}={-\frac{i(16+\kk^2)}{32\pi^2}}\,\frac{\beta_0}{u}=\frac{\kk-2i(\kk^2+2)\,Z+7\kk\,Z^2+4i\,Z^3}{32\pi^2\kk\,u}$ computed with $\beta_0$ of appendix \ref{sec:GM}.
Since $d\tilde{\omega}$ is exact, its integral over any closed cycle vanishes and it does not contribute to any period. Up to it, $\OmB$ is the combination of the first two terms. 
Both terms are needed and they play different roles. The first carries the poles: it is the second-kind differential \eqref{eq:defsecondkind}, and it is what makes $\Omega_B$ of the second kind. The second is holomorphic, so that it changes neither the poles nor the residues, but it fixes the period. Indeed, we have 
\begin{equation}
\oint_A\big(Z^2-i\kk Z\big)\frac{dZ}{u}=8\pi^2i\kk\,\Bhalf-2\pi^2i\,(\kk^2+4)\,\dk\la ,
\label{eq:Z2period}
\end{equation}
so that the first term gives the Bremsstrahlung function plus a holomorphic part that is cancelled by the $A$-period of the second term by \eqref{eq:ABperiods}. 
Since the coefficient of the first term does not vanish at any finite coupling, $dZ/u$ and $\OmB$ form a basis of $H^1_{\rm dR}(\Sigma_\kk)$, the space of differentials on the torus up to exact ones. The
$A$-periods of the two basis elements are, up to normalisation, $\dk\la$ and $\Bhalf$. The Bremsstrahlung function is the period that completes the holomorphic one into a period basis of the
curve. The second element could equally be normalised as $\partial_\kk^2\la$, which is, up to a constant, the $A$-period of
$\dk(dZ/u)$, but $\Bhalf$ is the natural choice. Indeed $\partial_\kk^2\la=-4\Bhalf/(16+\kk^2)$ diverges at the conifold
points $\kk=\pm4i$ only because of the factor $1/(16+\kk^2)$ that the derivative brings in, and the prefactor $16+\kk^2$ of
$\OmB$ in \eqref{eq:OmBdef} removes it: $\Bhalf$ is finite at every finite coupling, including the conifold, where
$\Bhalf(4i)=i/4\pi^2$ by \eqref{eq:BPVreduced}.

The same class can be built from the two planar loops which have the following differentials
\begin{equation}
\Omega_{1/6}:=\frac{(Z^2-1)\,dZ}{8\pi^2u},\qquad
\Omega_{1/2}:=\frac{\kk\,(4iZ+\kk)\,dZ}{32\pi^2u},
\label{eq:Om16Om12}
\end{equation}
with $A$-periods given by $\la\vev{W^{1/6}}$ and $\la\vev{W^{1/2}}$. Both have residues $\mp i\kk/8\pi^2$ at the two points over $Z=\infty$ which cancel in the following difference
\begin{equation}
\Omega_{1/6}-\Omega_{1/2}=\frac{1}{8\pi^2}\,\big(Z^2-i\kk Z\big)\frac{dZ}{u}-\frac{\kk^2+4}{32\pi^2}\,\frac{dZ}{u}
=i\kk\,\OmB^{\rm loop},
\label{eq:OmBloop}
\end{equation}
where $\OmB^{\rm loop}=\OmB-d\tilde{\omega}$ collects just the first two terms of \eqref{eq:OmBZbasis}. Up to the exact
differential $d\tilde{\omega}$, $\OmB$ is therefore the difference of the two loop differentials divided by $i\kk$.
Taking $A$-periods of \eqref{eq:OmBloop} gives $i\kk\Bhalf=\la\vev{W^{1/6}}-\la\vev{W^{1/2}}$, which is the polar
decomposition \eqref{eq:BJ}. Therefore, the identity $\OmB=\OmB^{\rm loop}$ up to exact differential contains the formula \eqref{eq:main}. Writing it for differentials has an important advantage: it holds on
every cycle, not only on $A$.

In particular $\OmB$ has a $B$-period which defines a second function of the coupling. Since the
$B$-period of $dZ/u$ is given by \eqref{eq:ABperiods}, the $B$-period of $\OmB$ follows from
\eqref{eq:OmBdef} with the dual coordinate in place of the coupling. We define the
dual Bremsstrahlung function by
\begin{equation}
\oint_B\OmB=\frac{i}{2}\,\BhalfD,\qquad \text{with}\qquad
\BhalfD:=-\frac14\,(16+\kk^2)\,\partial_\kk^2\la_D
\label{eq:OmBperiods}
\end{equation}
where the factor $i/2$ is the normalisation of the dual coordinate relative to the cycle $B$,
fixed by $\partial_\la F_0=4\pi^3\la_D$. The holomorphic differential has it as well as shown in \eqref{eq:ABperiods}.
So $(\Bhalf,\BhalfD)$ is the period vector of $\OmB$, exactly as $(\dk\la,\dk\la_D)$ is the period vector of $dZ/u$. The dual Bremsstrahlung $\BhalfD$ is not an independent observable: it is the same differential integrated
over the other cycle.

The two period vectors are not independent either. Continuing the coupling around a degeneration shifts the cycles
and the two vectors are transformed by the same matrix. Their antisymmetric
pairing is unchanged
because the shift preserves the intersection number of the cycles. It is therefore a single-valued function of the
coupling, and its value is a relation between the four periods that holds on every branch. On a torus this value is
given by the Riemann bilinear relation. For a holomorphic differential and a differential of
the second kind, it gives $2\pi i$ times the sum of the residues. For $dZ/u$ and $\Omega_{1/6}-\Omega_{1/2}$ the only poles are
at the two points at infinity, each of which contributes $1/8\pi^2$, so that
\begin{equation}
\oint_A\frac{dZ}{u}\,\oint_B\big(\Omega_{1/6}-\Omega_{1/2}\big)
-\oint_B\frac{dZ}{u}\,\oint_A\big(\Omega_{1/6}-\Omega_{1/2}\big)=\frac{i}{2\pi}
\label{eq:looplegendre}
\end{equation}
at every coupling. With \eqref{eq:ABperiods}, \eqref{eq:OmBloop} and \eqref{eq:OmBperiods} this is equivalent to
\begin{equation}
\Bhalf\,\dk\la_D-\BhalfD\,\dk\la=-\frac{1}{8\pi^3\kk},
\label{eq:Blegendre}
\end{equation}
so at every coupling and on every branch, the Bremsstrahlung function and its dual are tied to the two periods of the coupling by a constant of the curve. 

\subsection{The Picard-Fuchs equation of \texorpdfstring{$\Bhalf$}{B1/2}}
\label{sec:ode}
\label{sec:sing}
\label{sec:hyp}
\label{sec:weak}
\label{sec:conifold}
\label{sec:modularloc}
\label{sec:term}
\label{rem:termlit}
\label{rem:24}
\label{sec:wronskian}

The closed form \eqref{eq:BPVreduced} gives $\Bhalf$ at every coupling, but its behaviour as a function on the moduli
space, which form it takes there and how it continues around the degenerations of the curve is
read most directly from the linear differential equation that it obeys. This is the Picard-Fuchs equation of $\OmB$
which all its periods obey, so that it describes $\Bhalf$ and its dual $\BhalfD$ at once.

The equation follows from the Picard-Fuchs equation \eqref{eq:PFg} of the holomorphic periods. Using \eqref{eq:main} or \eqref{eq:OmBperiods}, we obtain
\begin{equation}
\dk\la=4\Big(\dk\Bhalf+\frac{\Bhalf}{\kk}\Big)\qquad\text{and}\qquad
\dk\la_D=4\Big(\dk\BhalfD+\frac{\BhalfD}{\kk}\Big),
\label{eq:gB}
\end{equation}
that is nothing else than \eqref{eq:Bint} differentiated  so that the coupling and the Bremsstrahlung function determine each other, suggesting also the integral representation for the dual Bremsstrahlung as follows
\begin{equation}
\BhalfD(\kk)=\frac{1}{\pi^2\kk}+\frac{1}{4\kk}\int_0^\kk \kk'\,\partial_{\kk'}\la_D(\kk')\,d\kk' .
\label{eq:BDint}
\end{equation}
Differentiating \eqref{eq:gB} once more it gives
\begin{equation}
\kk^2(16+\kk^2)\,B''+\kk(16+\kk^2)\,B'-16\,B=0 ,\quad\text{i.e.}\quad
(16+\kk^2)\,(\kk\dk)^2B=16\,B .
\label{eq:ode}
\end{equation}
which is the Picard-Fuchs of $\OmB$ where the 2 solutions are the 2 periods of it.

\begin{figure}[t]
\centering
\includegraphics[width=\textwidth]{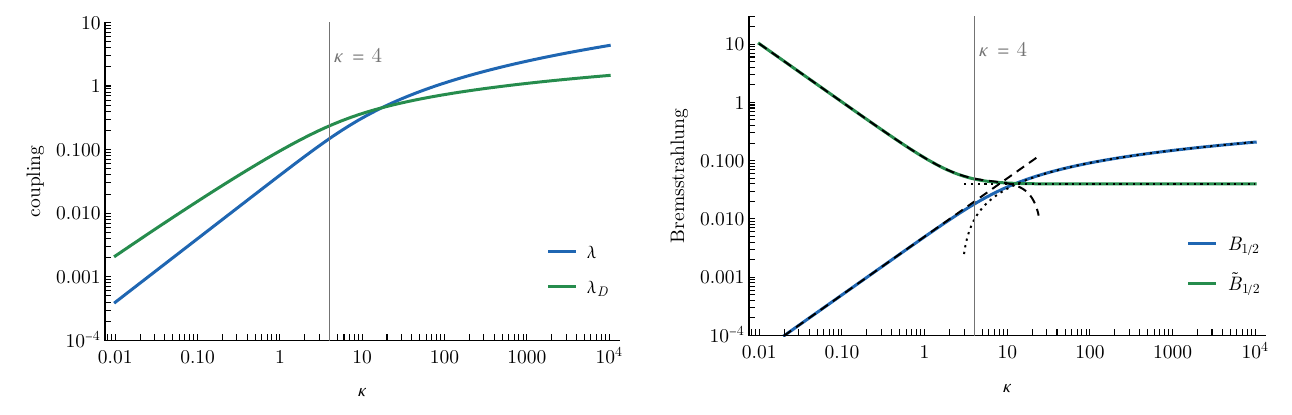}
\caption{The periods of the curve on the physical slice. Left: the 't~Hooft coupling $\la$ and the dual coupling $\la_D$, the
$A$- and $B$-periods of the differential \eqref{eq:lambdafed}. Right: the Bremsstrahlung function $\Bhalf$ and the dual one
$\BhalfD$ of \eqref{eq:dual}, the $A$- and $B$-periods of the Bremsstrahlung differential $\OmB$ of \eqref{eq:OmBdef}. Dashed and
dotted lines are the leading terms of the weak- and strong-coupling expansions. For $\Bhalf$ the dotted line,
$(\log\kk-1)/4\pi^2$, is exact up to worldsheet instantons. The
vertical line marks the self-dual point $\kk=4$, fixed by the duality $\kk\mapsto16/\kk$ that exchanges the two cycles, which
separates the regions where the weak- and strong-coupling series converge.}
\label{fig:periods}
\end{figure}

The leading coefficient of \eqref{eq:ode} is the discriminant \eqref{eq:discdef} of the curve, so
that its singular points are the degenerations of section \ref{sec:curve}: weak coupling $\kk=0$, the conifold points
$\kk=\pm4i$ and strong coupling $\kk=\infty$. In $\kk$ there are four singular points, but \eqref{eq:ode} depends on $\kk$ only through $\kk^2$
by the symmetry $\kk\to-\kk$ of the curve which exchanges the two conifold points. In the variable $z=-\kk^2/16$ the
singular points become three, and with $B=\tfrac{\kk}{64\pi}F(z)$, \eqref{eq:ode} becomes Gauss's hypergeometric
equation
\begin{equation}
z(1-z)\,F''+\big(2-2z\big)F'-\tfrac14\,F=0\,.
\label{eq:gauss}
\end{equation}
Its singular points $z=0,1,\infty$ are weak coupling,
the conifold and strong coupling, and its solution is \eqref{eq:BPVreduced}. However, equation \eqref{eq:gauss} has a second independent solution that develops a logarithmic divergence. Which is which is decided by the geometry. Indeed, the period on the shrinking cycle is the solution without logarithm, while the period on any other cycle acquires a logarithm. By section \ref{sec:curve}, the cycle that shrinks is $A$ at weak coupling, $2B\mp A$ at $\kk=\pm4i$, and $B$ at strong
coupling. So the $A$-period $\Bhalf$ is the solution without logarithm at weak coupling, while the
$B$-period $\BhalfD$ is the one at strong coupling (figure~\ref{fig:periods}).

The solutions without logarithm are written with complete elliptic integrals of the modulus that vanishes at their
singular point. The dual function follows
from $\dk\la_D$ of \eqref{eq:ABperiods} and reads
\begin{equation}
\BhalfD=\frac{1}{4\pi^2}\,E\left(-\frac{16}{\kk^2}\right)
=\frac{1}{8\pi}\;{}_2F_1\left(\genfrac{}{}{0pt}{}{-\frac12,\ \frac12}{1}\,\Biggl|\,-\frac{16}{\kk^2}\right).
\label{eq:dual}
\end{equation}
where $E$ is the complete elliptic integral of the second kind.
This is the modulus we use for the dual function. It is a power series in $1/z$, it vanishes at strong coupling and it is
the solution of \eqref{eq:gauss}. 

At the conifold $\kk=4i$ the shrinking cycle
is $2B-A$, whose period is $i\BhalfD-\Bhalf$. The vanishing modulus is $1+\kk^2/16=1-z$, such that
\begin{equation}
i\BhalfD-\Bhalf=\frac{i}{16\pi^2\kk}\Big[\kk^2\,K\Big(1+\frac{\kk^2}{16}\Big)+16\,E\Big(1+\frac{\kk^2}{16}\Big)\Big].
\label{eq:Bvanish}
\end{equation}
with $\BhalfD$ continued from the physical slice through $\text{Re}\kk > 0$.
This is a check of the two closed
forms. Indeed, since $K(0)=E(0)=\pi/2$ the bracket vanishes at $\kk=4i$. At the same time, \eqref{eq:BPVreduced} gives $\Bhalf(4i)=i/4\pi^2$ and \eqref{eq:dual} gives
$\BhalfD(4i)=1/4\pi^2$, so that the left-hand side vanishes as well.

The two solutions are tied together by their Wronskian $W_B$. The Abel's identity for the equation \eqref{eq:ode} gives $W_B'/W_B=-1/\kk$ so that
\begin{equation}
W_B(\kk):=\Bhalf\,\dk\BhalfD-\BhalfD\,\dk\Bhalf=\frac{C_B}{\kk}
\label{eq:WB}
\end{equation}
for a constant $C_B$. Applying \eqref{eq:gB} immediately fixes the constant to be $C_B=-1/32\pi^3$ since it relates the Wronskian to \eqref{eq:Blegendre}, so we have
\begin{equation}
\Bhalf\,\dk\BhalfD-\BhalfD\,\dk\Bhalf=-\frac{1}{32\pi^{3}\kk},
\label{eq:CB}
\end{equation}
the Wronskian of the Picard-Fuchs equation is the Riemann bilinear relation of the curve.

At weak coupling \eqref{eq:ode} reduces to $(\kk\dk)^2B=B$. The solution regular there is
$\Bhalf$, which vanishes with the coupling. Writing $\Bhalf=\sum_{n\ge0}a_n\kk^{2n+1}$, \eqref{eq:ode} gives the
two-term recursion
\begin{equation}
a_n=-\frac{(2n-1)^2}{64\,n(n+1)}\,a_{n-1}\quad(n\ge1),\qquad a_0=\frac1{64\pi},
\label{eq:recursion}
\end{equation}
with $a_0$ fixed by \eqref{eq:main}. Since $\la$
is odd in $\kk$, and so is its inverse, the weak-coupling expansion of $\Bhalf$ is a power series in the odd powers of
the 't~Hooft coupling as well
\begin{equation}
\Bhalf=\frac{\la}8-\frac{\pi^2\la^3}{48}+\frac{\pi^4\la^5}{60}-\frac{841\,\pi^6\la^7}{40320}+O(\la^9),
\label{eq:Blamweak}
\end{equation}
whose first two terms are those of the three-loop computation of \cite{Bianchi:2017svd}. The dual function carries the
logarithm, with a coefficient proportional to $\Bhalf$ as the shrinking of $A$ requires. Expanding \eqref{eq:dual} at
large argument gives it to all orders,
\begin{equation}
\begin{aligned}
\BhalfD+\frac2\pi\,\Bhalf\,\log\frac\kk{16}
&=\frac1{\pi^2\kk}+\frac2\pi\sum_{n\ge0}a_n\,\kk^{2n+1}\Big[\frac1{2n+2}-2\big(H_{2n}-H_n\big)\Big]\\
&=\frac{1}{\pi^{2}\kk}+\frac{\kk}{64\pi^{2}}+\frac{3\kk^3}{16384\pi^{2}}+O(\kk^{5}) ,
\end{aligned}
\label{eq:dualweak}
\end{equation}
with $H_n$ the harmonic numbers. The left-hand side is therefore an odd Laurent series in $\kk$. The scale $16$ in the logarithm is the one for which this
Laurent series has rational coefficients, up to a factor $\pi^{-2}$.

At the conifold $\kk=4i-32i\delta$ and the discriminant $\Delta$ vanishes linearly. The other conifold point $-4i$ behaves similarly because of the symmetry $\kk\to-\kk$. The solution without logarithm is the period
\eqref{eq:Bvanish} of the vanishing cycle while $\Bhalf$ carries the logarithm
\begin{equation}
\Bhalf=\frac{i}{4\pi^2}+\frac{i(1+\log\delta)}{\pi^2}\delta+\frac{i(11+6\log\delta)}{\pi^2}\delta^2+O(\delta^3\log\delta).
\label{eq:conifoldlog}
\end{equation}
So $\Bhalf$ is finite at the conifold, but its derivative diverges logarithmically. As at weak coupling, the coefficient of the logarithm is the solution without logarithm. Then we can write
\begin{equation}
\Bhalf-\frac i\pi\,\big(i\BhalfD-\Bhalf\big)\log\delta=\frac{i}{4\pi^2}+\frac{i\delta}{\pi^2}+\frac{11i\delta^2}{\pi^2}+O(\delta^3),
\label{eq:conifoldexact}
\end{equation}
where now the remainder is a power series in $\delta$.

At strong coupling \eqref{eq:ode} reduces to $(\kk\dk)^2\Bhalf=0$. Since the $B$-cycle shrinks, $\BhalfD$ is the solution
without logarithm while $\Bhalf$ carries the logarithm such that 
\begin{equation}\begin{split}
\Bhalf(\kk)-\frac2\pi\BhalfD \log\kk=&\frac2\pi\sum_{n\ge0}\tilde a_n\,\kk^{-2n}\Big[\frac1{2n-1}-2\big(H_{2n}-H_n\big)\Big]\\
=&-\frac{1}{4 \pi ^2}+\frac{5}{2 \pi ^2 \kappa ^4}-\frac{62}{3 \pi ^2 \kappa ^6}+O\left(\frac{1}{\kappa^8 }\right)
\label{eq:strongkappa4}
\end{split}\end{equation}
where $\tilde a_n$ is the coefficient of the series of $\BhalfD$ as $a_n$ is for $\Bhalf$ in \eqref{eq:recursion}.
The power series starts with the constant $-1/4\pi^2$, and since the rational part of the bracket
vanishes at $n=1$, its next term is of order $\kk^{-4}$. 

To read \eqref{eq:strongkappa4} in the 't~Hooft coupling we need the mirror map at large $\kk$. Closing the
Mellin-Barnes contour of the ${}_3F_2$ in \eqref{eq:lamkappaintro} to the right gives \cite{Drukker:2010nc}
\begin{equation}
\la=\frac{\log^{2}\kk}{2\pi^{2}}+\frac1{24}
-\frac{1}{2\pi^{2}}\sum_{n\ge1}\frac{(-1)^{n}}{n\,\kk^{2n}}\binom{2n}{n}^{\!2}
\Big[\log\kk+\frac1{2n}-2\big(H_{2n}-H_{n}\big)\Big],\qquad \kk>4 .
\label{eq:lamconv}
\end{equation}
At large $\kk$ the periods are polynomials in $\log\kk$, up to corrections in powers of $\kk^{-2}$. Dropping these
corrections, the mirror map \eqref{eq:lamconv} is quadratic and by
\eqref{eq:strongkappa4}, since $\BhalfD=1/8\pi$ from \eqref{eq:dual}, the Bremsstrahlung function is linear
$\Bhalf=(\log\kk-1)/4\pi^2$. In the shifted coupling $\lah=\la-\tfrac1{24}$, \eqref{eq:lamconv} gives simply
$\log\kk=\pi\sqrt{2\lah}$ that inserted in $\Bhalf$ gives
\begin{equation}
\Bhalf^{\rm pert}=\frac{x}{4\pi}-\frac{1}{4\pi^2},\qquad x:=\sqrt{2\lah}.
\label{eq:term}
\end{equation}
This expansion stops after two terms. All the dropped corrections are powers of $\kk^{-2}\simeq e^{-2\pi x}$.
This is the weight of a worldsheet instanton
of action $2\pi\sqrt{2\lah}$ \cite{Drukker:2010nc}, a string wrapping $\mathbb{CP}^1\subset\mathbb{CP}^3$ in
the type IIA dual \cite{Drukker:2011zy}. In the following section, the modular frame will make evident that the truncation of the perturbative series at strong coupling is consequence of the quasimodularity of the Bremsstrahlung function.

\subsection{The modular frame}
\label{sec:modular}

The local analysis of section \ref{sec:ode} describes $\Bhalf$ around each degeneration separately in the variable
adapted to it, and each of its expansions converges only up to the conifold. However, it is possible to introduce a single coordinate that treats the three
degenerations on the same footing. This is the shape of the period lattice, namely the ratio of the two periods of the
holomorphic differential. For the ABJM curve it is the second derivative of the planar free energy as follows
\begin{equation}
\tau:=i\,\frac{\dk\la_D}{\dk\la}=\frac{i}{4\pi^3}\,\partial_\la^2F_0 ,
\label{eq:tauSG}
\end{equation}
the effective coupling of the special geometry of section \ref{sec:curve}, as in Seiberg-Witten theory
\cite{Seiberg:1994rs}. In this coordinate the local statements of section \ref{sec:ode} become exact: $\Bhalf$ is a
quasimodular form, the Legendre relation \eqref{eq:Blegendre} is its transformation law, the logarithms of the two
physical ends are $\tau$ and $-1/\tau$, and the strong-coupling expansion has exactly two perturbative terms, every
further correction being a worldsheet instanton.

\begin{figure}[t]
\centering
\includegraphics[width=\textwidth]{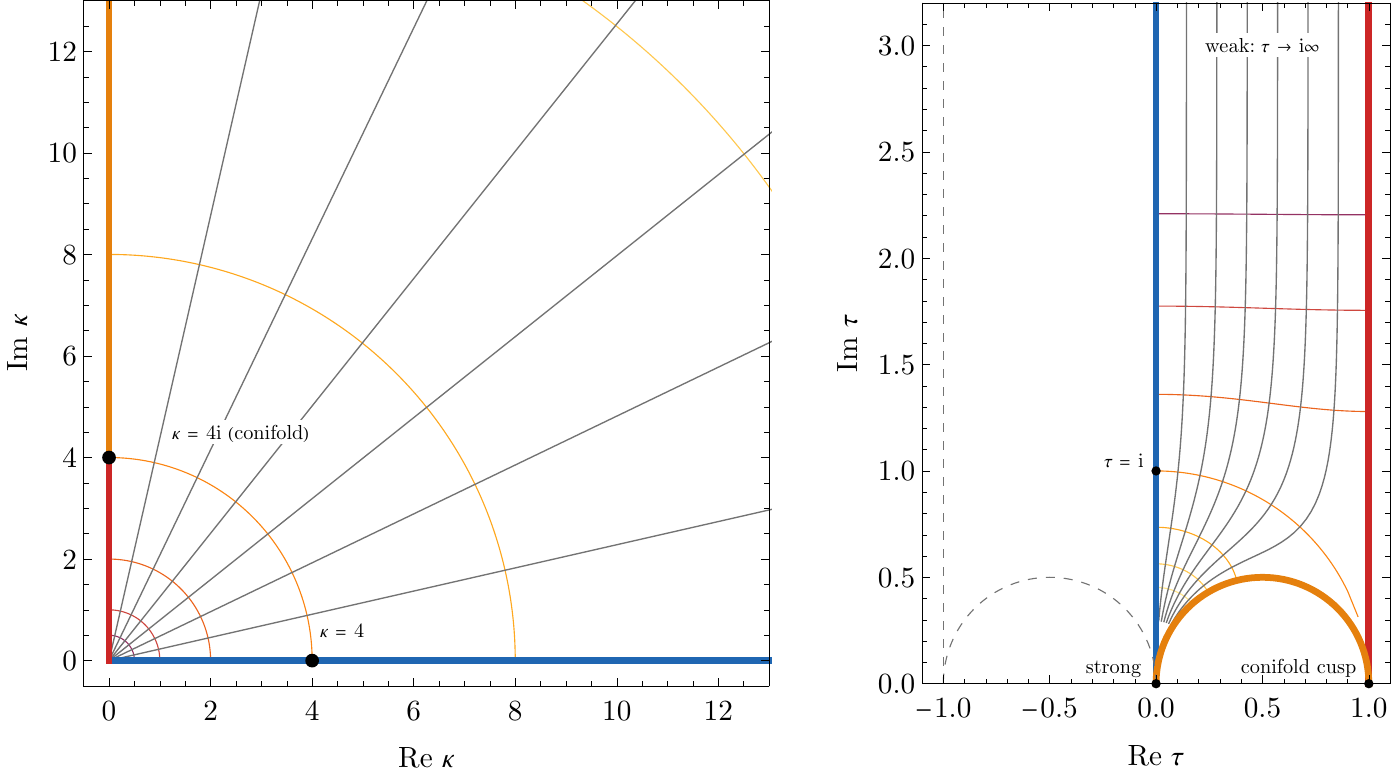}
\caption{The modular frame. Left: the quadrant $\re\kk,\im\kk\ge0$ of the complex modulus, with circles of constant
$|\kk|$ (darker for smaller $|\kk|$) and rays of constant $\arg\kk$ (grey). Right: their images under the map
\eqref{eq:frame}. The physical slice $\kk>0$ (blue) is mapped to the imaginary axis, from weak coupling at
$\tau\to i\infty$ to strong coupling at $\tau\to0$, through the self-dual point $\kk=4$ at $\tau=i$. The imaginary axis
of $\kk$ is mapped to the line $\re\tau=1$ below the conifold (red) and to the semicircle $|\tau-\tfrac12|=\tfrac12$
above it (orange). The conifold $\kk=4i$ is mapped to the cusp $\tau=1$. The quadrant fills half of a fundamental domain of $\Gamma(2)$. The dashed lines bound the other half. It is
the image of the quadrant $\im\kk<0$, and also the image of the first half under $S:\tau\mapsto-1/\tau$, which acts as
$\kk\mapsto16/\kk$.}
\label{fig:frame}
\end{figure}

As there, $\tau$ lives in the upper half plane and is defined only up to the choice of the two
cycles, which changes it by a duality transformation $\tau\mapsto(a\tau+b)/(c\tau+d)$ with integer entries. For our
curve the allowed transformations are those with $b$ and $c$ even, which do not exchange the branch points. They form
the group $\Gamma(2)$, the same as for the Seiberg-Witten curve of pure $SU(2)$. The three degenerations are the points
where $\tau$ reaches the boundary of the upper half plane, called \emph{cusps}: $\tau\to i\infty$ at weak coupling, $\tau\to0$
at strong coupling and $\tau\to1$ at the conifold. Near a cusp the natural expansion variable
is an exponential of $\tau$ known as the \emph{nome}: $q:=e^{i\pi\tau}$ at weak coupling with $\sqrt q\simeq\pi\la/2$, and
$\tilde q:=e^{i\pi\tilde\tau}$ with $\tilde\tau:=-1/\tau$ at strong coupling, where $\tilde q\simeq e^{-2\pi\sqrt{2\lah}}$
is the weight of a worldsheet instanton. The duality $S:\tau\mapsto-1/\tau$ exchanges weak and strong coupling and
fixes the conifold. On the physical slice it acts as $\kk\mapsto16/\kk$, with the self-dual coupling $\kk=4$ at
$\tau=i$, and it exchanges the two cycles of the holomorphic differential.

The building blocks of the frame are the Jacobi theta constants $\thh2,\thh3,\thh4$ of nome $q$, with
$\thh3^4=\thh2^4+\thh4^4$, and the Eisenstein series $\Eis2(\tau)=1-24\sum_{n\ge1}\sigma_1(n)\,q^{2n}$ with $\sigma_1(n)$ the sum of all positive divisors of $n$. Under $S$ the
theta constants behave as forms of weight one: $\thh2^2$ and $\thh4^2$ are exchanged and $\thh3^2$ is mapped to
itself, each multiplied by $-i\tau$. The Eisenstein series instead picks up an inhomogeneous term, its anomaly that reads
\begin{equation}
\Eis2\!\left(-\frac1\tau\right)=\tau^2\Eis2(\tau)-\frac{6i\tau}{\pi}.
\label{eq:E2anom}
\end{equation}
Polynomials in $\Eis2$ with modular coefficients are called \emph{quasimodular}, and their degree in $\Eis2$ is their depth. The dictionary with the curve follows from \eqref{eq:ABperiods} so that
\begin{equation}
\tau=i\,\frac{K(1-\hmod)}{K(\hmod)},\qquad\text{with}\qquad \hmod:=\frac{\kk^2}{16+\kk^2}=\frac{\thh2^4}{\thh3^4},
\label{eq:frame}
\end{equation}
and with the data of the curve rational in the theta constants
\begin{equation}
\kk=4\,\frac{\thh2^2}{\thh4^2},\qquad
\frac{16+\kk^2}{16}=\frac{\thh3^4}{\thh4^4},\qquad
4\pi^2\dk\la=K\!\left(-\tfrac{\kk^2}{16}\right)=\frac\pi2\,\thh4^2 .
\label{eq:framedict}
\end{equation}
This is the frame of \cite{Drukker:2010nc}. On the physical slice $\hmod$ goes from $0$ to $1$ and $\tau$ from $i\infty$ to $0$ along the imaginary axis (figure~\ref{fig:frame}). 

Equation \eqref{eq:main} becomes a statement about modular forms once the derivative in $\kk$ is written through $D=\tfrac{1}{2\pi i}\partial_\tau$. In order to do so, we need to exploit the  Ramanujan system of identities for $\Gamma(2)$ that reads
\begin{equation}
D\thh2^2=\tfrac{\thh2^2}{12}\big(\Eis2-\thh2^4+2\thh3^4\big),\quad
D\thh3^2=\tfrac{\thh3^2}{12}\big(\Eis2+2\thh2^4-\thh3^4\big),\quad
D\thh4^2=\tfrac{\thh4^2}{12}\big(\Eis2-\thh2^4-\thh3^4\big).
\label{eq:ramanujan}
\end{equation}
Using that $\dk=(D\kk)^{-1}D$, substituting \eqref{eq:framedict} in  \eqref{eq:main} and performing the derivatives with \eqref{eq:ramanujan}, we find that the Bremsstrahlung function becomes remarkably simple
\begin{equation}
\Bhalf=\frac{\thh2^4+\thh3^4-\Eis2(\tau)}{24\pi\,\thh2^2} .
\label{eq:E2form}
\end{equation}
Everything in it is modular except $\Eis2$ which enters linearly. As a consequence, $\Bhalf$ is quasimodular of weight one and depth
one. This is the modular counterpart of the statement that $\Bhalf$ is a period of the second kind. In the modulus $\hmod$, \eqref{eq:BPVreduced} and \eqref{eq:dual} read
\begin{equation}
\Bhalf=\frac{K(\hmod)-E(\hmod)}{4\pi^2\sqrt{\hmod}} ,\qquad \text{and}\qquad \BhalfD=\frac{E(1-\hmod)}{4\pi^2\sqrt{\hmod}} ,
\label{eq:BKEmhat}
\end{equation}
where the integral of the first kind is modular, while that of the second kind is quasimodular. 

Quasimodularity of the Bremsstrahlung function is an extremely powerful statement since one can reconstruct the whole function from few perturbative orders.  Indeed, the quasimodular forms of weight two and depth one for
$\Gamma(2)$ such as $\thh2^2\Bhalf$ are regular at the cusps and span a three-dimensional space generated by $\Eis2$, $\thh2^4$ and
$\thh3^4$. Writing $24\pi\thh2^2\Bhalf=a\Eis2+b\,\thh2^4+c\,\thh3^4$ and matching with the nome expansion of the basis, one can generate 3 equations just using the vanishing of $\Bhalf$ at zero coupling and its one- and three-loop coefficients known perturbatively from
\cite{Bianchi:2017svd}. As a result $(a,b,c)=(-1,1,1)$, which reconstructs \eqref{eq:E2form}. Thus, quasimodularity and two orders of perturbation theory fix the exact planar Bremsstrahlung
function. However, without the curve, a reason for quasimodularity on the side of the defect is not known.

The relation \eqref{eq:Blegendre} is the transformation law of the pair $(\Bhalf,\BhalfD)$. In the modulus
$\hmod$ it is literally Legendre's relation $EK'+E'K-KK'=\pi/2$. Divided by $\dk\la$,
with \eqref{eq:tauSG} and \eqref{eq:framedict}, it reads
\begin{equation}
\BhalfD=-i\tau\,\Bhalf+\frac{1}{4\pi^2\,\thh2^2(\tau)}.
\label{eq:Bdualtau}
\end{equation}
Written in terms of periods, \eqref{eq:Bdualtau} says that the $B$-period of $\OmB$ is $\tau/2$ times its $A$-period, as for the holomorphic differential $dZ/u$, plus an extra term that accounts for the anomaly of $\Eis2$.
Inserting \eqref{eq:E2form} in \eqref{eq:Bdualtau} and passing to $\tilde\tau$ with \eqref{eq:E2anom}, this extra term cancels the image of the anomaly exactly, leading to
\begin{equation}
\BhalfD=\frac{\thh3^4(\tilde\tau)+\thh4^4(\tilde\tau)+\Eis2(\tilde\tau)}{24\pi\,\thh4^2(\tilde\tau)} ,
\label{eq:Bdualmod}
\end{equation}
which is \eqref{eq:E2form} with $\thh2$ and $\thh4$ exchanged and the sign of $\Eis2$ reversed. In other words, the dual function is
the $S$-transform of $\Bhalf$ without its anomaly. Solved for $\Bhalf$, \eqref{eq:Bdualtau} gives the same relation
seen from the strong-coupling cusp
\begin{equation}
\Bhalf=-i\tilde\tau\,\BhalfD-\frac{1}{4\pi^2\,\thh4^2(\tilde\tau)} .
\label{eq:Sform}
\end{equation}
These two relations are the exact form of the logarithmic subtractions of section \ref{sec:ode}. At weak coupling
$i\tau=\tfrac2\pi\log\tfrac{\kk}{16}+O(\kk^2)$, and \eqref{eq:Bdualtau} is the closed form of \eqref{eq:dualweak}, while at strong coupling $i\tilde\tau=-\tfrac2\pi\log\kk+O(\kk^{-2}\log\kk)$, and \eqref{eq:Sform} is the closed form of
\eqref{eq:strongkappa4}.

The two relations have a direct reading in the gauge theory. Indeed, the relation \eqref{eq:Blegendre} divided by $\dk\la$ gives the dependence on the coupling of the $\tfrac12$BPS circle as follows
\begin{equation}
\la\,\partial_\la\log\vev{W^{1/2}}=8\pi^3\la\,\big(\BhalfD-\Bhalf\,\partial_\la\la_D\big)-1
=\frac{2\pi\la}{\thh2^2(\tau)}-1 .
\label{eq:circleN4}
\end{equation}
In $\mathcal N=4$ super Yang-Mills the same derivative is the Bremsstrahlung function itself,
$\la\,\partial_\la\log\vev{W}=2\pi^2B_{\mathcal N=4}$. In ABJM it is not, and the modular frame
says why. Since $\partial_\la\la_D=-i\tau$, the bracket is $\BhalfD+i\tau\Bhalf=1/4\pi^2\thh2^2$ by \eqref{eq:Bdualtau},
a modular form in which the $\Eis2$ of $\Bhalf$ is cancelled by that of $\BhalfD$. The circle sees $\Bhalf$ only
through this combination, and since $\Bhalf$ is quasimodular it cannot fix it alone: the dual function supplies the
missing $\Eis2$.

At the strong-coupling cusp \eqref{eq:Sform} is an exact expansion.
Defining $\tilde\tau=it$, the Bremsstrahlung $\Bhalf$ is exactly a polynomial of degree one in $t$ whose coefficients are series in $\tilde q$, and at
$\tilde q=0$ only $\tfrac t{8\pi}-\tfrac1{4\pi^2}$ terms survive. The degree is the depth of \eqref{eq:E2form}. Indeed, under $S$ its modular part of weight one comes back with one power of $-i\tilde\tau=t$, while $\Eis2$ adds the
inhomogeneous term of \eqref{eq:E2anom}, which carries no power of $t$. That term is the whole constant. The anomaly
$-6i\tilde\tau/\pi$ of $\Eis2(\tau)$ enters the numerator of \eqref{eq:E2form} with a minus sign, and the denominator
is $24\pi\thh2^2(\tau)=-24\pi i\tilde\tau\,\thh4^2(\tilde\tau)$, so that $\tilde\tau$ cancels
\begin{equation}
\frac{6i\tilde\tau/\pi}{-24\pi i\,\tilde\tau\,\thh4^2(\tilde\tau)}=-\frac{1}{4\pi^2\,\thh4^2(\tilde\tau)}
\;\xrightarrow{\ \tilde q\to0\ }\;-\frac{1}{4\pi^2},
\label{eq:anomcusp}
\end{equation}
so the shift is the anomaly of $\Eis2$ evaluated at the cusp. Since $t=2x=2\sqrt{2\hat\lambda}$ up to instantons by \eqref{eq:lamconv},
these are the two terms of \eqref{eq:term}: the classical string term is the modular part, the constant is the anomaly, and every power of $\tilde q=e^{-2\pi x}\big(1+O(e^{-2\pi x})\big)$
is a worldsheet instanton.

\begin{figure}[t]
\centering
\includegraphics[width=\textwidth]{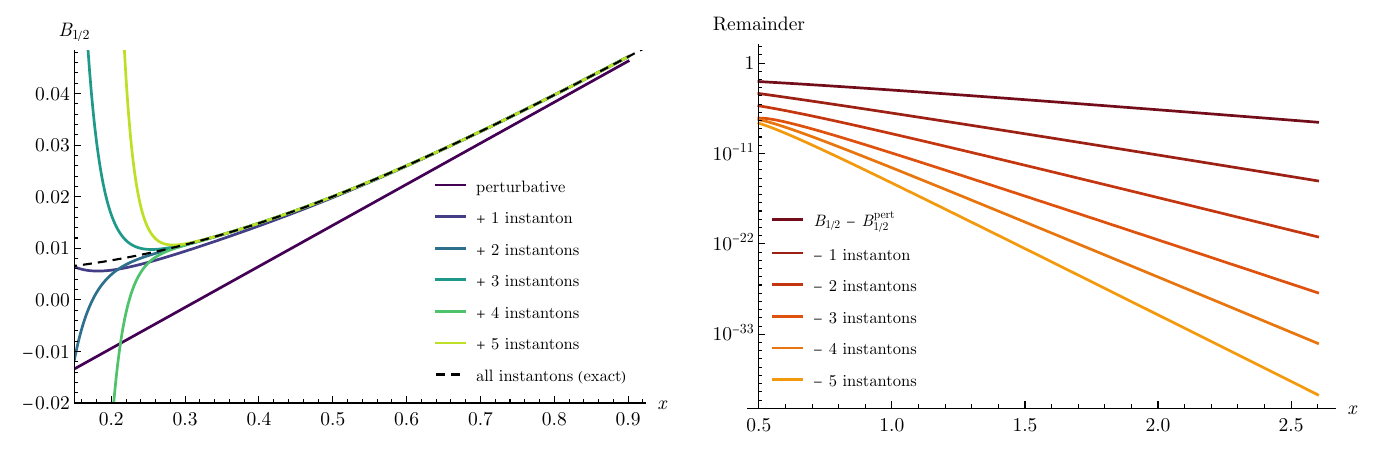}
\caption{Worldsheet instantons in $\Bhalf$ against $x=\sqrt{2\lah}$. Left: the perturbative part \eqref{eq:term} and
the partial sums of \eqref{eq:threeinst} with the first $N=1,\dots,5$ instantons \eqref{eq:C1}, compared with the exact
$\Bhalf$ (dashed). One instanton already brings the perturbative part close to the exact curve, and from two on the curves
cannot be told apart above $x\simeq0.35$. Right: the absolute value of the remainder, namely $\Bhalf$
minus its perturbative part and the first $N=0,\dots,5$ instantons, on a logarithmic scale. For $N=0$ it decays as
$\Iinst_1(\pi x)\,e^{-2\pi x}/\pi^2$, with no power-law tail, so the perturbative series stops after two terms. Each
further instanton makes the decay faster by a factor $e^{-2\pi x}$.}
\label{fig:instantons}
\end{figure}

The instanton corrections follow from \eqref{eq:Sform} expanding at fixed $x$. To make the expansion coefficients simpler, we rescale the variable by $\pi$ defining $y:=\pi x$ and we define 
\begin{equation}
\Bhalf-\Bhalf^{\rm pert}=\frac{1}{\pi^2}\sum_{n\ge1}\Iinst_n(y)\,e^{-2 n y},
\label{eq:threeinst}
\end{equation}
where the overall factor of $1/\pi^2$ is needed to have the instanton coefficients $\Iinst$ to be rational. The first few
are given by
\begin{equation}\small\begin{split}\raisetag{16ex}
\Iinst_1(y)&=y-\frac1{2}+\frac1{4y}\\
\Iinst_2(y)&=y-\frac5{4}+\frac{23}{16y}-\frac1{2y^2}
-\frac1{8y^3},\\
\Iinst_3(y)&=2y-\frac{17}{6}+\frac{16}{9y}-\frac{21}{8y^2}
+\frac{17}{16y^3}+\frac{3}{4y^4}+\frac1{8y^5},\\
\Iinst_4(y)&=-3y-\frac{265}{72}+\frac{7247}{576y}-\frac{89}{144y^{2}}
+\frac{3751}{1152y^{3}}-\frac{49}{12y^{4}}-\frac{123}{32y^{5}}-\frac{5}{4y^{6}}-\frac{5}{32y^{7}},\\
\Iinst_5(y)&=29 y+\frac{1061}{360}-\frac{9097}{300y}
-\frac{16375}{288y^{2}}-\frac{225}{64y^{3}}+\frac{2125}{192y^{4}}+\frac{3475}{128y^{5}}+\frac{1075}{48y^{6}}+\frac{305}{32y^{7}}
+\frac{35}{16y^{8}}+\frac{7}{32y^{9}} .
\label{eq:C1}
\end{split}\normalsize
\end{equation}
At planar level this is not
an asymptotic trans-series but a re-expansion of convergent $\tilde q$-series (figure~\ref{fig:instantons}), and the prefactors $\Iinst_n$ are
predictions for a semiclassical computation around the string worldsheet wrapping $\mathbb{CP}^1$ and ending on the
line. 

Finally, on the physical slice $S$ acts as $\kk\mapsto\tilde\kk:=16/\kk$, which exchanges strong and weak coupling and
fixes the self-dual point $\kk=4$. Written through the functions of the dual coupling, \eqref{eq:Sform} and \eqref{eq:dual}
read
\begin{equation}
\Bhalf(\kk)=t\,\BhalfD(\kk)-\frac{1}{32\pi^3\,\dk\la(\tilde\kk)},\qquad
\BhalfD(\kk)=\Big(1+\frac{\tilde\kk^2}{16}\Big)\dk\la(\tilde\kk)-\frac{\tilde\kk}{4}\,\Bhalf(\tilde\kk)\,.
\label{eq:Sdual}
\end{equation}
The weak-coupling series of $\Bhalf$, $\dk\la$ and the nome $q$, which converge for $\kk<4$, give $\Bhalf$ also for $\kk>4$ with $t$ given by $e^{-\pi t}=\tilde q(\kk)=q(\tilde\kk)$.

This completes the description of the fermionic function. $\Bhalf$ is the $A$-period of a second-kind differential, a
single Gauss hypergeometric function, a quasimodular form of weight one and depth one, the solution regular at weak coupling of a hypergeometric equation whose second solution is its
$B$-period. By \eqref{eq:B13rel} the four Bremsstrahlung
functions of the $\tfrac13$BPS line are fixed by $\Bhalf$, so they are given by \eqref{eq:main} as well. The bosonic
functions are not: by \eqref{eq:BPVrel} they need, besides $\Bhalf$, the derivative at $\nu=1$ of the phase
$\PhB(\nu)$ of the bosonic latitude at framing $\nu$, the one datum that does not live on the great circle.

\section{Winding Wilson loops and logarithmic periods}
\label{sec:logperiods}

By \eqref{eq:BPVrel}, the bosonic Bremsstrahlung function is the product of $\Bhalf$ and of one more number, the slope
$\partial_\nu\PhB(\nu)|_{\nu=1}$ of the phase of the bosonic latitude at framing $\nu$. Away from $\nu=1$ the phase is known through three loops at weak coupling \cite{Bianchi:2018scb} and, through
the fermionic latitude, at strong coupling \cite{Bianchi:2018bke}. At finite planar coupling, one would need the planar solution of the latitude matrix model, and the method of section~\ref{sec:setup} does not apply to it directly. The planar solution of
section~\ref{sec:setup} rests on the fact that in \eqref{eq:MM} the eigenvalues interact through rational functions of
$Z=e^{\mu}$ alone, which is what makes the planar resolvent a function on an algebraic curve. In the latitude case, they interact through rational functions of $e^{\sqrt\nu\mu}$ and of $e^{\mu/\sqrt\nu}$ at the
same time. For generic $\nu$ no single variable makes the interaction rational, and the resolvent of the latitude model
is not a function on an algebraic curve.

However, at first order around the great circle, the deformation drops out of the measure and the latitude
changes only through the inserted operator $\partial_\nu e^{\sqrt\nu\mu}|_{\nu=1}=\tfrac12\,\mu\,e^{\mu}
=\tfrac12\,\partial_ne^{n\mu}|_{n=1}$. This is the operator of the loop wound $n$ times differentiated at $n=1$, so one can conclude that
\begin{equation}
\partial_\nu\log\vev{W_B(\nu)}_\nu\Big|_{\nu=1}=\frac12\,\partial_n\log\vev{W^{1/6}_n}\Big|_{n=1}.
\label{eq:latwindexact}
\end{equation}
Then, to first order around the great circle, the latitude is the circle wound $\sqrt\nu$ times! The identity uses nothing
but the symmetry of the measure, so that it holds for any ranks and at
any $N$ and $k$. Its real part, through \eqref{eq:LM} and \eqref{eq:B16lat}, is the relation $\Bsixth=\tfrac12\Bsix$ of
\eqref{eq:BthetaBphi}. Its imaginary part is the derivative we are after
\begin{equation}
\partial_\nu\PhB(\nu)\Big|_{\nu=1}=\frac12\,\partial_n\arg\vev{W^{1/6}_n}\Big|_{n=1} .
\label{eq:dPhinu}
\end{equation}
Thus, the two exact prescriptions for the bosonic function are statements about a single object, the derivative of the
wound loop with respect to its winding number at $n=1$: \eqref{eq:LM} uses its real part and \eqref{eq:BPVrel} uses its
imaginary part through \eqref{eq:dPhinu}. To compute it we need the wound loop as an analytic function of the winding
around $n=1$, not only at the integers. In the matrix model this is natural. The average
$\vev{W^{1/6}_n}=\vev{\tfrac1N\sum_ie^{n\mu_i}}$ makes sense for every complex $n$, and its derivative
$\partial_n\vev{W^{1/6}_n}=\vev{\tfrac1N\sum_i\mu_i\,e^{n\mu_i}}$ is the loop with one insertion of the eigenvalue. On
the curve the eigenvalue is $\mu=\log Z$, so that the derivative inserts a logarithm into the differential of the wound
loop. We first solve the wound loops at every integer winding. After it, we study the analytic continuation in $n$. Then, we introduce the logarithmic periods that the derivative in the winding produces, count the
independent ones, derive their Gauss-Manin system and solve it, as section~\ref{sec:setup} did for the ordinary periods. The solution gives the derivative at $n=1$ in closed form. It contains $\Bhalf$ once more, now as
a winding derivative of the $\tfrac12$BPS loop, and a single new function of the coupling, the winding velocity of the
wound loop, and with the two of them the slope of the latitude phase. 

\subsection{The wound loops at integer winding}
\label{sec:intwinding}
\label{sec:tower}

The logic that led to the planar loops in closed form applies, essentially unchanged, to the loops that wind $n$ times around the circle. On the matrix model side, winding the contour $n$ times replaces $e^{\mu_i}$ by $e^{n\mu_i}$, that is $Z$ by $Z^n$, in the eigenvalue average \eqref{eq:MMloops}. The $n$-wound bosonic loop $\vev{W^{1/6}_n}$ is therefore the $n$-th moment of the eigenvalue density of the first node: the standard loop is the first moment, while the zeroth one is the normalisation. Since the matrix model \eqref{eq:MM} is invariant under the simultaneous reflection $\mu_i\to-\mu_i$, $\hat\mu_j\to-\hat\mu_j$ of all the eigenvalues, $\vev{W^{1/6}_{-n}}=\vev{W^{1/6}_n}$, so that it is enough to consider $n\ge1$. The fermionic loops follow from the node relation \eqref{eq:node}. 

Similarly to the coupling \eqref{eq:lambdafed}, the moments are $A$-periods of the resolvent \cite{Marino:2009jd,Klemm:2012ii}, now with the weight $Z^{n-1}$ in place of $Z^{-1}$. As for $n=1$, the logarithm in the resolvent disappears after one integration by parts, because its derivative along the curve is algebraic. What remains are the integrals \eqref{eq:Jdef} along the cut
\begin{equation}
\vev{W^{1/6}_n}=\frac{J_{n+1}-J_{n-1}}{4\pi^2 n\,\la}=\frac{1}{8\pi^2 n\,\la}\oint_A\big(Z^{n+1}-Z^{n-1}\big)\frac{dZ}{u}
\label{eq:Wn}
\end{equation}
which for $n=1$ is \eqref{eq:W1}. 
If instead we differentiate the resolvent in $\kk$ before integrating by parts, the cycle does not move and \eqref{eq:loginsertion} replaces $\varrho_0$ by $iZ/u$, so that
\begin{equation}
\dk\big(\la\vev{W^{1/6}_n}\big)=\frac{i}{4\pi^2}\,J_n \,,\qquad
\dk\big(\la\vev{W^{1/2}_n}\big)=\frac{1}{4\pi^2}\,\text{Re}(i^{2-n}J_n)\,.
\label{eq:dWn}
\end{equation}
This extends \eqref{eq:dWJ1} to any winding: the $\kk$-derivative of the loop of winding $n$ is the $n$-th moment of the holomorphic differential $dZ/u$. Together, \eqref{eq:Wn} and \eqref{eq:dWn} say that the moments obey a recursion in the winding number which is of first order in the coupling
\begin{equation}
\dk\big(J_{n+1}-J_{n-1}\big)=i\,n\,J_n ,\qquad n\ge1 ,
\label{eq:Jrec}
\end{equation}
which at $n=1$ is the difference of the rows \eqref{eq:GM2} and \eqref{eq:GM0} of the Gauss-Manin system. As will become clear later, this is all that is needed to solve the tower in closed form.

From the discussion of section~\ref{sec:curve}, we know that the five-term relation \eqref{eq:contig} expresses every $J_m$ through the three basic periods $J_{0,1,2}$ and so through $\dk\la$ and $\partial_\kk^2\la$ with polynomial coefficients in a unique way. Thus, every wound loop is fixed by the data that we have already computed, and no new constant of integration can appear. For every integer $n\ge1$ we have 
\begin{equation}\begin{split}
\vev{W^{1/6}_n}&=\frac{i^{\,n-1}}{8\pi n\la}\left[R_n(\kk)-2\pi i{(16+\kk^2)}\Big(S_n(\kk)\,\dk\la+T_n(\kk)\,\partial_\kk^2\la\Big)\right]\,,\\
\vev{W^{1/2}_n}&=\frac{1}{8\pi n\la}\,R_n(\kk)\,,
\label{eq:towerclosed}
\end{split}\end{equation}
where $R_n$, $S_n$ and $T_n$ are polynomial in $\kk$ of degree $n$, $n-1$ and $n$ respectively. Their leading terms are $R_n=\kk^n+\dots$, $T_n=H_n\kk^n+\dots$ and $S_n=(H_n-1)\kk^{n-1}+\dots$, where $H_n$ are the harmonic numbers. The first members of the tower are
\begin{center}
\renewcommand{\arraystretch}{1.15}
\begin{tabular}{@{}c|l|l|l@{}}
$n$ & $R_n$ & $S_n$ & $T_n$\\\hline
1 & $\kk$ & $0$ & $\kk$\\
2 & $\kk^2$ & $\tfrac12\kk$ & $\tfrac32\kk^2$\\
3 & $\kk^3-3\kk$ & $\tfrac56\kk^2$ & $\tfrac16\kk(11\kk^2-2)$\\
4 & $\kk^4-8\kk^2$ & $\tfrac1{12}\kk(13\kk^2-12)$ & $\tfrac1{12}\kk^2(25\kk^2-52)$\\
5 & $\kk^5-15\kk^3+5\kk$ & $\tfrac1{60}\kk^2(77\kk^2-234)$ & $\tfrac1{60}\kk(137\kk^4-758\kk^2+12)$
\end{tabular}
\end{center}
At $n=1$, where $R_1=T_1=\kk$ and $S_1=0$, \eqref{eq:towerclosed} is \eqref{eq:Wexact} and the structure found there persists at every winding. The polynomials are real on the physical slice, so that the bracket has real part $R_n$ and imaginary part proportional to $\dk\la$ and $\partial^2_\kk\la$. The cohomological relation \eqref{eq:node} keeps exactly the real part, and the fermionic loop is a polynomial in $\kk$. The transcendental part is carried by the two periods of the compact torus of section~\ref{sec:threeperiods} and it is multiplied by the conifold factor $16+\kk^2$ of the discriminant \eqref{eq:discdef}. The same architecture was found at finite $N$ and $k$ in \cite{Boldis:2025yll} where the authors solved the grand-canonical wound loops of the Fermi gas exactly with a Tracy-Widom analysis.

The derivation of \eqref{eq:towerclosed} is given in appendix~\ref{app:integertower}: the five-term relation \eqref{eq:contig}
becomes a first-order differential equation for the generating function of the moments, which is solved for every $n$ at once.
The three polynomials are built from the Taylor coefficients $q_n(\kk)$ of
\begin{equation}
\frac{1}{\sqrt{\mathcal P(z)}}=\sum_{n\ge0}q_n(\kk)\,z^n,\qquad
\mathcal P(z):=z^4P\big(\tfrac iz\big)=1-2\kk z+(\kk^2+2)z^2+2\kk z^3+z^4 ,
\label{eq:qn}
\end{equation}
as
\begin{equation}
R_n=q_n+q_{n-2},\qquad
T_n=\kk\sum_{j=1}^{n}\frac{R_j'\,R_{n-j}}{j^2},\qquad
S_n=\kk\Big[\sum_{j=1}^{n-1}\frac{R_j'\,R_{n-j}'}{j(n-j)}-\sum_{j=1}^{n}\frac{R_j''\,R_{n-j}}{j^2}\Big],
\label{eq:ST-convolutions}
\end{equation}
with $q_{-1}=q_{-2}=0$, $R_0=1$ and primes denoting derivatives in $\kk$. They reproduce the table above, and they make the
structure of the tower evident. Everything follows from polynomial arithmetic, since $R_n'=n\,q_{n-1}$ and the $q_n$ obey a
recursion with polynomial coefficients. The polynomials of the fermionic loop have an algebraic generating function,
$\sum_{n\ge0}R_nz^n=\mathcal R(z)$, which is the derivative of the resolvent along the curve,
\begin{equation}
Z\,\partial_Z\varrho_0(Z)=1-\frac{Z^2-1}{u(Z)}=1-\mathcal R(z) ,\qquad
\mathcal R(z):=\frac{1+z^2}{\sqrt{\mathcal P(z)}}=\frac{Z^2-1}{u(Z)}\,\bigg|_{Z=i/z} ,
\label{eq:dZrho}
\end{equation}
that is the factor $(Z^2-1)/u$ of the differential \eqref{eq:Wn} after collecting $Z^{n-1}$. The polynomials $S_n$ and $T_n$,
which multiply the transcendental periods, are instead finite convolutions of $R_n$ and of its derivatives in $\kk$.

\subsection{The hypergeometric tower}
\label{sec:hyptower}

The convolutions \eqref{eq:ST-convolutions} are essentially the simplest form of $S_n$ and $T_n$ taken separately. However, the wound loop never needs $S_n$ and $T_n$ separately: it only contains the combination $S_n\dk\la+T_n\partial^2_\kk\la$, and this combination turns out to be as elementary as $R_n$. To find it in closed form we go back to the moments $J_n$, from which the loops descend through \eqref{eq:dWn}, and we compute them directly from the recursion \eqref{eq:Jrec} and from the residues of section~\ref{sec:threeperiods}.

At $\kk=0$ the branch points $a$ and $a^{-1}$ merge at $Z=1$, the cycle $A$ can be taken to be a small circle around this point, and the curve becomes $u=\pm(Z^2-1)$, as in the computation of the flat invariant. Every moment then reduces to the same residue
\begin{equation}
J_n(0)=-i\pi\Res_{Z=1}\frac{Z^n}{Z^2-1}=-\frac{i\pi}{2},\qquad n\ge0, 
\label{eq:Jn0}
\end{equation}
that we use to normalise their expansion in $\kk$ as follows 
\begin{equation}
J_n=-\frac{i\pi}{2}\sum_{k\ge0}\beta_k(n)\,(i\kk)^k,\qquad \beta_0(n)=1 .
\label{eq:Jexp}
\end{equation}
The recursion \eqref{eq:Jrec} contains an explicit factor $i$, which is absorbed by expanding in $i\kk$, the variable in which the curve \eqref{eq:curve} has real coefficients. Inserting it into \eqref{eq:Jrec}, order by order it gives
\begin{equation}
(k+1)\big[\beta_{k+1}(n+1)-\beta_{k+1}(n-1)\big]=n\,\beta_k(n) .
\label{eq:ladder}
\end{equation}
It is a first-order difference equation in the winding number with boundary conditions fixed by the first two moments \eqref{eq:J0im} and \eqref{eq:J1}, which give $\beta_k(0)$ and $\beta_k(1)$. 
Solutions are perfect squares in $n$ with double zeros in $n=k-1, k-3,...,1-k$ that can be written as follows
\begin{equation}
\beta_k(n)=\binom{\frac{n+k-1}{2}}{k}^{\!2}\,.
\label{eq:betasol}
\end{equation}
Notice that the double zeros make every coefficient with $k>n$ and $k-n$ odd vanish identically. As a consequence, the polynomial part of \eqref{eq:towerclosed} is exactly the sector that the zeros of $\beta_k(n)$ make terminate. Plugging it in \eqref{eq:Jexp}, we obtain
\begin{equation}
J_n=-\frac{i\pi}{2}\sum_{k\ge0}\binom{\frac{n+k-1}{2}}{k}^{\!2}(i\kk)^k ,\qquad |\kk|<4 .
\label{eq:Jnseries}
\end{equation}
The terms with even and odd $k$ are the imaginary and the real part of $J_n$ for real $\kk$. Grouping and summing them gives the following exact closed formula for the periods
\begin{equation}
J_n=\frac{-i\pi}2{}_4F_3\left(\genfrac{}{}{0pt}{}{\frac{1-n}2,\frac{1-n}2,\frac{1+n}2,\frac{1+n}2}{\frac12,\ \frac12,\ 1}\,\Biggl|\,\frac{-\kk^2}{16}\right)
+\frac{\pi n^2\kk}{8}{}_4F_3\left(\genfrac{}{}{0pt}{}{\frac{2-n}2,\frac{2-n}2,\frac{2+n}2,\frac{2+n}2}{1,\ \frac32,\ \frac32}\,\Biggl|\,\frac{-\kk^2}{16}\right)
\label{eq:Jn4F3}
\end{equation}
where ${}_pF_q$ are generalised hypergeometric functions and $n$ is any integer $n\ge 0$.
The series in \eqref{eq:Jn4F3} converge only for $|\kappa|<4$, but this is a property of the expansion around $\kappa=0$ and not of the function it represents. Being periods of the curve, $J_n$ and the loops are analytic in $\kappa$ except where two branch points collide and pinch the integration contour. In the variable $x=-\kappa^2/16$, the whole real $\kappa$ line is mapped onto $x\le0$ and never meets the cut. Both sides of \eqref{eq:Jn4F3} are therefore analytic on a connected domain containing the real axis and coincide for $|\kappa|<4$, hence they coincide for all real $\kappa$.

Once the periods $J_n$ are computed, one can reconstruct the wound Wilson loops using \eqref{eq:dWn}, so that
\begin{equation}
\vev{W^{1/6}_n}=\frac{i}{4\pi^2\la}\int_0^{\kk}J_n(\kk')\,d\kk' .
\label{eq:WintJ}
\end{equation}
What the integration does is to raise the power of $\kk$ by one in the series representation of the hypergeometric that translates in replacing the lower parameter $\frac12$ by $\frac32$ in the first and the lower parameter $1$ by $2$ in the second one. The result is the closed form of every wound loop for every $n\ge1$
\begin{equation}
\vev{W^{1/6}_n}=\frac{\kk}{8\pi\la}\,{}_4F_3\left(\genfrac{}{}{0pt}{}{\frac{1-n}2,\frac{1-n}2,\frac{1+n}2,\frac{1+n}2}{\frac12,\ 1,\ \frac32}\,\Biggl|\,\frac{-\kk^2}{16}\right)
+\frac{i\,n^2\kk^2}{64 \pi\la}\,{}_4F_3\left(\genfrac{}{}{0pt}{}{\frac{2-n}2,\frac{2-n}2,\frac{2+n}2,\frac{2+n}2}{\frac32,\ \frac32,\ 2}\,\Biggl|\,\frac{-\kk^2}{16}\right).
\label{eq:tower4F3}
\end{equation}
This is a formula into which $n$ enters only through the parameters, and no table, sum or recursion is left. Both hypergeometric functions have real coefficients, so that for real $\kk$ the first one is the real part of the bosonic wound loop and the second one is its imaginary part. The phase $i^{\,n-1}$ of \eqref{eq:towerclosed} has been absorbed. For odd $n$ the first series terminates, and for even $n$ the second one does. As a consequence, given the cohomological relation \eqref{eq:node} or equivalently \eqref{eq:dWn}, the real part of $i^{1-n}\vev{W^{1/6}_n}$ corresponds to the fermionic wound loop where only one of the two series survives, and it is always the one that terminates. Then, for odd $n$ it is given by the real part of \eqref{eq:tower4F3}, while for even $n$ by the imaginary one. 
For integer $n$, the result admits the following compact representation
\begin{equation}
\vev{W^{1/2}_n}=\frac{R_n}{8\pi n\la}
=\frac{\kk^n}{8\pi n \la}\,
{}_4F_3\left(
\genfrac{}{}{0pt}{}{\frac{-n}2,\frac{1-n}2,\frac{1-n}2,\frac{2-n}2}{1,1-n,1-n}
\,\Biggl|\,\frac{-16}{\kk^2}\right).
\label{eq:tower4F3fermionic}
\end{equation}
Although the denominator contains the Pochhammer symbols $(1-n)_j$, which vanish for integer $n$ when $j\geq n$, the hypergeometric series terminates before these singular terms are reached. Indeed, the numerator parameters enforce termination at $j=\lfloor n/2\rfloor$, so the expression is well defined for every positive integer $n$.

This representation makes the strong-coupling structure particularly transparent. The terminating series in \eqref{eq:tower4F3fermionic} is naturally organised in descending powers of $\kk$, so that the expression itself provides an exact, finite strong-coupling expansion in the variable $-16/\kk^2$. Since $\kk\simeq e^{\pi\sqrt{2\hat\lambda}}$ at strong coupling, the leading term reproduces the expected contribution of an $n$-times-wound string
$\vev{W^{1/2}_n}\simeq
\frac{e^{n\pi\sqrt{2\hat\lambda}}}{8\pi n\lambda}$,
while the remaining $\lfloor (n-1)/2\rfloor$ terms give corrections suppressed by successive powers of $\kk^{-2}\simeq e^{-2\pi\sqrt{2\hat\lambda}}$, the characteristic exponential scale of worldsheet instantons.
Both closed forms \eqref{eq:tower4F3} and \eqref{eq:tower4F3fermionic} reproduce the known weak- and strong-coupling behaviour of the multiply wound Wilson loops \cite{Bianchi:2016gpg,Bianchi:2014laa,Klemm:2012ii,Drukker:2008zx,Chen:2008bp,Rey:2008bh}.

The fact that the fermionic loop is elementary has a geometric reason, which \eqref{eq:dZrho} makes explicit. With $z=i/Z$ it says that $\mathcal R(z)=1-Z\partial_Z\varrho_0(Z)$ on the sheet where $u\simeq Z^2$, and integrating
\begin{equation}
\varrho_0(Z)=2\log2+8\pi\la\sum_{n\ge1}\vev{W^{1/2}_n}\Big(\frac iZ\Big)^{n} .
\label{eq:rhoinfty}
\end{equation}
The $\tfrac12$BPS loops are thus the coefficients of the planar resolvent at infinity \cite{Drukker:2010nc,Klemm:2012ii}. A contour around $Z=\infty$ encloses both cuts and the fermionic loop is precisely the combination of the two cuts that such a contour computes. The bosonic loops are moments of a single cut and need the transcendental periods. Moreover, since $\kk=8\pi\la\vev{W^{1/2}}$ by \eqref{eq:Wexact}, every $\tfrac12$BPS loop is a polynomial in the singly wound one, with coefficients polynomial in $\la$. For instance 
\begin{equation}\begin{split}
\vev{W^{1/2}_2}&=4\pi\la\vev{W^{1/2}}^2\\
\vev{W^{1/2}_3}&=\frac{64\pi^2\la^2}{3}\vev{W^{1/2}}^3-\vev{W^{1/2}}\\
\vev{W^{1/2}_4}&=128 \pi^3\la^3\vev{W^{1/2}}^4-16\pi\la\vev{W^{1/2}}^2
\end{split}\label{eq:W12poly}\end{equation}
This was also noticed in \cite{Boldis:2025yll} at finite $N$ and $k$.

\begin{figure}[t]
\centering
\includegraphics[width=0.6\textwidth]{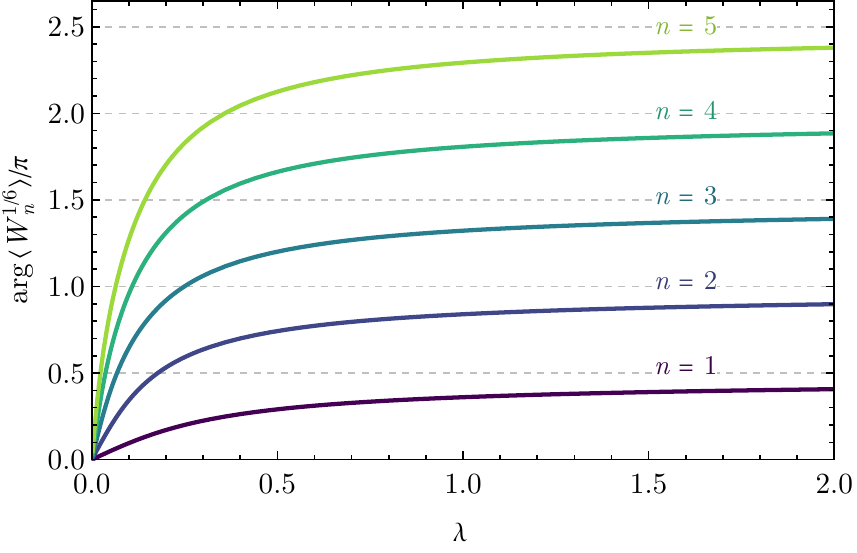}
\caption{The phase \eqref{eq:phasen} of the wound bosonic loop for $n=1,\dots,5$. It vanishes at weak coupling and
grows with the coupling towards $n\pi/2$ (dashed), the phase $i^n$ of the $n$-wound string.}
\label{fig:phases}
\end{figure}

The non-terminating series is the transcendental part. Comparing \eqref{eq:tower4F3} with \eqref{eq:towerclosed} in the same way,
\begin{equation}
\frac{\Delta}{\kk^2}\big(S_n\,\dk\la+T_n\,\partial^2_\kk\la\big)=
\begin{cases}
(-1)^{\frac{n+1}2}\,\dfrac{n^3\kk^2}{16\pi}\;{}_4F_3\left(\genfrac{}{}{0pt}{0}{\frac{2-n}2,\frac{2-n}2,\frac{2+n}2,\frac{2+n}2}{\frac32,\ \frac32,\ 2}\,\Biggl|\,-\dfrac{\kk^2}{16}\right), & n\ \text{odd},\\[4mm]
(-1)^{\frac{n+2}{2}}\,\dfrac{n\kk}{2\pi}\;{}_4F_3\left(\genfrac{}{}{0pt}{0}{\frac{1-n}2,\frac{1-n}2,\frac{1+n}2,\frac{1+n}2}{\frac12,\ 1,\ \frac32}\,\Biggl|\,-\dfrac{\kk^2}{16}\right), & n\ \text{even}.
\end{cases}
\label{eq:ST4F3}
\end{equation}
Therefore, the two polynomial sequences $S_n$ and $T_n$ are the price of expanding a single hypergeometric function against the basis $\{\dk\la,\partial^2_\kk\la\}$ of the periods of the torus. The expansion is what exhibits the geometry, the conifold factor and the two kinds of differentials. The ratio of the two series of \eqref{eq:tower4F3} is the framing phase of the wound loop
\begin{equation}
\tan\left(\arg\vev{W^{1/6}_n}\right)=\frac{n^2\kk}{8}\;
\frac{{}_4F_3\left(\genfrac{}{}{0pt}{0}{\frac{2-n}2,\frac{2-n}2,\frac{2+n}2,\frac{2+n}2}{\frac32,\ \frac32,\ 2}\,\Biggl|\,-\dfrac{\kk^2}{16}\right)}
{{}_4F_3\left(\genfrac{}{}{0pt}{0}{\frac{1-n}2,\frac{1-n}2,\frac{1+n}2,\frac{1+n}2}{\frac12,\ 1,\ \frac32}\,\Biggl|\,-\dfrac{\kk^2}{16}\right)} ,
\label{eq:phasen}
\end{equation}
which at $n=1$ is \eqref{eq:tanPhB} of the singly wound loop. At strong coupling it tends to $n\pi/2$, the phase $i^n$ of the $n$-wound string (figure~\ref{fig:phases}).

\subsection{The analytic continuation and the large winding limit}

The closed form \eqref{eq:tower4F3} was derived for integer winding, but actually nothing in it requires $n$ to be an integer. The winding only enters the upper parameters of the two ${}_4F_3$ functions, while their lower parameters and their argument do not depend on it. Every Taylor coefficient in $\kk^2$ is therefore a product of Pochhammer symbols, i.e. a polynomial in $n^2$ and the series converge uniformly on compact sets of $n$.

This is not an arbitrary interpolation of the integer values, it is the continuation defined by the matrix model and it is unique. 
In the matrix model, the winding enters only through $\vev{W^{1/6}_n}=\int d\mu\,\rho(\mu)\,e^{n\mu}$, with $\rho$ the planar eigenvalue density, and it can assume any complex value. At weak coupling the eigenvalues lie within a distance of order $\sqrt\kk$ from $\mu=0$, so expanding $e^{n\mu}$ shows that each order in $\kk$ is a polynomial in $n$. The same holds for \eqref{eq:tower4F3}, and two polynomials that agree at the integers are identical. Hence \eqref{eq:tower4F3} holds for any complex $n$, and so do \eqref{eq:Jn4F3}, \eqref{eq:Wn} and the first relation in \eqref{eq:dWn}, with $Z^n=e^{n\mu}$ along the cut. It is also the only continuation with this property, since a polynomial is fixed by its values at the integers.

The fermionic formula \eqref{eq:tower4F3fermionic}, instead, cannot be continued as it stands. There, the integer value of $n$ is used to terminate the series: one of the upper parameters becomes a non-positive integer, and so do the lower parameters $1-n$, so that the formula only makes sense because the numerator vanishes before the denominator does. For non-integer $n$ nothing terminates. The series becomes an infinite expansion in $-16/\kk^2$, which does not even converge at weak coupling. What \eqref{eq:tower4F3fermionic} really encodes is a selection rule, namely which of the two series of \eqref{eq:tower4F3} survives according to the parity of $n$, and a parity cannot be continued by itself. The fermionic loop has to be continued instead through the node relation \eqref{eq:node}, which writes it in terms of the bosonic loops of the two nodes, each continued as above, and of the sign $(-1)^n$ due to the antiperiodicity of the fermionic couplings. This sign is the only new ingredient but the integer values do not fix it. Therefore a prescription is needed. The choice $(-1)^n\to e^{i\pi n}$, with $i^{1-n}=e^{i\pi(1-n)/2}$, keeps the loop real, then we obtain
\begin{equation}
\begin{split}
\vev{W^{1/2}_n}=\frac{\kk}{8\pi\la}\sin\frac{\pi n}{2}&\;{}_4F_3\left(\genfrac{}{}{0pt}{}{\frac{1-n}2,\frac{1-n}2,\frac{1+n}2,\frac{1+n}2}{\frac12,\ 1,\ \frac32}\,\Biggl|\,\frac{-\kk^2}{16}\right)\\
&-\frac{n^2\kk^2}{64\pi\la}\cos\frac{\pi n}{2}\;{}_4F_3\left(\genfrac{}{}{0pt}{}{\frac{2-n}2,\frac{2-n}2,\frac{2+n}2,\frac{2+n}2}{\frac32,\ \frac32,\ 2}\,\Biggl|\,\frac{-\kk^2}{16}\right).
\end{split}
\label{eq:W12cont}
\end{equation}
At integer $n$ each trigonometric factor vanishes exactly where its series does not terminate, and \eqref{eq:tower4F3fermionic} is recovered. Off the integers both series contribute. The integers are special for the periods as well. There, \eqref{eq:contig} reduces every $J_n$ to the three basic periods $J_0$, $J_1$ and $J_2$. For non-integer $n$ this reduction fails, because $Z^n$ has a branch point at $Z=0$ and $Z^n\,dZ/u$ is no longer a meromorphic differential on the curve. Accordingly, the derivative in the winding brings down a $\log Z$ and gives a logarithmic period, of the same type as the coupling \eqref{eq:lambdafed}, which we study in section~\ref{sec:logGM}.

At large winding, the loop probes the edge of the eigenvalue distribution. Indeed, in the moments \eqref{eq:Jdef}, the factor $Z^n$ localizes the integral near the endpoint of largest modulus, namely the outer branch point $a^{-1}$, since $|a|<1$. Writing $a^{-1}=e^{\mu_*}$, the relation $a+a^{-1}=2+i\kk$ gives
\begin{equation}
\sinh\frac{\mu_*}{2}=\frac{\sqrt{i\kk}}{2}.
\label{eq:mustar}
\end{equation}
Near this endpoint,
$u^2\simeq 8e^{\mu_*}\sinh\mu_*(Z-e^{\mu_*})$,
so the inverse curve has a square-root singularity. The resulting endpoint asymptotics give
\begin{equation}
J_n\simeq-\frac{i}{4}\sqrt{\frac{2\pi}{n\sinh\mu_*}}\,
e^{n\mu_*}\qquad\Rightarrow\qquad
\vev{W^{1/6}_n}\simeq
-\frac{i}{\la}\sqrt{\frac{\sinh\mu_*}{32\pi^3}}\,
\frac{e^{n\mu_*}}{n^{3/2}}
\left(1+O(n^{-1})\right),
\label{eq:Wlargen}
\end{equation}
where the Wilson loop scaling is computed using \eqref{eq:Wn}.
The asymptotics \eqref{eq:Wlargen} are those of the $n$-wound circle in $\mathcal N=4$ SYM, $\frac{2}{n\sqrt\la}I_1(n\sqrt\la)\sim e^{n\sqrt\la}(n\sqrt\la)^{-3/2}$ \cite{Erickson:2000af,Drukker:2000rr}, with $\sqrt\la$ replaced by the endpoint $\mu_*$. The exponential is set by the endpoint itself, and the power $n^{-3/2}$ by the square-root branch point there: the endpoint integral gives $n^{-1/2}$ and \eqref{eq:Wn} an extra $1/n$. The endpoint interpolates between $\mu_*\simeq\sqrt{8\pi i\la}$ at weak coupling and $\mu_*\simeq\pi\sqrt{2\lah}+\frac{i\pi}{2}$
at strong coupling, where $e^{n\mu_*}\simeq i^n e^{n\pi\sqrt{2\lah}}$ reproduces the exponential growth of the $n$-wound string, including its phase $i^n$.

However, the large-winding and strong-coupling limits do not commute. At fixed winding the strong-coupling expansion has corrections of order $n^2/\kk^2$, as in $R_n=\kk^n-n(n-2)\kk^{n-2}+\dots$, so the string regime is $n\ll\kk$, while \eqref{eq:Wlargen} requires $n\gg\kk$. The reason is geometric: at strong coupling the outer endpoints $a^{-1}$ and $-b^{-1}$ of the two cuts are only $O(1/\kk)$ apart in $\mu=\log Z$, while a loop of winding $n$ resolves distances of order $1/n$. Only for $n\gg\kk$ does it separate the two and probe the edge of a single cut. The crossover $n\sim\kk\simeq e^{\pi\sqrt{2\lah}}$ is exponentially large at strong coupling.

At weak coupling, the crossover occurs when $n\sqrt{\kk}\sim1$. It is captured by the double-scaling limit $n\to\infty$, $\kk\to0$, with $n^2\kk$ fixed. In this limit, the coefficients of \eqref{eq:betasol} simplify to $\beta_k(n)\to(n/2)^{2k}/k!^2$, and the expressions \eqref{eq:Jnseries} and \eqref{eq:WintJ} reduce to
\begin{equation}
J_n\longrightarrow-\frac{i\pi}{2}\,
I_0\big(n\sqrt{i\kk}\big),
\qquad\text{and}\qquad
\vev{W^{1/6}_n}\longrightarrow
\frac{2I_1\big(n\sqrt{i\kk}\big)}
{n\sqrt{i\kk}},
\label{eq:Wbessel}
\end{equation}
up to relative corrections of order $n^{-2}$. Thus the two hypergeometric series in \eqref{eq:tower4F3} combine into the real and imaginary parts of a modified Bessel function of the first kind. The result resums the leading terms $n^{2j}\la^j$ in the weak-coupling expansion to all orders.

The Bessel-function expression also explains why the $\mathcal N=4$ SYM result \cite{Erickson:2000af,Drukker:2000rr} emerges in this limit. The dominant eigenvalues scale as $\mu\sim1/n$, allowing the hyperbolic functions in the matrix model \eqref{eq:MM} to be expanded. The first node then reduces to a Gaussian matrix model, while the bifundamental matter and the second node contribute only at relative order $\la\sim n^{-2}$. The resulting Wilson loop is precisely the $n$-wound circular loop of $\mathcal N=4$ SYM, with the effective coupling
$\la_{\mathcal N=4}=i\kk\simeq8\pi i\la$.
The coupling is imaginary because the Gaussian weight inherited from Chern-Simons theory is oscillatory. This hierarchy is also visible in perturbation theory: at order $\la^j$, the leading power $n^{2j}$ multiplies $\la_1^j$ alone, whereas contributions involving the matter and the second node carry lower powers of $n$ \cite{Bianchi:2016gpg}. 

For the fermionic loop, the node relation \eqref{eq:node} gives the large-winding asymptotics
\begin{equation}
\vev{W^{1/2}_n}\simeq
\frac{1}{\la n^{3/2}}\,
\re\left[
\sqrt{\frac{\sinh\mu_*}{32\pi^3}}\,
e^{n(\mu_*-\frac{i\pi}{2})}
\right].
\label{eq:W12largen}
\end{equation}
Unlike the bosonic loop, the fermionic one receives contributions from both cuts, with opposite signs as in \eqref{eq:rhoinfty}. Their outer endpoints have the same modulus, so the two contributions are equally large and \eqref{eq:W12largen} oscillates in $n$ with frequency $\omega=\frac\pi2-\im\mu_*$, half the angle between the endpoints. At weak coupling $\omega\to\frac\pi2$, giving the alternating pattern of the weak-coupling expansion, with even windings suppressed. At strong coupling $\omega\simeq2/\kk$, so the oscillations only set in for $n\gtrsim\kk$, the same crossover found above for the bosonic loop.

\subsection{Logarithmic periods}
\label{sec:logGM}

Before evaluating directly the derivative at $n=1$ of the analytically continued Wilson loops, we study the logarithmic periods that it produces as
we did in section~\ref{sec:setup} for the ordinary ones.  By \eqref{eq:Wn}, the bosonic multiply wound loop is proportional to the difference $J_{n+1}-J_{n-1}$ of the periods. The derivative in the winding inserts a $\log Z$ in the differential. Then, we can define the following quantity
\begin{equation}
\widehat J_n(\kk):=\partial_nJ_n =\int_{a^{-1}}^{a}Z^n\log Z\,\frac{dZ}{u}.
\label{eq:Itilde}
\end{equation}
that we define \emph{logarithmic period}.
In terms of them, the derivatives in the winding of \eqref{eq:Wn} and of the first relation \eqref{eq:dWn} read
\begin{equation}
\partial_n\big(4\pi^2n\la\vev{W^{1/6}_n}\big)=\widehat J_{n+1}-\widehat J_{n-1},\qquad
\dk\,\partial_n\big(\la\vev{W^{1/6}_n}\big)=\frac{i}{4\pi^2}\,\widehat J_n ,
\label{eq:Wnmoment}
\end{equation}
so that for  $n=1$ we have
\begin{equation}
\partial_n\log\vev{W^{1/6}_n}\Big|_{n=1}=-1+\frac{\widehat J_2-\widehat J_0}{J_2-J_0},
\label{eq:dlogW}
\end{equation}
where the $-1$ derives from the explicit dependence on $n$ of the normalisation of \eqref{eq:Wn}. In the matrix model the logarithm is the eigenvalue itself,
and by the second relation \eqref{eq:Wnmoment} the logarithmic period $\widehat J_n$ is, up to the factor in front,
the $\kk$-derivative of the $n$-th moment of the eigenvalue density with one more insertion of the eigenvalue
$\la\vev{\tfrac1N\sum_i\mu_i\,e^{n\mu_i}}$.

The integrals \eqref{eq:Itilde} are defined for every complex $n$. The cut, the support of the first node's density, is an arc from $a^{-1}$ to $a$ through $Z=1$. It avoids $Z=0$ and $Z=\infty$ and is mapped to itself by the reflection $Z\to1/Z$. On it $\log Z=\mu$ is single valued, and $\widehat J_s$ is half the integral around $A$. Since $\log Z$ branches over $Z=0$ and $Z=\infty$, the logarithmic periods are periods of the curve with these points removed, the setting of the flat coordinates of local mirror symmetry \cite{Aganagic:2002wv,Marino:2009jd}.

As for the ordinary ones, only a few logarithmic periods are independent, and two mechanisms reduce them. The first
is a symmetry of the curve. Indeed, the curve \eqref{eq:curveexp} gives $P(1/Z)=Z^{-4}P(Z)$, so that
$(Z,u)\mapsto(1/Z,u/Z^2)$ is a symmetry of $\Sigma_\kk$. It exchanges $a$ with $a^{-1}$ and $-b$ with
$-b^{-1}$ reversing the cycle $A$. Under it, for any complex $n$ the periods transform as follows
\begin{equation}
J_{-n}=J_n,\qquad \widehat J_{-n}=-\widehat J_n,\qquad\text{in particular}\qquad \widehat J_0=0 .
\label{eq:iotarel}
\end{equation}
In the matrix model this is the reflection of all the eigenvalues used in section \ref{sec:tower}, which makes the
eigenvalue density even.

The second mechanism is the five-term relation \eqref{eq:contig} that holds for any complex $\ell=n$. Differentiating it in $n$, we obtain
\begin{equation}
\sum_{j=0}^{4}p_j\,(2n+j)\,\widehat J_{n-1+j}=-2\sum_{j=0}^{4}p_j\,J_{n-1+j} ,
\label{eq:logcontig}
\end{equation}
Since the coefficient $2n+4$ of the highest index does not vanish, every $\widehat J_m$ with $m\ge3$
is a combination of $\widehat J_0$, $\widehat J_1$, $\widehat J_2$ and of ordinary periods. For instance, considering that $\widehat
J_0=0$, we have
\begin{equation}
\widehat J_3=\frac{\kk^2+2}{2}\,\widehat J_1+\frac{3i\kk}{2}\,\widehat J_2+\frac{3i\kk}{4}\,J_0
+\frac{\kk^2}{4}\,J_1+\frac{i\kk}{4}\,J_2 .
\label{eq:Jt3}
\end{equation}
Exactly as the ordinary periods, the logarithmic ones obey a Gauss-Manin system. The reduction of appendix~\ref{sec:GM} writes the derivative of $Z^m\,dZ/u$ as the combination
$\sum_jc^{(m)}_jZ^j\,dZ/u$ plus an exact differential $d(\beta_m/u)$, with $\beta_m$ a polynomial. On the ordinary periods the exact term integrates to zero, while in the logarithmic case it does
not. Integrating by parts on the cycle
$\oint_A\log Z\,d(\beta_m/u)=-\oint_A(\beta_m/u)\,dZ/Z$, turning the exact differential into ordinary periods obtaining
\begin{equation}
\dk\widehat J_m=\sum_jc^{(m)}_j\,\widehat J_j+s_m,\qquad\text{with}
\qquad s_m:=-\int_{a^{-1}}^a\frac{\beta_m(Z)}{Z}\,\frac{dZ}{u} \,,
\label{eq:logGM}
\end{equation}
with $\beta_m$ the polynomials of appendix~\ref{sec:GM}. Simplifying the sources with \eqref{eq:GM0} and
using the flat invariant \eqref{eq:J1}, the three rows read
\begin{align}
\dk\widehat J_0
&={-\frac{\kk(\kk^2+4)}{\Dsc}\widehat J_0-\frac{4i\kk^2}{\Dsc}\widehat J_1+\frac{4\kk}{\Dsc}\widehat J_2}
\;+\;\dk J_0{-\frac{2\pi\kk^2}{\Dsc}},
\label{eq:logGM0}\\
\dk\widehat J_1
&={\frac{3i\kk^2}{\Dsc}\widehat J_0+\frac{\kk^3}{\Dsc}\widehat J_1+\frac{i\kk^2}{\Dsc}\widehat J_2}
\;+\;{\frac{7i\kk^2}{\Dsc}J_0+\frac{2\kk^3}{\Dsc}J_1+\frac{i\kk^2}{\Dsc}J_2},
\label{eq:logGM1}\\
\dk\widehat J_2
&={-\frac{\kk(\kk^2+4)}{\Dsc}\widehat J_0+\frac{i\kk^2(\kk^2+12)}{\Dsc}\widehat J_1+\frac{4\kk}{\Dsc}\widehat J_2}
\;+\;\dk J_0{-\frac{2\pi\kk^2}{\Dsc}}+iJ_1 .
\label{eq:logGM2}
\end{align}
The logarithmic periods thus extend the system of section \ref{sec:curve} by itself. In other words, the logarithmic periods obey the same Gauss-Manin system as the ordinary ones, but with a source. Gauss-Manin systems with a source are familiar from the relative periods of open topological strings
\cite{Lerche:2002ck,Morrison:2007bm}. Here the source is fixed by the curve itself.

We now solve this system in the steps of section \ref{sec:threeperiods}. There, one period was known from the start in terms of the coupling. Here, one logarithmic period is known as well from \eqref{eq:iotarel} and it vanishes. Since its derivative vanishes too, the first row \eqref{eq:logGM0} is not a differential equation but a linear relation which reads
\begin{equation}
\widehat J_2-i\kk\,\widehat J_1=\pi\kk-\big(J_2-J_0\big) \,,
\label{eq:LJ}
\end{equation}
where it was used also \eqref{eq:J2J0}. The second step was to use the flat invariant $\mathcal Q=J_1-\tfrac{i\kk}4J_0$, annihilated by the connection. Its logarithmic counterpart is not annihilated because of the sources. However, the sources combine into the
holomorphic period such that
\begin{equation}
\dk\Big(\widehat J_1-\frac{i\kk}{4}\,\widehat J_0\Big)=\frac{i}{2}\,J_0=2\pi^2\,\dk\la .
\label{eq:logflat}
\end{equation}
Using that the logarithmic periods vanish at $\kk=0$ and that $\widehat J_0=0$ we have
\begin{equation}
\widehat J_1=2\pi^2\la .
\label{eq:lamL1}
\end{equation}
The flat invariant of the ordinary periods is a residue, while its logarithmic counterpart is the flat coordinate. At $n=0$ the first
relation \eqref{eq:Wnmoment} reads $4\pi^2\la\vev{W^{1/6}_0}=\widehat J_1-\widehat J_{-1}=2\widehat J_1$, so that
\eqref{eq:lamL1} is equivalent to the statement $\vev{W^{1/6}_0}=1$. The logarithmic period $\widehat J_1$ counts the eigenvalues, as the
$A$-period of the resolvent does in \eqref{eq:lambdafed}. The two statements are in fact the same. Indeed, integrating
\eqref{eq:lambdafed} by parts, with the derivative \eqref{eq:dZrho} of the resolvent along the curve, the logarithm
moves from the resolvent onto $Z$.

Finally, the difference of \eqref{eq:logGM2} and \eqref{eq:logGM0}
$\dk\big(\widehat J_2-\widehat J_0\big)=i\widehat J_1+iJ_1$ is equivalent to the derivative in the winding at $n=1$ of the
recursion \eqref{eq:Jrec}. This can be solved just substituting the results and using that $\dk(J_2-J_0)=iJ_1$. The
system is therefore solved with no integration other than that of \eqref{eq:logflat}, and the three basic
logarithmic periods
are expressed through the coupling and its derivatives with coefficients polynomial in $\kk$ as follows
\begin{equation}\begin{split}
\widehat J_0&=0,\\
\widehat J_1&=2\pi^2\la,\\
\widehat J_2&=\frac{\pi\kk}{2}+2\pi^2i\kk\,\la+i\pi^2\kk(16+\kk^2)\,\partial_\kk^2\la ,
\label{eq:logbasic}
\end{split}\end{equation}
Comparing with the ordinary periods at the end of section \ref{sec:threeperiods}, the list contains one new transcendental function: the flat coordinate $\la$ itself. By \eqref{eq:logcontig}, the same holds for every logarithmic period of integer index.
This is precisely the information that the derivative in the winding adds to the integer tower. The loops of integer
winding \eqref{eq:towerclosed} need only $1$, $\dk\la$ and $\partial_\kk^2\la$, while
their derivative in $n$ brings in $\la$ through $\widehat J_1$.

\subsection{\texorpdfstring{$\Bhalf$}{B1/2} and the winding velocity from the winding derivative}
\label{sec:B12wind}

Once the logarithmic Gauss-Manin system is solved, one can easily perform derivatives with respect to the winding in terms of known data. We first recall from section \ref{sec:tower} that the sum of the
loops of the two nodes is elementary $\vev{W^{1/6}_1}+\vev{\hat{W}^{1/6}_1}
=2\vev{W^{1/2}}=\kk/4\pi\la$. Then, it is easy to recognise that the right-hand side of \eqref{eq:LJ} is the period of the loop of the second node such that
\begin{equation}
\widehat J_2-i\kk\,\widehat J_1=4\pi^2\la\vev{\hat{W}^{1/6}_1}=\frac{\pi\kk}{2}\,\big(1-8\pi i\,\Bhalf\big) :
\label{eq:LJhalf}
\end{equation}
where we used \eqref{eq:BJ} to write the difference of ordinary periods in terms of the Bremsstrahlung function. At $n=1$ the first relation \eqref{eq:Wnmoment} reads $4\pi^2\la\big(\vev{W^{1/6}_1}+\partial_n\vev{W^{1/6}_n}|_{n=1}\big)
=\widehat J_2-\widehat J_0=\widehat J_2$. With \eqref{eq:lamL1}, the derivative of
the wound loop at $n=1$ is then
\begin{equation}
\partial_n\vev{W^{1/6}_n}\Big|_{n=1}=\vev{\hat{W}^{1/6}_1}-\vev{W^{1/6}_1}+\frac{i\kk}{2}
=4\pi i\,\big(\la-4\Bhalf\big)\,\vev{W^{1/2}} ,
\label{eq:windphase}
\end{equation}
where the second form follows from \eqref{eq:W16W12tanphi}, \eqref{eq:framing} and $\vev{W^{1/2}}=\kk/8\pi\la$. On the physical slice $\vev{\hat{W}^{1/6}_1}=\overline{\vev{W^{1/6}_1}}$, and the derivative is purely imaginary. This statement is equivalent to the fact that the sum of the loops of the two nodes is stationary in the winding, namely
$\partial_n\log\big(\vev{W^{1/6}_n}+\vev{\hat W^{1/6}_n}\big)|_{n=1}=0$. This is the first term of \eqref{eq:B12split} written
for the winding in \cite{Bianchi:2014laa}, where it was observed to vanish at weak and strong coupling. The elliptic curve gives an independent exact derivation.

The $\tfrac12$BPS loop follows immediately. Continued to real winding through the node relation
$\vev{W^{1/2}_n}=\re\big(i^{1-n}\vev{W^{1/6}_n}\big)$ as in \eqref{eq:W12cont}, its derivative at $n=1$ has two
contributions. The one proportional to the derivative of the $\tfrac16$BPS vanishes since \eqref{eq:windphase} is purely imaginary, while the other one together with \eqref{eq:framing} and \eqref{eq:LJhalf} gives
\begin{equation}
\Bhalf=\frac1{4\pi^2}\,\partial_n\log\vev{W^{1/2}_n}\Big|_{n=1}
=-\frac{1}{4\pi^2\kk}\,\im\big(\widehat J_2-i\kk\,\widehat J_1\big) .
\label{eq:windphase2}
\end{equation}
Then the fermionic Bremsstrahlung function is a winding derivative of the $\tfrac12$BPS loop and it can be written through logarithmic periods as well. A similar mechanism appears in 
$\mathcal N=4$ super Yang-Mills. There the wound circle depends on the winding only
through $n^2\la$, so that $\tfrac1{4\pi^2}\partial_n\log\vev{W_n}|_{n=1}=\tfrac1{2\pi^2}\la\partial_\la\log\vev{W}=B$. In ABJM the derivatives in the winding and in the coupling do not coincide. 

\begin{figure}[t]
\centering
\includegraphics[width=0.6\textwidth]{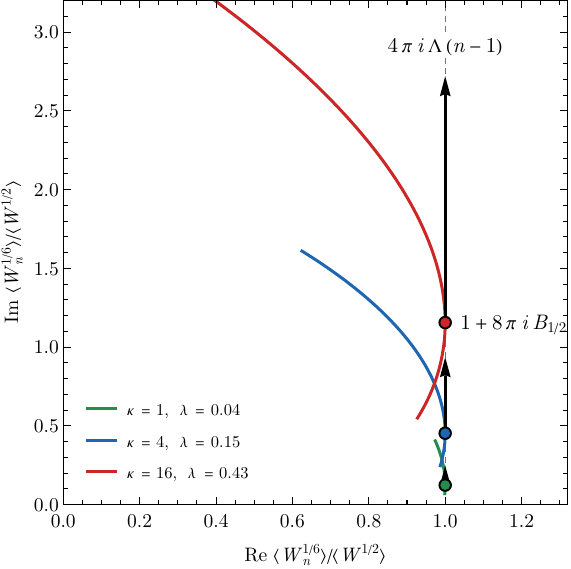}
\caption{The wound bosonic loop $\vev{W^{1/6}_n}/\vev{W^{1/2}}$ in the complex plane for $\tfrac34\le n\le\tfrac32$,
at three couplings. The winding grows upwards along each curve. At $n=1$ (dots) the loop sits at $1+8\pi i\Bhalf$, at
the height fixed by the framing phase, on the vertical line through $1$ (dashed). There it moves orthogonally to the
real axis, as in \eqref{eq:Wnear1}. The arrows are the first-order term $4\pi i\Lcl\,(n-1)$ of \eqref{eq:Wnear1} at
$n=\tfrac32$: their tips are where the loop would be if it kept its initial velocity. At weak coupling the loop follows
the arrow, while at strong coupling it bends away.}
\label{fig:traj}
\end{figure}

\begin{figure}[t]
\centering
\includegraphics[width=\textwidth]{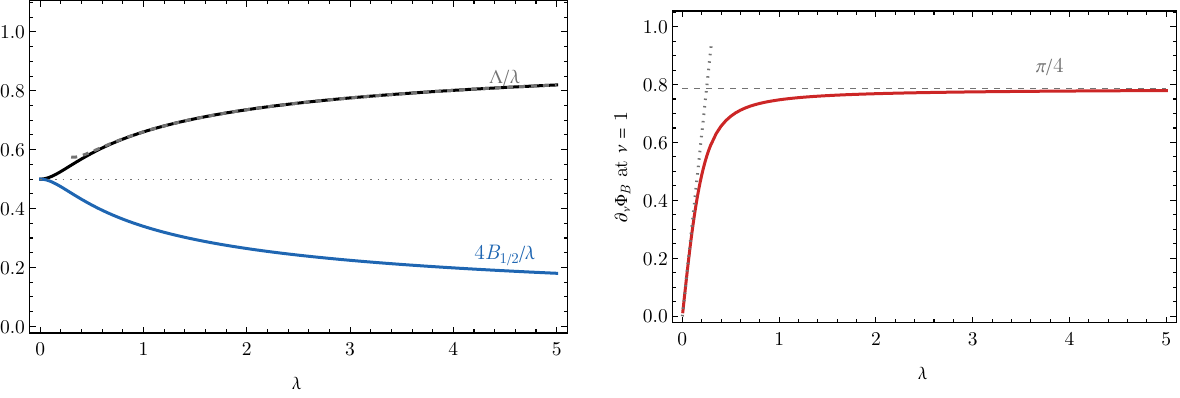}
\caption{Left: the winding velocity $\Lcl$ (black) and $4\Bhalf$ (blue) in units of $\la$. By \eqref{eq:Lamthm} the
two curves add up to one. Both start from $\tfrac12$ (dotted) at weak coupling, and at strong coupling the winding
velocity takes over, approaching its strong-coupling form $\Lcl\simeq\la-\tfrac1\pi\sqrt{2\lah}+\tfrac1{\pi^2}$
(dashed). Right: the slope \eqref{eq:dPhinumain} of the phase of the bosonic latitude at the great circle. It starts
from its weak-coupling value $\pi\la$ (dotted) and tends to $\pi/4$ at strong coupling (dashed), which is half the rate
$\pi/2$ at which the phase $i^n$ of the $n$-wound string grows with the winding.}
\label{fig:windvel}
\end{figure}

The derivative \eqref{eq:windphase} is fixed by a single real function of the coupling. We name it the
\emph{winding velocity} $\Lcl$ of the wound loop defined by
\begin{equation}
\partial_n\vev{W^{1/6}_n}\Big|_{n=1}=4\pi i\,\Lcl\,\vev{W^{1/2}} ,
\label{eq:windveldef}
\end{equation}
so that, comparing with \eqref{eq:windphase}, it is simply given by
\begin{equation}
\Lcl=\la-4\Bhalf .
\label{eq:Lamthm}
\end{equation}
Besides $\Bhalf$, it is the second datum of the wound loop on the great circle, and it is worth understanding it in some
detail. Near $n=1$, by \eqref{eq:W16W12tanphi} the bosonic multiply wound Wilson loop can be written as follows
\begin{equation}
\vev{W^{1/6}_n}=\vev{W^{1/2}}\Big[1+8\pi i\,\Bhalf+4\pi i\,\Lcl\,(n-1)+O\big((n-1)^2\big)\Big] .
\label{eq:Wnear1}
\end{equation}
In the complex plane the loop starts from the point fixed by the framing phase and, as the winding grows it moves
orthogonally to the real axis with speed $4\pi\Lcl$ in units of the $\tfrac12$BPS circle (figure~\ref{fig:traj}). The name is borrowed from $\mathcal N=4$ SYM, where the derivative in the winding at $n=1$
measures the rate at which the effective coupling $n^2\la$ moves with the winding. At weak coupling it gives the following expansion in terms of $\la$
\begin{equation}
\Lcl=\frac\la2+\frac{\pi^2\la^3}{12}-\frac{\pi^4\la^5}{15}+\frac{841\,\pi^6\la^7}{10080}+O(\la^9),
\label{eq:Lamweak}
\end{equation}
while at strong coupling \eqref{eq:term} gives $\Lcl=\la-\tfrac1\pi\sqrt{2\lah}+\tfrac1{\pi^2}$ up to worldsheet
instantons (figure~\ref{fig:windvel}, left).

The winding velocity is also the observable anticipated in section \ref{sec:mainthm}: indeed by \eqref{eq:meanbody} it is the running
mean of the coupling $\Lcl=\Cop_\kk\la$. This can be read off the logarithmic periods directly. Using the second relation
\eqref{eq:Wnmoment} at $n=1$ together with \eqref{eq:lamL1}, we have
\begin{equation}
\dk\,\partial_n\big(\la\vev{W^{1/6}_n}\big)\Big|_{n=1}=\frac{i}{4\pi^2}\,\widehat J_1=\frac{i\la}{2} ,
\label{eq:mixed}
\end{equation}
while $\la\,\partial_n\vev{W^{1/6}_n}|_{n=1}=\tfrac{i\kk}2\Lcl$ by \eqref{eq:windveldef} and \eqref{eq:Wexact}. Hence
\begin{equation}
\dk\big(\kk\Lcl\big)=\la,\qquad\text{i.e.}\qquad \Lcl=\frac1\kk\int_0^\kk\la(\kk')\,d\kk'=\Cop_\kk\la .
\label{eq:Lam}
\end{equation}
The two basic logarithmic periods give the winding velocity twice: $\widehat J_1$ through the mixed derivative
\eqref{eq:mixed} as the running mean of the coupling, and $\widehat J_2$ through \eqref{eq:windphase} as
$\la-4\Bhalf$. Equating the two gives back the running-mean form \eqref{eq:meanbody} of $\Bhalf$, and with it
\eqref{eq:Bint} and \eqref{eq:main}, without evaluating the framed loop. Equation
\eqref{eq:Lam} is also equivalent to
\begin{equation}
\Bhalf=\frac\kk4\,\dk\Lcl\,,
\end{equation}
so one can read the framing phase as the derivative of
the wound loop velocity. 

In hypergeometric form, the running mean
multiplies the coefficient of $\kk^{2j+1}$ of the mirror map \eqref{eq:lamkappaintro} by $1/(2j+2)$, which raises its
lower parameter $1$ to $2$ and halves the prefactor giving
\begin{equation}
\Lcl=\frac{\kk}{16\pi}\;{}_3F_2\left(\genfrac{}{}{0pt}{}{\frac12,\ \frac12,\ \frac12}{\frac32,\ 2}\,\Biggl|\,-\frac{\kk^2}{16}\right).
\label{eq:Lam3F2}
\end{equation}
The same results follow directly from the closed form \eqref{eq:tower4F3}, which holds at any complex winding. We write it as
\begin{equation}
\vev{W^{1/6}_n}=\frac{\kk}{8\pi\la}\Big[\mathcal F_1(n)+\frac{i\,n^2\kk}{8}\,\mathcal F_2(n)\Big],
\label{eq:W4F3short}
\end{equation}
where $\mathcal F_1$ and $\mathcal F_2$ are the two ${}_4F_3$ functions. Both simplify at $n=1$. In $\mathcal F_1$ the upper parameter $(1-n)/2$ appears twice, so every term of the series except the first has a double zero at $n=1$, so that
\begin{equation}
\mathcal F_1(1)=1,\qquad\text{and}\qquad \partial_n\mathcal F_1(n)\big|_{n=1}=0\,.
\end{equation}
In $\mathcal F_2$, two of the upper parameters become $3/2$ that cancel against the lower ones. Together with \eqref{eq:BPVreduced} and \eqref{eq:Lam3F2}, this gives
\begin{align}
\mathcal F_2(1)&={}_2F_1\left(\genfrac{}{}{0pt}{}{\frac12,\ \frac12}{2}\,\Biggl|\,-\frac{\kk^2}{16}\right)=\frac{64\pi}{\kk}\,\Bhalf,\label{eq:F12atone}\\
\partial_n\big[n^2\mathcal F_2(n)\big]_{n=1}&=2\;{}_3F_2\left(\genfrac{}{}{0pt}{}{\frac12,\ \frac12,\ \frac12}{\frac32,\ 2}\,\Biggl|\,-\frac{\kk^2}{16}\right)=\frac{32\pi}{\kk}\,\Lcl.\nonumber
\end{align}
Since $\mathcal F_1$ is flat at $n=1$, both data of the wound loop sit in the second series: $\Bhalf$ and $\Lcl$ are the value and the slope at $n=1$ of the single function $\kk\,n^2\mathcal F_2(n)/64\pi$. Near $n=1$ this function is the framing phase \eqref{eq:phasen},
\begin{equation}
\frac1{8\pi}\tan\left(\arg\vev{W^{1/6}_n}\right)=\Bhalf+\frac{\Lcl}{2}\,(n-1)+O\big((n-1)^2\big),
\end{equation}
in agreement with \eqref{eq:Wnear1}.

With $\Bhalf$ and $\Lcl$ at hand, we can compute the slope that this section set out to find. By \eqref{eq:dPhinu}, the slope of the phase of the latitude is half the rate at which the phase of the wound loop grows with the winding. By \eqref{eq:Wnear1} the tangent of this phase grows at the rate $4\pi\Lcl$, so the phase itself grows at the rate $4\pi\Lcl\cos^2\PhB$, then it gives
\begin{equation}
\partial_\nu\PhB(\nu)\Big|_{\nu=1}=2\pi\,\Lcl\cos^2\PhB=\frac{2\pi\big(\la-4\Bhalf\big)}{1+64\pi^2\Bhalf^2} .
\label{eq:dPhinumain}
\end{equation}
This slope, shown on the right of figure~\ref{fig:windvel}, is the last ingredient required by the relation \eqref{eq:BPVrel}.

\section{The bosonic Bremsstrahlung function from the curve}
\label{sec:bosonic}

Section~\ref{sec:logperiods} has supplied the one datum that the bosonic function needs beyond the great circle. The
derivative of the wound loop at $n=1$ is fixed by two real functions of the coupling, $\Bhalf$ and the winding velocity
$\Lcl$, and with them the slope \eqref{eq:dPhinumain} of the phase of the bosonic latitude. In this section we assemble
the bosonic Bremsstrahlung function and study it as a function of the coupling.

\subsection{\texorpdfstring{$\Bsix$}{B1/6} is a ratio of periods}
\label{sec:B16}

The winding prescription \eqref{eq:LM} to compute the bosonic Bremsstrahlung function takes the modulus of the wound loop because its phase depends on the framing
\cite{Lewkowycz:2013laa,Bianchi:2018scb}. Since the logarithm of the modulus is the real part of the logarithm, we just rewrite it as
\begin{equation}
\Bsix=\frac{1}{4\pi^2}\,\re\,\partial_n\log\vev{W^{1/6}_n}\Big|_{n=1},
\label{eq:B16def}
\end{equation}
and all that is needed is the derivative of the multiply winding bosonic Wilson loop at $n=1$. Section~\ref{sec:logperiods} gives it in two independent ways.

The first is the Gauss-Manin system of the logarithmic periods. By \eqref{eq:dlogW} the derivative is a ratio of a
logarithmic and ordinary periods, and the solution \eqref{eq:logbasic} together with \eqref{eq:BJ} gives
\begin{equation}
\partial_n\log\vev{W^{1/6}_n}\Big|_{n=1}=-1+\frac{\widehat J_2-\widehat J_0}{J_2-J_0}
=\frac{4\pi i\,\Lcl}{1+8\pi i\,\Bhalf}.
\label{eq:ratio}
\end{equation}
where $\Lambda=\la-4\Bhalf$ is the winding velocity defined in \eqref{eq:windveldef}.
The second needs no logarithmic periods at all. It differentiates the wound loop itself, in the closed form computed in \eqref{eq:tower4F3} that holds at every complex winding. There the real part is stationary at $n=1$, while
 the imaginary one has value $\Bhalf$ and slope $\tfrac12\Lcl$. (section \ref{sec:B12wind}), namely
$\vev{W^{1/6}_1}=(1+8\pi i\Bhalf)\vev{W^{1/2}}$ and $\partial_n\vev{W^{1/6}_n}|_{n=1}=4\pi i\Lcl\vev{W^{1/2}}$. The derivative of the logarithm is just their ratio that indeed gives again \eqref{eq:ratio}.

Now, taking the real part of \eqref{eq:ratio}, we can reconstruct the exact closed form of bosonic Bremsstrahlung function to be
\begin{equation}
\Bsix=\frac{8\,\Bhalf\,\Lcl}{1+64\pi^2\Bhalf^{2}}=\frac{\Lcl}{2\pi}\,\sin2\PhB ,\qquad\text{with}\qquad \Lcl=\la-4\Bhalf ,
\label{eq:B16}
\end{equation}
where $1+64\pi^2\Bhalf^2=\sec^2\PhB$ and $8\pi\Bhalf=\tan\PhB$. 
Since $\sin2\PhB\le1$, the bosonic function is bounded by the winding velocity $\Bsix\le\Lcl/2\pi$ with equality only at $\PhB=\pi/4$, that corresponds to $\Bhalf=1/8\pi$.
The imaginary part of \eqref{eq:ratio} $4\pi\Lcl\cos^2\PhB$ is twice the
slope \eqref{eq:dPhinumain} of the latitude phase, as \eqref{eq:dPhinu} requires. Another way to compute the same quantity starts from that slope.
Inserted in the relation \eqref{eq:BPVrel}, it gives $\tfrac4\pi\Bhalf\cdot2\pi\Lcl\cos^2\PhB$, which is \eqref{eq:B16}
again. The two exact prescriptions of section \ref{sec:sota:brem} therefore use the real and the imaginary part of the same
derivative \eqref{eq:ratio}, and they agree because the derivative of the wound loop itself is purely imaginary.

As written, the closed form is a projection. By \eqref{eq:Wnear1} the wound loop sits at the
angle $\PhB$ and moves orthogonally to the real axis as the winding grows with speed $4\pi\Lcl$ in units of the
$\tfrac12$BPS circle. The radial component of this motion is $4\pi\Lcl\sin\PhB\cos\PhB$ (which is
$4\pi^2\Bsix$) and the angular one is $4\pi\Lcl\cos^2\PhB$, that is twice the slope of the latitude phase. In other words, the bosonic Bremsstrahlung function is the radial part of the motion and needs both data: the angle at which the loop sits and the speed at which it moves. Moreover, using the relation between winding velocity and $\Bhalf$, formula \eqref{eq:B16} can be seen as a trigonometric polynomial in the framing phase as follows
\begin{equation}
\Bsix=\frac{\la}{2\pi}\,\sin2\PhB-\frac{1}{2\pi^2}\,\sin^2\PhB ,
\label{eq:B16sincos}
\end{equation}
so that, given the framing phase of the circle, the bosonic function needs only the coupling and no derivative, as the
fermionic one does in \eqref{eq:framing}.

Given the hypergeometric forms \eqref{eq:BPVreduced} of
$\Bhalf$ and \eqref{eq:Lam3F2} of $\Lcl$, one can express \eqref{eq:B16} as the following explicit rational function
\begin{equation}
\Bsix=\frac{\kk^2}{2\pi^2}\;\frac{{}_2F_1\left(\genfrac{}{}{0pt}{}{\frac12,\ \frac12}{2}\,\Biggl|\,\frac{-\kk^2}{16}\right)\,{}_3F_2\left(\genfrac{}{}{0pt}{}{\frac12,\ \frac12,\ \frac12}{\frac32,\ 2}\,\Biggl|\,\frac{-\kk^2}{16}\right)}{64+\kk^2\;\,{}_2F_1\left(\genfrac{}{}{0pt}{}{\frac12,\ \frac12}{2}\,\Biggl|\,\frac{-\kk^2}{16}\right)^2}.
\label{eq:B16hyp}
\end{equation}
that is exactly the same result one obtains applying the prescription \eqref{eq:B16def} to the analytically continued quantity \eqref{eq:tower4F3}. The periods of the curve $\dk\la$ and $\Bhalf$ are Gauss functions ${}_2F_1$ of $-\kk^2/16$, while $\la$ and $\Lcl$ are ${}_3F_2$'s because they are
primitives of periods: $\Bsix$ is a rational function of one function of each kind. Therefore, it loses the two
properties that characterise $\Bhalf$ in section~\ref{sec:deriv}, a second-order equation and quasimodularity. First, $\Bsix$ obeys a nonlinear equation of order three, because
$\Bhalf$, $\dk\Bhalf$ and $\Lcl$ are algebraically independent. Second, $\la$ is a primitive in $\tau$ of the weight-three modular form $\thh2^2\thh3^4$: the fermionic function lives in the quasimodular world while the
bosonic one leaves it through the winding velocity $\Lcl$.

At weak coupling, since $\Bhalf\simeq\la/8$ and $\Lcl\simeq\la/2$, the bosonic function starts one order later than the fermionic and its expansion reads
\begin{equation}
\Bsix=\frac{\la^{2}}2-\frac{\pi^{2}\la^{4}}2+\frac{47\pi^{4}\la^{6}}{72}-\frac{17\pi^{6}\la^{8}}{18}
+\frac{73667\,\pi^{8}\la^{10}}{50400}-\frac{3252239\,\pi^{10}\la^{12}}{1360800}+O(\la^{14})
\label{eq:B16weak}
\end{equation}
that reproduces the known planar series of \cite{Lewkowycz:2013laa}.

\subsection{Radiation and entropy along the defect flow}
\label{sec:B16bounds}

The two Bremsstrahlung functions belong to two lines that are related by a defect renor\-mal\-i\-sa\-tion group flow. ABJM has a
family of fermionic loops that interpolates between the bosonic $\tfrac16$BPS line and the fermionic $\tfrac12$BPS line,
with the bosonic line at the ultraviolet end of the flow and the fermionic one at the infrared end
\cite{Castiglioni:2022yes,Castiglioni:2023uus,Drukker:2019bev}. Two quantities compare the ends of such a flow: the defect
entropy, which decreases along it, and the radiation. We first obtain the drop of the entropy $\Delta s$ in closed form from the polar
decomposition \eqref{eq:W16W12tanphi}, and then compare the radiation through the closed form \eqref{eq:B16}.

A conformal line defect carries a $g$-function, the finite part of its contribution to the partition function on the sphere, which for a Wilson line is the expectation value of the circular loop. Its scale-dependent generalisation is the defect entropy. It
decreases along any defect flow in a reflection-positive theory \cite{Cuomo:2021rkm,Kobayashi:2018lil}, so that
$g_{\rm UV}>g_{\rm IR}$. At the infrared end $g_{\rm IR}=\vev{W^{1/2}}$. At the ultraviolet end the localisation result is
complex, with a phase that depends on the framing. The $g$-function is the circle in a scheme that preserves parity (framing zero), where by parity it is real and equal to the framing-independent modulus
$g_{\rm UV}=\bigl|\vev{W^{1/6}}\bigl|$ \cite{Bianchi:2016yzj,Bianchi:2017svd,Bianchi:2018scb}.
By \eqref{eq:W16W12tanphi} the two $g$-functions are the modulus and the real part of the same complex
number, and
\begin{equation}
\Delta s:=\log\frac{g_{\rm UV}}{g_{\rm IR}}=-\log\cos\PhB=\frac12\log\big(1+64\pi^2\Bhalf^2\big),\qquad
\log\frac{\vev{W^{1/6}}}{\vev{W^{1/2}}}=\Delta s+i\,\PhB .
\label{eq:gdrop}
\end{equation}
The drop is positive and grows monotonically with the coupling, from $0$ to $\infty$, as $\Bhalf$
does. 
The inequality $g_{\rm UV}>g_{\rm IR}$ is automatic once the two $g$-functions are the modulus and the real part of one number, and what \eqref{eq:gdrop} adds is the value of the drop. It also shows that the framing phase, usually regarded as a regularisation artefact, is the imaginary partner of the entropy drop, which does not depend on the scheme. Since the drop depends only on $\Bhalf^2$, it is positive as soon as $\Bhalf\neq0$. At weak coupling, inserting the series \eqref{eq:recursion} of $\Bhalf$ and the mirror map into \eqref{eq:gdrop} gives
\begin{equation}
\Delta s=\frac{\pi^2}{2}\,\la^2-\frac{5\pi^4}{12}\,\la^4+\frac{173\pi^6}{360}\,\la^6-\frac{661\pi^8}{1008}\,\la^8+O(\la^{10}),
\label{eq:gweak}
\end{equation}
whose leading term, which follows from $\Bhalf\simeq\la/8$, is the two-loop result with which \cite{Castiglioni:2022yes} established the inequality.

Given the closed form \eqref{eq:B16}, it is natural to ask which of the two lines radiates more, and how
does the answer depend on the coupling?
The ratio of the two functions is
\begin{equation}
\frac{\Bsix}{\Bhalf}=8\Lcl\cos^2\PhB=1-\Upsilon\cos^2\PhB,\qquad \text{with}\qquad \Upsilon:=1+64\pi^2\Bhalf^2-8\Lcl\,.
\label{eq:B16ratio}
\end{equation}
Since $\Bhalf$ grows with the coupling, $\cos^2\PhB$ decreases and the ratio grows monotonically as soon as $\Upsilon$ is positive and decreasing. To show that, we differentiate $\Upsilon$. By the running mean \eqref{eq:Lam} $\dk\Lcl=4\Bhalf/\kk$, so that
\begin{equation}
\dk\Upsilon=\frac{32\Bhalf}{\kk}\,\big(4\pi^2\kk\,\dk\Bhalf-1\big).
\label{eq:Dprime}
\end{equation}
Its sign is fixed by the slope $\kk\dk\Bhalf$ of $\Bhalf$ as a function of $\log\kk$. The Picard-Fuchs equation \eqref{eq:ode} gives
\begin{equation}
(\kk\dk)^2\Bhalf=\frac{16\,\Bhalf}{16+\kk^2}>0 ,
\end{equation}
so this slope grows monotonically. At weak coupling $\Bhalf$ and $\Lcl$ vanish and $\Upsilon=1$. At strong coupling the slope tends to $1/4\pi^2$ and it stays below this value at every finite coupling. Hence $\dk\Upsilon<0$.  Now, using the definition of winding velocity and the termination formula of $\Bhalf$ at strong coupling, we can compute the minimum value of $\Upsilon$ that reads
\begin{equation}
\Upsilon_\infty=1-\frac13-\frac4{\pi^2}=\frac23-\frac4{\pi^2}\simeq0.261 .
\label{eq:Dinf}
\end{equation}
So $\Upsilon$ decreases from $1$ to $\Upsilon_\infty$, and the ratio \eqref{eq:B16ratio} increases monotonically from $0$ at weak coupling to $1$ at strong coupling. Finally, since $\cos^2\PhB=e^{-2\Delta s}$ by \eqref{eq:gdrop}, the bounds $\Upsilon_\infty<\Upsilon<1$ can be written in terms of the entropy drop as follows
\begin{equation}
0<\Bsix<\Bhalf,\qquad
\Big(\frac23-\frac4{\pi^2}\Big)\,e^{-2\Delta s}\ <\ 1-\frac{\Bsix}{\Bhalf}\ <\ e^{-2\Delta s},
\label{eq:Bineq}
\end{equation}
at every coupling. The lower bound is approached at strong coupling, the upper one at weak coupling.

\begin{figure}[t]
\centering
\includegraphics[width=0.6\textwidth]{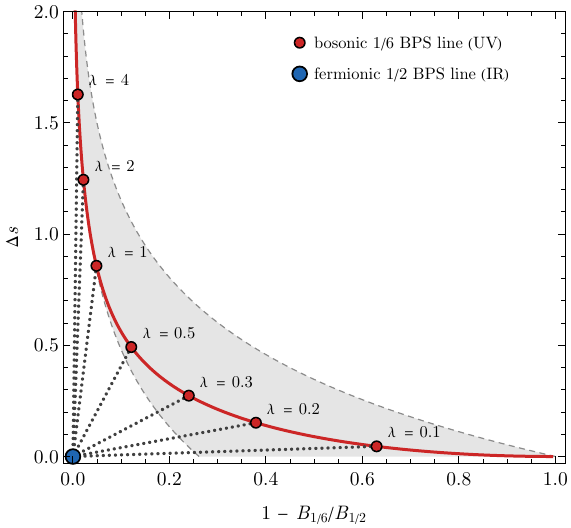}
\caption{The two ends of the defect flow from the bosonic $\tfrac16$BPS line (UV, red dots) to the fermionic $\tfrac12$BPS
line (IR, blue dot). The coordinates are measured from the $\tfrac12$BPS line, so the infrared line sits at the origin,
while the ultraviolet line moves along the red curve as the coupling grows. The dotted segments only pair the two ends
at the same coupling. They are not trajectories: the flow in between is not computed. At weak coupling the segments are horizontal: the two lines differ in
their radiation but not yet in their entropy. At strong coupling they are vertical: their radiation has merged, while the
entropy keeps separating them. The grey band is the region allowed by the bounds \eqref{eq:Bineq}. The red curve starts
on its upper edge and approaches the lower one at strong coupling.}
\label{fig:flow}
\end{figure}

The bosonic line radiates less than the fermionic one at every coupling. Along the flow the defect entropy decreases while
the radiation grows, and \eqref{eq:Bineq} ties the two (figure~\ref{fig:flow}): the relative difference of the two Bremsstrahlung functions is
bounded on both sides by multiples of the square of the ratio $g_{\rm IR}/g_{\rm UV}=e^{-\Delta s}$ of the two
$g$-functions. At weak
coupling the two lines differ maximally because the bosonic line starts to radiate one loop later and so $\Bsix/\Bhalf\simeq4\la$.
At strong coupling they become indistinguishable by their radiation since $\Bsix/\Bhalf\simeq 1- \Upsilon_\infty/8\la$, while their
$g$-functions move apart as $\Delta s\simeq\log\big(2\sqrt{2\la}\big)$, so that the two lines remain distinguished by their entropy long
after their radiation has merged. Since $B=2h_w$ for both lines \cite{Lewkowycz:2013laa,Bachas:2025brp}, \eqref{eq:B16ratio} is also the ratio of their
stress-tensor one-point functions, which merge as well. The same
ordering holds at one loop for the flow of $\mathcal N=4$ super Yang-Mills
from the Wilson loop without scalars to the Maldacena-Wilson loop \cite{Polchinski:2011im}: by the one-loop cusp anomalous
dimension \cite{Polyakov:1980ca,Korchemsky:1987wg} the first has $B=\la/24\pi^2$, below the $\la/16\pi^2$ of the second
\cite{Correa:2012at}. We do not know whether it is a general property of defect flows.

\begin{figure}[t]
\centering
\includegraphics[width=\textwidth]{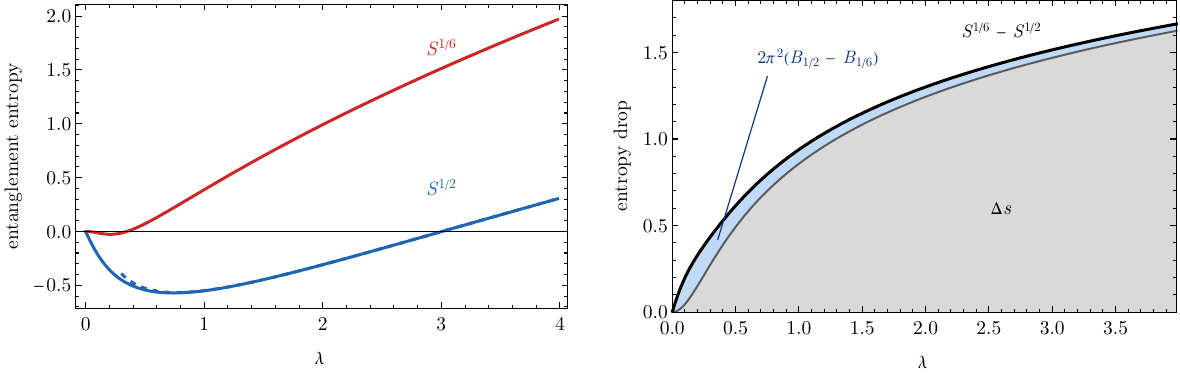}
\caption{Left: the entanglement entropies \eqref{eq:SEE} of the two lines, with the strong-coupling form
\eqref{eq:S12} of $S^{1/2}$ (dashed). Right: the drop of the entanglement entropy along the flow (black), split as in
\eqref{eq:SEEdrop} into the drop $\Delta s$ of the defect entropy (grey) and the excess radiation of the fermionic line
$2\pi^2(\Bhalf-\Bsix)=\tfrac\pi8\,\Upsilon\sin2\PhB$ (blue). At weak coupling the radiation term dominates, being of
one-loop order while $\Delta s$ starts at two loops. At strong coupling it fades and the two drops merge.}
\label{fig:SEE}
\end{figure}

The entanglement entropy of a disk centred on the line is a third quantity at the two ends of the flow. In three
dimensions it is defined as $S=\log g-4\pi^2h_w$ \cite{Lewkowycz:2013laa} and considering that $B=2h_w$, for the two lines is given by 
\begin{equation}
S^{1/6}=\log\big|\vev{W^{1/6}}\big|-2\pi^2\Bsix,\qquad S^{1/2}=\log\vev{W^{1/2}}-2\pi^2\Bhalf ,
\label{eq:SEE}
\end{equation}
up to a constant from the normalisation of the trace, which cancels in the difference. 
Both entanglement entropies follow from \eqref{eq:SEE} in closed form. The bosonic one was computed perturbatively in \cite{Lewkowycz:2013laa} and \eqref{eq:SEE} reproduces that result and extends it to every coupling. The fermionic one, that could not be obtained with the squashed-sphere method, is given by 
\begin{equation}
S^{1/2}=\log\frac{\kk}{8\pi\la}-2\pi^2\Bhalf\simeq
\begin{cases}
-\dfrac{\pi^2}{4}\,\la+\dfrac{\pi^2}{3}\,\la^2+\dfrac{\pi^4}{24}\,\la^3, & \la\ll1,\\[2ex]
\dfrac\pi2\sqrt{2\la-\tfrac{1}{12}}+\dfrac12-\log(8\pi\la), & \la\gg1,
\end{cases}
\label{eq:S12}
\end{equation}
where the strong-coupling form follows from the termination \eqref{eq:term} and holds up to worldsheet instantons (figure~\ref{fig:SEE}, left).

Along the flow, the two entropies drop by different amounts. By \eqref{eq:gdrop} and \eqref{eq:B16ratio},
\begin{equation}
S^{1/6}-S^{1/2}=\Delta s+\frac\pi8\,\Upsilon\,\sin2\PhB\simeq
\begin{cases}
\dfrac{\pi^2}{4}\,\la, & \la\ll1,\\[2ex]
\log\big(2\sqrt{2\la}\big), & \la\gg1.
\end{cases}
\label{eq:SEEdrop}
\end{equation}
The second term is positive because the infrared line radiates more. The entanglement entropy therefore drops by more than the defect entropy (figure~\ref{fig:SEE}, right). At weak coupling it is of one-loop order since the fermionic line stores energy at one loop and the bosonic one only at two, while $\Delta s$ starts at two loops. At strong coupling $\sin2\PhB\to0$ and the drop of the entanglement entropy reduces to the drop of the defect entropy.

\subsection{Strong coupling}
\label{sec:B16strongsec}

At strong coupling the fermionic function terminates up to worldsheet instantons. Since the bosonic function is not linear, it cannot terminate. However, it inherits the
termination in another form. It becomes elementary and it reads
\begin{equation}\begin{split}
B^{\varphi,\rm pert}_{1/6}
&=\frac{x}{4\pi}-\frac1{4\pi^2}
-\frac{\pi^2-6}{24\pi^3}\,\frac1x-\frac{\pi^2-6}{24\pi^4}\,\frac1{x^2}
+\frac{\pi ^4-10 \pi ^2+24}{96 \pi ^5}\frac1{x^3}+O(x^{-4})\\
&=\left(\frac{x}{4\pi}-\frac1{4\pi^2}\right)\left[1-\frac{\Upsilon_\infty}{1+4\big(x-\tfrac1\pi\big)^2}\right]\,,
\label{eq:B16strongclosed}
\end{split}\end{equation}
where $x=\sqrt{2\hat{\la}}$ and $\Upsilon_\infty=\frac23-\frac4{\pi^2}$ is \eqref{eq:Dinf}.

Two features of these results are worth stressing. First, the two lines differ only at order $1/\sqrt\la$ as follows
\begin{equation}
\Bsix-\Bhalf=-\Upsilon\,\Bhalf\cos^2\PhB=-\frac{\pi^2-6}{24\pi^3}\,\frac1x+O(x^{-2}).
\label{eq:B16minusB12}
\end{equation}
The classical string and its one-loop correction reproduce the first two terms of both functions \cite{Aguilera-Damia:2014qgy,Aguilera-Damia:2014bqa} and do not distinguish the lines. Equation \eqref{eq:B16minusB12} is therefore a prediction for a two-loop computation in the string sigma model, comparing the string dual to the bosonic line  with the string dual to the fermionic line.

Second, worldsheet instantons push the two lines apart. At fixed $\la$ they enter only through $\Bhalf$ and, since $\partial_{\Bhalf}\Bsix\to-1$ at strong coupling, the first instanton shifts the two functions by opposite amounts. In the normalisation of \eqref{eq:threeinst} this gives
\begin{equation}
\Bsix-B^{\varphi,\rm pert}_{1/6}=\frac1{\pi^2}\,\Iinst^\varphi_1(\pi x)\,e^{-2\pi x}+O(e^{-4\pi x}),\qquad \text{with}\qquad\Iinst^\varphi_1=\Iinst_1\,\partial_{\Bhalf}\Bsix ,
\end{equation}
so that $\Iinst^\varphi_1(y)\simeq-y$ while $\Iinst_1(y)\simeq y$. The entropy drop, whose perturbative part is elementary as well
\begin{equation}
\Delta s^{\rm pert}=\frac12\log\Big[1+4\big(x-\tfrac1\pi\big)^2\Big]=\log2x-\frac1{\pi x}+O(x^{-2}),
\label{eq:gstrongpert}
\end{equation}
follows $\Bhalf$ and grows as follows
\begin{equation}
\Delta s-\Delta s^{\rm pert}=\frac{16\,(\pi x-1)\,\Iinst_1(\pi x)}{\pi^2\big[1+4\big(x-\frac1\pi\big)^2\big]}\,e^{-2\pi x}
+O(e^{-4\pi x})\ \longrightarrow\ 4\,e^{-2\pi x}.
\label{eq:ginst}
\end{equation}

\subsection{ABJ theory}
\label{sec:ABJ}

The bosonic Bremsstrahlung function closed form extends to unequal ranks. Its derivation used the ABJM curve only
through its Gauss-Manin system that survives the deformation to ABJ theory $U(N_1)_k\times U(N_2)_{-k}$ \cite{Aharony:2008gk} if adapted to the $N_1\neq N_2$ case.
In the planar limit with $\la_i=N_i/k$ fixed, the ABJ matrix model has the spectral curve
\cite{Drukker:2010nc}
\begin{equation}
u^2=\big(Z^2-i\kk e^{i\Theta}Z+1\big)^2-4e^{4i\Theta}Z^2,\qquad \Theta:=\pi(\la_1-\la_2),
\label{eq:ABJcurve}
\end{equation}
that at $\Theta=0$ reduces to \eqref{eq:curve}. The branch points are $a^{\pm1}$ and $-b^{\pm1}$, with
$a+a^{-1}=2e^{2i\Theta}+i\kk e^{i\Theta}$ and $b+b^{-1}=2e^{2i\Theta}-i\kk e^{i\Theta}$. The physical slice is $\kk$ real at fixed $\Theta$ beginning at $\kk_\Theta=4|\sin\Theta|$. Along it $\la_1-\la_2$ is constant and $\dk\la_1=\dk\la_2$ is real and positive. 
The ingredients of sections~\ref{sec:logGM}-\ref{sec:B12wind} carry over line by line. 
The resolvent identity \eqref{eq:dZrho}
holds verbatim for \eqref{eq:ABJcurve}, so that the $n$-wound $\tfrac16$BPS loop on the first node is
\begin{equation}
\vev{W^{1/6}_n}=\frac{J_{n+1}-J_{n-1}}{4\pi^2\la_1\,n}.
\label{eq:ABJWn}
\end{equation}
where $\la$ is replaced by the coupling of the first node $\la_1$.
What changes is the Gauss-Manin system. The Griffiths-Dwork reduction of section \ref{sec:GM} gives the following two
equations for the singly wound combination
\begin{equation}\begin{split}
\partial_\zeta (J_2-J_0)&=J_1,\\
[e^{4i\Theta}\Delta_{\Theta}\,\partial_\zeta^2+(4e^{4i\Theta}-\zeta^2-4)(\zeta\,\partial_\zeta -1)]\,(J_2-J_0)&=-4\zeta^2J_0,
\label{eq:ABJGM}
\end{split}\end{equation}
where $\zeta=i \kk e^{i \Theta}$ and 
\begin{equation}
\Delta_{\Theta}(\kk)=\big(\kk^2-16\sin^2\Theta\big)\big(\kk^2+16\cos^2\Theta\big)
\label{eq:ABJdisc}
\end{equation}
is the discriminant of \eqref{eq:ABJcurve} and it reduces to the one of the ABJM curve at $\Theta=0$, so $\Delta_{\Theta=0}=\Dsc$. Its zeros are the four degenerations: the collapse of one cut at $\kk=\pm4\sin\Theta$, which at
$\Theta=0$ merges into the double zero $\kk^2$ of the weak-coupling point, and the conifold at $\kk=\pm4i\cos\Theta$. The
holomorphic period $J_0=-4\pi^2i\,e^{-i\Theta}\dk\la$ obeys the following Picard-Fuchs equation\footnote{Recently it was studied also in \cite{Suyama:2026nzo}.}
\begin{equation}
\kk\,\Delta_{\Theta}\,\partial_\kk^3\la+\big(\kk\,\Delta_{\Theta}'-\Delta_{\Theta}\big)\,\partial_\kk^2\la+\kk^3\,\dk\la=0 ,
\label{eq:ABJPF}
\end{equation}
which at $\Theta=0$ is \eqref{eq:PFg}. However, there is no analogue of the flat invariant \eqref{eq:defQ} since
$J_1-\tfrac\zeta4J_0$ is purely imaginary but not constant, and the reduced system has rank three rather than two.

The $\tfrac12$BPS loop is the supertrace over the two nodes, and normalised by $N_1+N_2$ its planar value is \cite{Drukker:2010nc}
\begin{equation}
\vev{W^{1/2}}=\frac{\kk\,e^{i\Theta}}{4\pi(\la_1+\la_2)}\,.
\label{eq:ABJW12}
\end{equation}
On the physical slice we write the first-node loop as
\begin{equation}
\vev{W^{1/6}}=\frac{\kk\,e^{i\Theta}}{8\pi\la_1}\,\big(w+8\pi i\,\widehat B\big)=\frac{\la_1+\la_2}{2\la_1}\,\big(w+8\pi i\,\widehat B\big)\vev{W^{1/2}} ,
\label{eq:ABJpolar}
\end{equation}
where in the second relation we used \eqref{eq:ABJW12}.
The two quantities $w$ and $8\pi\widehat B$ are the real and imaginary parts of the rescaled bosonic loop. This is the analogue of \eqref{eq:W16W12tanphi}, where the flat invariant fixes the real part to $w=1$. The prefactor is the share of the first node in the supertrace. In ABJ the Gauss-Manin system fixes both quantities as follows
\begin{equation}
\widehat B=-\frac{\big(\kk^2-16\sin^2\Theta\big)\big(\kk^2+16\cos^2\Theta\big)}{4\kk^2}\,\partial_\kk^2\la_1,\qquad
\partial_\kk w=-\frac{16\pi\sin2\Theta}{\kk^2}\,\dk\la_1,
\label{eq:ABJmain}
\end{equation}
Integrating the second relation from the endpoint where $w(\kk_\Theta)=1+\operatorname{sgn}\Theta$, we have
\begin{equation}
w=1+\operatorname{sgn}\Theta-16\pi\sin2\Theta\int_{\kk_\Theta}^{\kk}\frac{\partial_t\la_1}{t^2}\,dt\ \simeq\
\begin{cases}
\dfrac{2\la_1}{\la_1+\la_2}, & \la_i\ll1,\\[2ex]
1+2(\la_1-\la_2), & \la_i\gg1.
\end{cases}
\label{eq:ABJw}
\end{equation}
At weak coupling this is just equivalent to $\vev{W^{1/6}}\to1$, while at strong coupling the real part is exact up to worldsheet instantons.
The value at the endpoint follows from the loop itself. At $\kk_\Theta$ one node is empty and the theory reduces to pure Chern-Simons theory on the other. For $\Theta>0$ the empty node is the second one, the first-node loop coincides with the $\tfrac12$BPS one, and \eqref{eq:ABJpolar} gives $w=2$ and $\widehat B=0$. For $\Theta<0$ the first node is empty and $w=0$. At $\Theta=0$ \eqref{eq:ABJmain} reduces to \eqref{eq:main}. In both cases the prefactor of the Bremsstrahlung function is the discriminant divided by the square of the modulus. In ABJM this division removes the whole weak-coupling factor $\kk^2$ of the discriminant. In ABJ that factor has split into $\kk^2-16\sin^2\Theta$, which vanishes where one cut collapses, and it survives. Finally, exchanging the nodes reverses the sign of $\Theta$ and leaves $\dk\la_1=\dk\la_2$ unchanged, so that $\widehat B$ is invariant while $w\to2-w$.

The bosonic Bremsstrahlung function follows as in ABJM. The logarithmic Gauss-Manin system reduces to
\begin{equation}
\begin{split}
\widehat J_0&=0,\\
\widehat J_1&=2\pi^2\la_1,\\
\widehat J_2&=2\pi^2\zeta\,\la_1+2\zeta J_1+\tfrac12\big(4e^{4i\Theta}-\zeta^2-2\big)J_0-J_2 ,
\end{split}
\label{eq:ABJlogGM}
\end{equation}
and by \eqref{eq:ABJWn}, the first winding derivative of the loop
needs only $\widehat J_2$ and \eqref{eq:ABJlogGM} trades it for ordinary periods. By \eqref{eq:ABJGM}
these are the loop, its derivative in the modulus and the holomorphic period, and with \eqref{eq:ABJmain} and
\eqref{eq:ABJPF} the result collapses to
\begin{equation}
\partial_n\vev{W^{1/6}_n}\Big|_{n=1}=e^{2i\Theta}\,\overline{\vev{ W^{1/6}}}-\vev{W^{1/6}}+\frac\zeta2 .
\label{eq:ABJLJ}
\end{equation}
Using the polar decomposition \eqref{eq:ABJpolar} and \eqref{eq:ABJW12} we obtain
\begin{equation}
\partial_n\log\vev{W^{1/6}_n}\Big|_{n=1}=\frac{4\pi i\,\big(\la_1-4\widehat B\big)}{w+8\pi i\widehat B},
\label{eq:ABJlogW}
\end{equation}
that given the prescription \eqref{eq:B16def} gives the bosonic Bremsstrahlung function of the first node in ABJ
\begin{equation}
B^{\varphi}_{1/6}=\frac{8\,\widehat B\,\big(\la_1-4\widehat B\big)}{w^2+64\pi^2\widehat B^{2}} .
\label{eq:ABJB16}
\end{equation}
where the second node one can be reconstructed exchanging $\la_1\leftrightarrow\la_2$.
We deduce then that the winding velocities are $\Lambda_i=\la_i-4\widehat B$ where $i=1,2$ represents the node of the quiver and $w$ replaces the $1$ of the ABJM
denominator in \eqref{eq:ratio}.

The winding prescription used for $B^{\varphi}_{1/6}$ is safe at unequal ranks, because it involves one node at a time:
the bosonic loop is the trace of an ordinary $U(N_1)$ holonomy so the cusped line exponentiates in the standard way. Indeed
\eqref{eq:ABJB16} reproduces the four-loop series in \cite{Bianchi:2017afp}. The fermionic line is different. Its connection is a supermatrix whose
off-diagonal entries are the fermions, and there is no a priori reason for its cusp to exponentiate in the standard way. That
$\widehat B$ is the imaginary part of the polar form \eqref{eq:ABJpolar}, as $\Bhalf$ is in ABJM, does not by itself make it
the Bremsstrahlung function of the $\tfrac12$BPS line of ABJ theory. However, after defining the correct exponentiation, in the rest of this section we will see that it is.

\subsubsection{Fermionic cusp exponentiation prescription}

The resummation of ladders \cite{Bonini:2016fnc} made the relevant structure evident. Since the fermions sit in the
off-diagonal entries of the supermatrix, a fermion exchanged across the cusp switches block, so that the trace of the cusped
loop over one block is fed by the other. In the ladder limit, where these exchanges dominate, the Bethe-Salpeter equation is a
coupled system for the two block traces, and its solution is a sum of two exponentials, which only at equal ranks reduces to
one. The structure is general, and it can be derived as follows. The trace normalised by $N_1+N_2$ is the sum of the traces over
the two blocks
\begin{equation}
\vev{W_{\rm cusp}}=\frac{N_1W_1+N_2W_2}{N_1+N_2},
\label{eq:ABJblocks}
\end{equation}
with $W_a$ the cusped loop traced over the block $a$ and normalised by $N_a$. The divergences that depend on the angles come from
the region near the cusp while those along the edges being removed by the normalisation to the straight line. Near the cusp, a fermion
exchanged across it, with one end on each edge, carries the index that flows through the cusp from one block to the other. In
$W_1$ it closes an index loop in the second block, with colour factor $N_2$, and leaves the structure of $W_2$, and vice versa.
The gauge fields, the scalar bilinears and the fermions exchanged along one edge do not change the block. Nothing else can
appear, since gauge invariance allows only two constant operators at the junction of the two lines, the projectors on the
blocks. The divergences of each block trace are therefore combinations of $W_1$ and $W_2$, and renormalisation acts on the pair
by a $2\times2$ matrix, $W^{\rm bare}_a=Z_{ab}W_b$. In a conformal theory its logarithmic derivative is a constant matrix of
anomalous dimensions, $\widehat\Gamma(\varphi,\theta)$, and with $\ell=\log(\Lambda_{\rm UV}/\Lambda_{\rm IR})$ the pair obeys
\begin{equation}
\frac{d}{d\ell}\begin{pmatrix}W_1\\W_2\end{pmatrix}=-\,\widehat\Gamma\begin{pmatrix}W_1\\W_2\end{pmatrix},
\qquad W_1=W_2=1\ \ \text{at}\ \ \ell=0,
\label{eq:ABJcuspmix}
\end{equation}
since for coinciding cutoffs there is no room for the cusp divergence. The solution is 
\begin{equation}
\begin{pmatrix}W_1\\W_2\end{pmatrix}=e^{-\widehat\Gamma(\varphi,\theta)\,\log\frac{\Lambda_{\rm UV}}{\Lambda_{\rm IR}}}
\begin{pmatrix}1\\1\end{pmatrix}\,,
\end{equation}
up to finite terms. Diagonalising $\widehat\Gamma$, one obtains two eigenvalues $\Gamma_\pm$ that are the two cusp anomalous dimensions. We also define the spectral projectors
$\Pi_\mp=(\widehat\Gamma-\Gamma_\pm)/(\Gamma_\mp-\Gamma_\pm)$ so that
$e^{-\widehat\Gamma\ell}=\Pi_-e^{-\Gamma_-\ell}+\Pi_+e^{-\Gamma_+\ell}$. By \eqref{eq:ABJblocks} the trace is then a sum of two
exponentials. Their weights add up to one since $\Pi_-+\Pi_+=1$. Moreover, since
$\widehat\Gamma=\Gamma_-\Pi_-+\Gamma_+\Pi_+$ their mean exponent is
\begin{equation}
\bar\Gamma=\frac{N_1\big(\widehat\Gamma_{11}+\widehat\Gamma_{12}\big)+N_2\big(\widehat\Gamma_{21}+\widehat\Gamma_{22}\big)}{N_1+N_2}.
\label{eq:ABJGbar}
\end{equation}
This is the anomalous dimension of the trace at the cutoff: a row of $\widehat\Gamma$ gives the running of a block trace when
both start at $1$, and $\bar\Gamma$ averages the two with the weights of the trace, so that
$\vev{W_{\rm cusp}}=1-\bar\Gamma\,\ell+\dots$. The two conditions fix the weights leading to
\begin{equation}
\vev{W_{\rm cusp}}=\frac{\Gamma_+-\bar\Gamma}{\Gamma_+-\Gamma_-}
\;e^{-\Gamma_-\,\log\frac{\Lambda_{\rm UV}}{\Lambda_{\rm IR}}}
+\frac{\bar\Gamma-\Gamma_-}{\Gamma_+-\Gamma_-}
\;e^{-\Gamma_+\,\log\frac{\Lambda_{\rm UV}}{\Lambda_{\rm IR}}}.
\label{eq:ABJcuspexp}
\end{equation}
Since $\bar\Gamma$ is a weighted average rather than an eigenvalue, $\log\vev{W_{\rm cusp}}$ is not linear in $\ell$: its quadratic term is
the variance $(\bar\Gamma-\Gamma_-)(\Gamma_+-\bar\Gamma)$ of the two exponents, and the cusped loop exponentiates in the standard
way only when $\bar\Gamma$ equals an eigenvalue. This happens at equal ranks, where the exchange of the blocks makes the trace an
eigenvector. It also happens on the straight line and on the BPS cusps $\theta=\pm\varphi$, where the trace does not run and
$\Gamma_-=\bar\Gamma=0$. There $\Gamma_+$ is the dimension of the constant supermatrix operator of \cite{Gorini:2022jws}, the
difference of the two blocks inserted on the line.

At one loop the matrix is fixed by the single fermion exchange as follows
\begin{equation}
\widehat\Gamma=\begin{pmatrix}\la_2&-c\,\la_2\\-c\,\la_1&\la_1\end{pmatrix}+O(\la^2),\qquad\text{with}\qquad
c=\frac{\cos\frac\theta2}{\cos\frac\varphi2}.
\label{eq:ABJcusp1loop}
\end{equation}
The off-diagonal entries are the exchange across the cusp, with the rank of the other block as colour factor. The diagonal ones
are the exchange next to the cusp on one edge, whose weight is fixed by the straight line ($c=1$) on which $W_1=W_2=1$ must not
run. The eigenvalues are $\Gamma_\pm=\tfrac12\big[\la_1+\la_2\pm\sqrt{(\la_1-\la_2)^2+4c^2\la_1\la_2}\,\big]$ and
$\bar\Gamma=-2\la_1\la_2(c-1)/(\la_1+\la_2)$. Each check of \eqref{eq:ABJcuspexp} against known data is then immediate. At equal ranks $\Gamma_-=-\la(c-1)$ is the one-loop cusp of ABJM, and on the straight line $\Gamma_+=\la_1+\la_2$. At large $c$, 
$\Gamma_\pm\to\pm c\sqrt{\la_1\la_2}$ and the weights become $(\sqrt{\la_1}\pm\sqrt{\la_2})^2/2(\la_1+\la_2)$ that are the
exponents and the weights of the ladders \cite{Bonini:2016fnc}. 
Prescription \eqref{eq:ABJcuspexp} is also compatible with the 2-loop computation of \cite{Griguolo:2012iq}.  There, the authors proposed the double exponentiation $(N_2e^{V_{N_1}}+N_1e^{V_{N_2}})/(N_1+N_2)$ with each potential depending on one rank. Up to two loop it perfectly matches our prescription but it fails at the next order: at three loops the planar ladders, whose index loops alternate between the two blocks,
have colour factor $2\la_1^2\la_2^2/(\la_1+\la_2)$, as \eqref{eq:ABJcuspexp} gives, while the double exponentiation gives
$\la_1\la_2(\la_1^2+\la_2^2)/(\la_1+\la_2)$.

Defining the Bremsstrahlung function for the fermionic line is not trivial. Indeed, since the logarithm of the fermionic cusp is not linear in $\ell$, there is no clear definition of where the Bremsstrahlung hides. In the following we will find some arguments to pinpoint where the ABJ fermionic Bremsstrahlung originates. First of all we need to approach the near BPS limit $\theta=\pm\varphi$, where we set $c=1+\delta$ with $\delta\to0$ at fixed $\varphi$. In this limit $\Gamma_-\to0$, while $\Gamma_+$ tends to the dimension of the constant operator: the supermatrix ${\rm diag}(1_{N_1},-1_{N_2})$ inserted on the line, which is the only operator besides the identity that can sit at the cusp. At one loop, \eqref{eq:ABJcusp1loop} gives
\begin{equation}
\Gamma_-=-\frac{2\la_1\la_2}{\la_1+\la_2}\,\delta-\frac{\la_1\la_2(\la_1-\la_2)^2}{(\la_1+\la_2)^3}\,\delta^2+O(\delta^3),\quad
\frac{\bar\Gamma-\Gamma_-}{\Gamma_+-\Gamma_-}=\frac{\la_1\la_2(\la_1-\la_2)^2}{(\la_1+\la_2)^4}\,\delta^2+O(\delta^3),
\label{eq:ABJnearBPS}
\end{equation}
where the second expression is the weight of the second exponential in \eqref{eq:ABJcuspexp}. Two features follow. First, at order $\delta$ this weight vanishes, so that $\bar\Gamma=\Gamma_-$ and the cusp exponentiates in the standard way. The constant operator enters only at order $\delta^2$, through the weight and through a level repulsion that lowers $\Gamma_-$, both proportional to $(\la_1-\la_2)^2$ and hence absent at equal ranks. 
Second, in the infrared when $\ell\gg1$, the term with the larger exponent is suppressed and the cusp scales as a single power with exponent $\Gamma_-$, the lower eigenvalue. Therefore, the Bremsstrahlung function has to be defined by the lower eigenvalue at small angles $\delta\simeq(\varphi^2-\theta^2)/8$ as follows
\begin{equation}
\Gamma_-\simeq-(\varphi^2-\theta^2)\,B_{1/2}.
\label{eq:ABJcuspB}
\end{equation}
At one loop this gives $B_{1/2}=\la_1\la_2/4(\la_1+\la_2)$, the harmonic mean of the two couplings divided by eight. Each block alone would give $\la_2/8$ or $\la_1/8$, and $B_{1/2}$ is their average with the weights of the trace.

This definition holds beyond perturbation theory, and it identifies $B_{1/2}$ with the function $\widehat B$ found from the polar form. The argument has three steps:
\begin{enumerate}
\item At small angles the cusp is the straight line deformed by two operators: the displacement which bends it by $\varphi$, and the tilt which rotates its scalar couplings by $\theta$. The cusp divergence is the dimension of the lowest operator at the junction, which on the straight line is the identity. Second-order conformal perturbation theory then gives $\Gamma_-\simeq-(\varphi^2-\theta^2)\,C_D/12$, with $C_D$ the normalisation of the two-point function of the displacement. Supersymmetry fixes the normalisation of the tilt in terms of $C_D$, so that the two angles enter only through $\varphi^2-\theta^2$, as in $\mathcal N=4$ super Yang-Mills \cite{Correa:2012at}, and the constant operator contributes only through mixing, at higher order in the angles. Hence $B_{1/2}=C_D/12$ is a property of the straight line alone.
\item The Ward identities of the superalgebra $su(1,1|3)$ preserved by the line relate $C_D$ to the latitude \cite{Bianchi:2017ozk}. They depend only on the algebra, which is the same at unequal ranks. 
\item The cohomological equivalence at unequal ranks \cite{Bianchi:2014laa} writes the fermionic latitude through the bosonic latitudes of the two nodes, and the relation between latitude and winding derivatives turns the result into the imaginary part of the polar form \eqref{eq:ABJpolar}.
\end{enumerate}
All together, we obtain
\begin{equation}
B_{1/2}=\frac{C_D}{12}=\frac1{4\pi^2}\,\partial_\nu\log\big|\vev{W_F(\nu)}_\nu\big|\Big|_{\nu=1}=\widehat B\,,
\label{eq:ABJchain}
\end{equation}
at every coupling. 
At weak coupling \eqref{eq:ABJmain} gives
\begin{equation}
\widehat B=\frac{\la_1\la_2}{4(\la_1+\la_2)}-\frac{\pi^2\la_1^2\la_2^2}{24(\la_1+\la_2)}
-\frac{\pi^4\la_1^2\la_2^2\big(7\la_1^2-62\la_1\la_2+7\la_2^2\big)}{1440\,(\la_1+\la_2)}+O(\la^7),
\label{eq:ABJBweak}
\end{equation}
whose first term is the one-loop value above. Together with the ladder limit \cite{Bonini:2016fnc} and the two-loop cusp \cite{Griguolo:2012iq}, it checks the planar part of the prescription proposed in \cite{Bianchi:2018bke}, which the argument above derives. At strong coupling $\widehat B$ terminates up to worldsheet instantons as in ABJM
\begin{equation}
\widehat B=\frac{\sqrt{2\lah}}{4\pi}-\frac1{4\pi^2}+O\big(e^{-2\pi\sqrt{2\lah}}\big),\qquad
\lah=\frac{\la_1+\la_2}{2}-\frac1{24}-\frac{(\la_1-\la_2)^2}{2}.
\label{eq:ABJBstrong}
\end{equation}
The ranks enter only through the shifted coupling $\lah$, which reduces to the usual $\la-\tfrac1{24}$ at equal ranks. This has a direct meaning in M-theory. Indeed, it is the planar part of the M2-brane charge of the dual background divided by $k$ \cite{Aharony:2009fc}. The term $-\tfrac1{24}$ is the charge induced by the curvature, which already shifts the coupling in ABJM \cite{Bergman:2009zh}. The last term is the charge induced by the discrete torsion of fractional M2-branes \cite{Aharony:2008gk}. The same planar shift governs the Airy function of the ABJ partition function \cite{Honda:2013pea,Matsumoto:2013nya}.

Integrability gives a different coordinate on the same slice. The interpolating function proposed for ABJ theory in
\cite{Cavaglia:2016ide} was fixed by matching the quantum spectral curve to the localisation loop on this curve.
It is a reparametrisation rather than a new datum: it vanishes at the endpoint of the
slice, where one node is empty, and it makes the discriminant, (the prefactor of \eqref{eq:ABJmain}) a function of $h$ alone. What the closed form can test is the expectation, true in ABJM, that the Bremsstrahlung function depends on the ranks only through
$h$. It does not. At weak coupling $\widehat B$ is not the ABJM function of the interpolating function $h$ of \cite{Cavaglia:2016ide}. Since $h\simeq\sqrt{\la_1\la_2}$, the ABJM relation would give $\sqrt{\la_1\la_2}/8$, while \eqref{eq:ABJBweak} gives $\la_1\la_2/4(\la_1+\la_2)$, which is smaller unless the ranks are equal. At strong coupling, instead, the two agree up to worldsheet instantons, because $\widehat B$ and $h$ depend on the ranks only through the same shifted coupling. The proposal for $h$, which concerns the spectrum, is not affected: the radiation also depends on how the line couples to the two nodes, which $h$ does not see.

\section{Latitude Wilson loops and periods with two charges}
\label{sec:latexact}

The Bremsstrahlung functions use the latitude only to first order around the great circle, where it reduces to the wound loop by \eqref{eq:latwindexact}. Beyond this order the deformation no longer drops out of the measure and, as recalled at the beginning of section~\ref{sec:logperiods}, the latitude matrix model has no spectral curve. The Fermi-gas form of the matrix model removes this obstruction: the deformation becomes a canonical rescaling of phase space, and undoing it turns the latitude into a loop of the undeformed theory with two charges, one for the position and one for the momentum of the gas. In the planar limit these loops are again periods of $\Sigma_\kk$, and computing them in closed form gives the bosonic and fermionic latitudes at every $\nu$ and every coupling, beyond what the winding can give.

\subsection{Loops with two charges}
\label{sec:twoloops}

The authors of \cite{Marino:2011eh} showed that \eqref{eq:MM} is the partition function of $N$
non-interacting fermions in one dimension. The rewriting takes three steps. First, the identity
\begin{equation}
\frac{\prod_{i<j}2\sinh\frac{\mu_{ij}}2\;2\sinh\frac{\hat\mu_{ij}}2}{\prod_{i,j}2\cosh\frac{\mu_i-\hat\mu_j}2}
=\det_{i,j}\,\frac1{2\cosh\frac{\mu_i-\hat\mu_j}2}\,,
\label{eq:Cauchy}
\end{equation}
turns the one-loop factor of \eqref{eq:MM}, which is the square of the left-hand side, into the square of a determinant.
Second, one of the two determinants is expanded as a sum over permutations, while the other can be replaced by $N!$ times
its diagonal term, since the rest of the integrand is symmetric under permutations of the eigenvalues of each node. Third,
every factor $1/2\cosh$ is written through its Fourier transform, after which the integrals over the $2N$ eigenvalues are
Gaussian and can be done. What remains is an integral over $N$ new variables $x_i$ that reads
\begin{equation}
Z(N)=\frac1{N!}\sum_{\sigma\in S_N}(-1)^\sigma\int d^Nx\,\prod_{i=1}^N\rho(x_i,x_{\sigma(i)}) .
\label{eq:ZFermi}
\end{equation}
This is the canonical partition function of $N$ identical fermions. The signed sum over permutations antisymmetrises the
$N$-particle state, and the weight is a product of one-particle factors, so that the fermions do not interact. The function
$\rho(x,y)=\langle x|\hat\rho|y\rangle$ is the kernel of a one-particle density operator $\hat\rho$, which plays the role of
the Boltzmann factor $e^{-\hat H}$ of a single fermion
\begin{equation}
\hat\rho=\chi(\hat q)\,\chi(\hat p),\qquad \chi(u)=\frac1{2\cosh\frac u2},\qquad [\hat q,\hat p]=2\pi ik,\qquad
\rho(x,y)=\frac{\chi(x)}{4\pi k\cosh\frac{x-y}{2k}} ,
\label{eq:rhogas}
\end{equation}
with $\hat q$ and $\hat p$ the position and the momentum of the fermion. The Chern-Simons level plays the role of the Planck
constant $\hbar=2\pi k$, and at leading order in $\hbar$ the one-particle Hamiltonian is
$\hat H=\log2\cosh\frac{\hat q}2+\log2\cosh\frac{\hat p}2$, which grows linearly at large position and momentum. The natural
regime of this description is $N\to\infty$ at fixed $k$, the M-theory regime, in which \cite{Marino:2011eh} computed the
partition function as a series in $\hbar$. The 't~Hooft limit, $k\to\infty$ at fixed $\la=N/k$, is instead a limit of large
$\hbar$, in which the gas is far from classical (see appendix~\ref{sec:gasperiods}).

The eigenvalue of the first node is not a one-particle
observable of the gas, but its exponential is. The loop of the first node wound $n$ times 
$\frac1N\sum_ie^{n\mu_i}$ of \eqref{eq:MMloops} becomes the average of the one-body operator $e^{n(\hat q+\hat p)/k}$
\cite{Klemm:2012ii,Boldis:2025yll}. In this operator the position and the momentum enter with the same coefficient, but nothing in the gas
requires it. Separating the two coefficients, we define the loop with two charges
\begin{equation}
\vev{W(n_1,n_2)}:=\frac1N\Big\langle\sum_{i=1}^Ne^{(n_1\hat q_i+n_2\hat p_i)/k}\Big\rangle ,\qquad
\vev{W(n,n)}=\vev{W^{1/6}_n}.
\label{eq:Wtwodef}
\end{equation}
In the plane of the charges $(n_1,n_2)$, the wound loops of the first node lie on the diagonal $n_1=n_2$.

Two symmetries of the gas act on the charges. The first exchanges them, and it comes from the symmetry of
\eqref{eq:rhogas} between position and momentum. This exchange also replaces $\hat\rho$ with its transpose and since every average is built from traces, the loop does not change when
its two charges are exchanged. The second symmetry is complex conjugation. In the position representation it leaves the
position unchanged and reverses the momentum, and since the kernel of $\hat\rho$ is real it maps the gas to itself. Hence,
for real charges we have
\begin{equation}
\vev{W(n_1,n_2)}=\vev{W(n_2,n_1)},\qquad \overline{\vev{W(n_1,n_2)}}=\vev{W(n_1,-n_2)} .
\label{eq:twosym}
\end{equation}
In particular the anti-diagonal is the loop of the second node. Using \eqref{eq:node} we have 
\begin{equation}
\vev{W(n,-n)}=\overline{\vev{W^{1/6}_n}}=\vev{\hat W^{1/6}_n}.
\end{equation}
The latitude matrix model of \cite{Bianchi:2018bke,Griguolo:2021rke} computes
\begin{equation}
\vev{W_B(\nu)}_\nu=\Big\langle\frac1N\sum_ie^{\sqrt\nu\,\mu_i}\Big\rangle_\nu
\label{eq:latMM}
\end{equation}
with the measure obtained from \eqref{eq:MM} by the replacements
\begin{equation}
\Big[2\sinh\frac{x}{2}\Big]^2\ \to\ 2\sinh\frac{\sqrt\nu\,x}{2}\;2\sinh\frac{x}{2\sqrt\nu},\qquad
\Big[2\cosh\frac{x}{2}\Big]^2\ \to\ 2\cosh\frac{\sqrt\nu\,x}{2}\;2\cosh\frac{x}{2\sqrt\nu},
\label{eq:latmeasure}
\end{equation}
which for $\nu=1$ corresponds to \eqref{eq:MM}. For $\nu\ne1$ each squared one-loop factor splits into two factors of the same kind, one with the argument
rescaled by $\sqrt\nu$ and one by $1/\sqrt\nu$. This double rescaling is what defeats the eigenvalue methods, as recalled at
the beginning of section~\ref{sec:logperiods} but it does not spoil the Fermi gas. Indeed, the identity \eqref{eq:Cauchy}
applies to the two products separately and the only effect of the deformation is a rescaling of the position and of the momentum in opposite directions, a squeeze of phase space
$(\hat q,\hat p)\mapsto(\nu^{\pm1/2}\hat q,\nu^{\mp1/2}\hat p)$. Since it preserves the commutator and does not change traces, undoing it maps the deformed one-particle
density back to \eqref{eq:rhogas}. As a consequence, the partition function of the latitude model is the one of ABJM
and the inserted operator becomes $e^{(\hat q+\nu\hat p)/k}$ \cite{Bianchi:2018bke} such that
\begin{equation}
\vev{W_B(\nu)}_\nu=\vev{W(1,\nu)}\,.
\label{eq:latW}
\end{equation}
The deformation has moved from the measure to the operator. Explicitly, in the deformed gas the inserted operator is
$e^{\sqrt\nu(\hat q+\hat p)/k}$, the great circle with both charges equal to $\sqrt\nu$.
Undoing the squeeze divides the
first charge by $\sqrt\nu$ and multiplies the second by $\sqrt\nu$, which gives $(1,\nu)$. The latitude is the great
circle of a squeezed gas. Latitudes are loops of ABJM with unequal charges for the position and for
the momentum. In the plane of the charges they lie on the line $n_1=1$, which crosses the diagonal at the great circle
$\nu=1$.

For $N=1$ the gas has a single fermion and the trace can be done by hand, giving
\begin{equation}
\vev{W(n_1,n_2)}_{N=1}=\frac{e^{i\pi n_1n_2/k}}{\cos\frac{\pi n_1}k\,\cos\frac{\pi n_2}k},\qquad |n_1|,|n_2|<\frac k2 .
\label{eq:WtwoN1}
\end{equation}
The two charges talk to each other only through the phase $e^{i\pi n_1n_2/k}$ produced by their commutator. On the diagonal it gives $e^{i\pi n^2/k}\sec^2\frac{\pi n}k$ of \cite{Marino:2009jd}. On the corner, it gives the latitude
$\sec\frac\pi k\,\big(1+i\tan\frac{\pi\nu}k\big)$. 

The diagonal and the line of the latitudes meet at the great circle. Moving along the latitude
changes only the second charge, while moving along the diagonal changes both. At the diagonal the two contribute equally by the exchange symmetry \eqref{eq:twosym} such that $\partial_n\vev{W(n,n)}=(\partial_{n_1}+\partial_{n_2})\vev{W(n_1,n_2)}
=2\,\partial_{n_2}\vev{W(n_1,n_2)}$ at $n_1=n_2=n$. Hence
\begin{equation}
\partial_\nu\vev{W(1,\nu)}\Big|_{\nu=1}=\frac12\,\partial_n\vev{W^{1/6}_n}\Big|_{n=1},
\end{equation}
which is \eqref{eq:latwindexact}, namely the first latitude derivative is half the winding derivative. At second order this is no
longer enough. Along the latitude the second derivative is $\partial_{n_2}^2$, along the diagonal it is
$\partial_{n_1}^2+2\,\partial_{n_1}\partial_{n_2}+\partial_{n_2}^2$, and the mixed derivative $\partial_{n_1}\partial_{n_2}$
is not fixed by the wound loops. 

\subsection{Periods with two charges}
\label{sec:twoperiods}

In the planar limit the wound loops are periods because the eigenvalue of the first node is the coordinate $Z$ of
$\Sigma_\kk$. For two charges the position and the momentum of the gas are needed separately, and the curve provides them.
Given the two square roots of the resolvent \eqref{eq:resolvent}, we can define the following system of coordinates
\begin{equation}
\begin{gathered}
X:=\frac{R_++R_-}2,\qquad Y:=\frac{R_+-R_-}2,\\
XY=Z,\qquad X^2-Y^2=u,\qquad \big(X-X^{-1}\big)\big(Y-Y^{-1}\big)=i\kk .
\end{gathered}
\label{eq:XYcurve}
\end{equation}
such that the curve is symmetric under $X\leftrightarrow Y$ that exchanges
the two sides of the cut. Then, the resolvent is $\varrho_0=2\log2Y$ and the coupling
\eqref{eq:lambdafed} and the wound loops \eqref{eq:Wn} take the form
\begin{equation}
\la=\frac1{4\pi^2}\oint_A\log Y\,d\log X,\qquad
\la\vev{W^{1/6}_n}=\frac1{4\pi^2n}\oint_A(XY)^nd\log X ,
\label{eq:phasespace}
\end{equation}
the first since $d\log Z$ and $d\log^2Y$ are exact along $A$, the second since $2Z\partial_Z\log X=1+(Z^2-1)/u$ by
\eqref{eq:dZrho}. 
These two formulas have a simple meaning as integrals over a phase space with coordinates $(\log X,\log Y)$. By Stokes' theorem the first integral is the area of the region enclosed by the cycle $A$. Hence \eqref{eq:phasespace} says that the region has area $4\pi^2\la$ and that
$\vev{W^{1/6}_n}$ is the average of $(XY)^n$ over it.

This is what a classical Fermi gas would give with $X=e^{\hat q/k}$ and $Y=e^{\hat p/k}$. In the variables $\hat q/k$ and
$\hat p/k$ the commutator is $2\pi i/k$, so at large $k$ they behave as classical coordinates. The $N=k\la$ fermions then fill a region of area $4\pi^2\la$ (first formula) and
the wound loop is the average over this region of its operator $e^{n(\hat q+\hat p)/k}=(XY)^n$ (second formula).
In the same picture, the loop with two charges should be the average of
$e^{(n_1\hat q+n_2\hat p)/k}=X^{n_1}Y^{n_2}$ over the same region as follows
\begin{equation}
\vev{W(n_1,n_2)}=\frac1{4\pi^2n_2\la}\oint_A X^{n_1}Y^{n_2}\,d\log X ,
\label{eq:Wtwoplanar}
\end{equation}
which is symmetric in the charges by $X\leftrightarrow Y$ and an integration by parts, and reduces to \eqref{eq:phasespace}
on the diagonal. With no further discussion, since the gas is not classical in the 't~Hooft limit this classical picture is only a conjecture. Nevertheless, it is the exact planar limit of the loop with two charges, as we prove in appendix~\ref{sec:gasperiods}.

The idea of the proof is simple. Besides its Hamiltonian, the gas has a second operator that commutes with the density matrix,
whose classical value is the left-hand side of the curve. In the planar limit its commutators with the loops give a Ward
identity relating loops whose two charges differ by one unit. The periods obey the same identity, because that left-hand side is
constant along the cycle. Since at each order in the coupling both are polynomials in the charges that coincide on the
diagonal, where they are the wound loops, the identity carries the agreement to all charges.

As for the wound loops, a derivative in the coupling turns the loop with two charges into a period of the holomorphic
differential $dZ/u$. In \eqref{eq:Wtwoplanar} the coupling enters only through the curve, which fixes $Y$ as a function of
$X$. Differentiating the curve \eqref{eq:XYcurve} at fixed $X$ gives $\dk\log Y=i/\big((X-X^{-1})(Y+Y^{-1})\big)$, and the
same denominator appears in the holomorphic differential. Then we can write the derivative of \eqref{eq:Wtwoplanar} as follows
\begin{equation}
\dk\big(\la\vev{W(n_1,n_2)}\big)=\frac i{4\pi^2}\,J(n_1,n_2),\qquad
J(n_1,n_2):=\frac12\oint_AX^{n_1}Y^{n_2}\,\frac{dZ}u .
\label{eq:Jtwo}
\end{equation}
This is the analogue of \eqref{eq:dWn}: the $\kk$-derivative of the loop is a moment of $dZ/u$, now weighted by
$X^{n_1}Y^{n_2}$ instead of $Z^n$. On the diagonal $X^nY^n=Z^n$, and $J(n,n)=J_n$. We call $J(n_1,n_2)$ the \emph{periods with two
charges}.

Whether $J(n_1,n_2)$ is a period in the strict sense depends on the charges. The coordinates of the curve are $Z=XY$ and
$X/Y$, while $X$ and $Y$ separately are defined only up to a common sign. Writing
$X^{n_1}Y^{n_2}=Z^{\frac{n_1+n_2}2}(X/Y)^{\frac{n_1-n_2}2}$, the weight is a function on $\Sigma_\kk$ when both exponents are
integers, which for integer charges means $n_1+n_2$ even, and then $J(n_1,n_2)$ is an ordinary period of a meromorphic
differential. For other charges, and in particular for the latitude $(1,\nu)$, it is defined by continuation in the charges,
as the wound loops were continued in the winding.

The periods with two charges can be computed in one step because the curve \eqref{eq:XYcurve} separates the variables.
In terms of $t_1=X-X^{-1}$ and $t_2=Y-Y^{-1}$ it is $t_1t_2=i\kk$, and since $dt_1=(X+X^{-1})\,d\log X$ the
integrand of \eqref{eq:Jtwo} is a function of $t_1$ times a function of $t_2$ as follows
\begin{equation}
X^{n_1}Y^{n_2}\,\frac{dZ}u=\frac{2X^{n_1}}{X+X^{-1}}\,\frac{2Y^{n_2}}{Y+Y^{-1}}\,\frac{dt_1}{2t_1},\qquad\text{with}\qquad
\frac{2X^n}{X+X^{-1}}=\sum_{j\ge0}\binom{\frac{n+j-1}2}{j}t_1^j.
\label{eq:gfactor}
\end{equation}
where the same series in $t_2$ expands the factor of $Y$. Using that $t_2=i\kk/t_1$ and performing the integration, only the terms in which the powers of $t_1$ cancel are kept, then we can write
\begin{equation}
J(n_1,n_2)=-\frac{i\pi}{2}\sum_{j\ge0}\binom{\frac{n_1+j-1}2}{j}\binom{\frac{n_2+j-1}2}{j}(i\kk)^j ,\qquad |\kk|<4 .
\label{eq:Jtwoseries}
\end{equation}
On the diagonal the two binomial
coefficients coincide and \eqref{eq:Jtwoseries} is \eqref{eq:Jnseries}.
Grouping the terms of even and odd $j$ and integrating in $\kk$ as in
section \ref{sec:hyptower}, the loop with two charges is
\begin{equation}
\begin{split}
\vev{W(n_1,n_2)}={}\frac{\kk}{8\pi\la}\,&{}_4F_3\left(\genfrac{}{}{0pt}{}{\frac{1-n_1}2,\frac{1+n_1}2,\frac{1-n_2}2,\frac{1+n_2}2}{\frac12,\ 1,\ \frac32}\,\Biggl|\,\frac{-\kk^2}{16}\right)\\
&+\frac{i\,n_1n_2\,\kk^2}{64\pi\la}\,{}_4F_3\left(\genfrac{}{}{0pt}{}{\frac{2-n_1}2,\frac{2+n_1}2,\frac{2-n_2}2,\frac{2+n_2}2}{\frac32,\ \frac32,\ 2}\,\Biggl|\,\frac{-\kk^2}{16}\right),
\end{split}
\label{eq:Wtwo4F3}
\end{equation}
on the whole physical slice. It is \eqref{eq:tower4F3} with the two equal pairs of
upper parameters split between the two charges. It is symmetric in the charges and conjugated by $n_2\to-n_2$ as
\eqref{eq:twosym} requires.

The charges enter \eqref{eq:Wtwo4F3} as parameters, and not through a rescaling of the coupling, except at weak coupling and
large charges. In the double-scaling limit of \eqref{eq:Wbessel} the binomial coefficients become $(n/2)^j/j!$ and
\eqref{eq:Jtwoseries} becomes $-\frac{i\pi}2\,I_0\big(\sqrt{n_1n_2\,i\kk}\big)$: the two charges enter only through their
product, and the latitude wound $m$ times is the circle of $\mathcal N=4$ super Yang-Mills wound $m$ times at the coupling
$\nu\la_{\mathcal N=4}$. In this regime, where the second node and the matter decouple, the deformation is a rescaling of the
coupling, as it is for the latitude of $\mathcal N=4$ super Yang-Mills at every coupling \cite{Drukker:2006ga}, and more generally for the $\tfrac18$BPS loops on $S^2$, where it is rescaled by $A_1A_2/A^2$, with $A_{1,2}$ the two areas into which the loop divides the sphere of area $A$ \cite{Drukker:2007dw,Drukker:2007qr,Pestun:2009nn,Bonini:2014vta}.

The weak- and strong-coupling expansions of \eqref{eq:Wtwo4F3} follow as for the wound loops. At weak coupling, using the mirror
map \eqref{eq:lamkappaintro} we have
\begin{equation}
\begin{split}
\vev{W(n_1,n_2)}={}&1+i\pi\,n_1n_2\,\la+\frac{\pi^2}3\big(n_1^2+n_2^2-n_1^2n_2^2\big)\la^2\\
&+\frac{i\pi^3}{18}\,n_1n_2\big(4n_1^2+4n_2^2-4-n_1^2n_2^2\big)\la^3+O(\la^4),
\end{split}
\label{eq:Wtwoweak}
\end{equation}
which reduces to the expansion of the wound loops on the diagonal. At strong coupling, for $\kk>4$, each ${}_4F_3$ is continued as
a sum over its upper parameters. The two sums combine into four terms, one for each of the exponents $\kk^{\pm n_1}$ and
$\kk^{\pm n_2}$, as follows
\begin{equation}
\begin{gathered}
\vev{W(n_1,n_2)}=\sum_{s=\pm1}\Big[e^{\frac{i\pi}2sn_2}\,\mathcal W_{sn_1}(n_2)+e^{\frac{i\pi}2sn_1}\,\mathcal W_{sn_2}(n_1)\Big],\\
\mathcal W_m(n)=\frac{\Gamma\big(\frac{m+n}2\big)\,\Gamma\big(\frac{m-n}2\big)}{8\pi^2\la\,\Gamma(1+m)}\,\kk^m\,
{}_4F_3\left(\genfrac{}{}{0pt}{}{-\frac m2,\ \frac{1-m}2,\ \frac{1-m}2,\ \frac{2-m}2}{1-m,\ \frac{2-m+n}2,\ \frac{2-m-n}2}\,\Biggl|\,\frac{-16}{\kk^2}\right).
\end{gathered}
\label{eq:Wtwostrong}
\end{equation}
With $\kk\simeq e^{\pi\sqrt{2\lah}}$ the term $\mathcal W_m$ has weight $e^{\pi m\sqrt{2\lah}}$, and its series in $-16/\kk^2$ collects
the corrections of relative order $e^{-2\pi\sqrt{2\lah}}$, the worldsheet instantons. At $(n_1,n_2)=(1,\nu)$ these terms give the
two caps of the latitude of section~\ref{sec:latclosed}. The expansion holds for generic charges. At special values, for instance
on the diagonal, poles of the Gamma functions cancel between two terms and the limit produces powers of $\log\kk$, as in the
strong-coupling expansion of the wound loops.

\subsection{The bosonic and the fermionic latitudes}
\label{sec:latclosed}

The remaining step is to set the charges to those of the latitude $(n_1,n_2)=(1,\nu)$ by \eqref{eq:latW}. At $n_1=1$ the loop \eqref{eq:Wtwo4F3} simplifies in two ways. In the first ${}_4F_3$ the upper
parameter $\frac{1-n_1}2$ vanishes, so that the series stops at its first term that is equal to one. This is the selection rule that
made the real part of the wound loop elementary at odd winding, below \eqref{eq:tower4F3} and it fixes the real part of the
bosonic latitude to be the same at every $\nu$ as follows
\begin{equation}
\re\vev{W_B(\nu)}_\nu=\frac{\kk}{8\pi\la}=\vev{W^{1/2}} .
\end{equation}
This is, at genus zero, the exact property recalled below \eqref{eq:B12split}. In the second ${}_4F_3$ the parameter $\frac32$ appears above and below and cancels and the ${}_3F_2$ that remains gives the imaginary part.

Since the real part is fixed, both latitudes depend on $\nu$ only through the phase of the bosonic one. For the fermionic latitude this follows from the cohomological
relation \eqref{eq:cohomnu}. Since for equal ranks the loop of the second node is the complex conjugate, we have
\begin{equation}
\tan\PhB(\nu)=\tan\frac{\pi\nu}{2}\,\Big[1-\frac{\vev{W_F(\nu)}_\nu}{\vev{W^{1/2}}}\Big],
\label{eq:PhiWF}
\end{equation}
at any $N$ and $k$. In the planar limit the ${}_3F_2$ gives the phase, and the two latitudes are
\begin{align}
\vev{W_B(\nu)}_\nu=\vev{W^{1/2}}\big(1+i\tan\PhB(\nu)\big)&,\qquad
\vev{W_F(\nu)}_\nu=\vev{W^{1/2}}\Big(1-\cot\frac{\pi\nu}{2}\,\tan\PhB(\nu)\Big),\nonumber\\
\tan\PhB(\nu)=\frac{\nu\kk}{8}\;&{}_3F_2\left(\genfrac{}{}{0pt}{}{\frac12,\ 1-\frac\nu2,\ 1+\frac\nu2}{\frac32,\ 2}\,\Biggl|\,-\frac{\kk^2}{16}\right) ,
\label{eq:latexact}
\end{align}
with $\vev{W^{1/2}}=\kk/8\pi\la$ and $\kk$ related to $\la$ by the mirror map \eqref{eq:lamkappaintro}.

Four consequences follow from the form of \eqref{eq:latexact}.
First, the Bremsstrahlung function $\Bhalf$ and the winding velocity $\Lambda$ are its first two Taylor coefficients at the great circle. At $\nu=1$ the ${}_3F_2$
reduces to the ${}_2F_1$ of \eqref{eq:BPVreduced}, and \eqref{eq:latexact} is the framing relation \eqref{eq:framing} for
$\Bhalf$. Its derivative at $\nu=1$ is the ${}_3F_2$ of \eqref{eq:Lam3F2} $\partial_\nu\tan\PhB(\nu)|_{\nu=1}=2\pi\Lcl$.
Second, $\nu$ enters only through the parameters of the ${}_3F_2$, and not through the coupling. In $\mathcal N=4$ SYM the latitude is the great circle at the rescaled coupling $\la\cos^2\theta_0$
\cite{Drukker:2006ga}. In ABJM this happens only in the double-scaling limit of section~\ref{sec:twoperiods}.

Third, like the periods, the phase obeys a linear differential equation in the coupling. Since the real part is elementary at
$n_1=1$, the equation is of third order and reads
\begin{equation}
\kk(16+\kk^2)\,T'''+(48+5\kk^2)\,T''+(4-\nu^2)\,\kk\,T'=0 ,\qquad T=\tan\PhB(\nu)/2\pi,
\label{eq:PFnu}
\end{equation}
with primes denoting derivatives in $\kk$. At $\nu=1$ the operator is
$\big(\dk+\frac1\kk\big)\frac1\kk\big[\kk^2(16+\kk^2)\partial_\kk^2+\kk(16+\kk^2)\dk-16\big]$, and the equation contains the
equation \eqref{eq:ode} of the Bremsstrahlung function.
Fourth, the latitude contains information that the wound loops do not, and it organises the logarithmic periods of
section~\ref{sec:logGM}. By \eqref{eq:Jtwo} $T$ is the running mean in the coupling of the real part of
the period $J(1,\nu)$. The $m$-th derivative in $\nu$ of $J(1,\nu)$ at $\nu=1$ is the period of $Z\,dZ/u$ times a polynomial of degree $m$ in $\log Z$ and $\log(X/Y)$. Therefore, the Taylor coefficients of $T$ at the
great circle are running means of periods with more and more logarithms: $T(1)=4\Bhalf$ comes from the period $J_1$,
$T'(1)=\Lcl$ from the logarithmic period $\widehat J_1$ and $T''(1)$
from the period of $\frac14Z\big(\log^2Z+\log^2\frac XY\big)\frac{dZ}u$ with two logarithms and so on. The first two are data of the
wound loops, through their derivatives in the winding. The third is not, since it needs the logarithm of the second
coordinate $X/Y$ of the curve, which no derivative in the winding produces. The ${}_3F_2$ of \eqref{eq:latexact} resums the
whole tower.

With the mirror map \eqref{eq:lamkappaintro} the phase at weak coupling is given by
\begin{equation}
\begin{split}
\tan\PhB(\nu)={}&\pi\nu\la+\frac{\pi^3\la^3}6\,\nu(\nu^2-2)+\frac{\pi^5\la^5}{60}\,\nu\big(\nu^4-10\nu^2+17\big)\\
&\qquad+\frac{\pi^7\la^7}{5040}\,\nu\big(5\nu^6-140\nu^4+1106\nu^2-1812\big)+O(\la^9),
\end{split}
\label{eq:tanPhiweak}
\end{equation}
that reproduces up to order $\lambda^3$ the perturbative latitude at framing $\nu$ of \cite{Bianchi:2018bke} and the phase guessed in \cite{Bianchi:2018scb}. 
At $\nu=1$ the slope of the
phase is \eqref{eq:dPhinumain}, which section~\ref{sec:logperiods} obtained from the wound loops alone. 
For $\kk>4$ the series in \eqref{eq:latexact} diverge and the latitude is better
written as an expansion in $-16/\kk^2$, given by \eqref{eq:Wtwostrong} at $(n_1,n_2)=(1,\nu)$. At strong coupling, the fermionic latitude contains only the terms
$\kk^{\pm\nu}$ as follows
\begin{equation}
\vev{W_F(\nu)}_\nu\!=\!\frac{G(\nu)\!-\!G(-\nu)}{8\pi\la\,\sin\frac{\pi\nu}2},
\quad\text{with}\quad
G(\nu)\!=\!\frac{\Gamma\big(\frac{1+\nu}{2}\big)\;\kk^\nu}{\Gamma(1\!+\!\nu)\,\Gamma\big(\frac{3-\nu}{2}\big)}{}_3F_2\!\left(\genfrac{}{}{0pt}{}{-\frac\nu2,\frac{1-\nu}2,\frac{2-\nu}2}{1-\nu,\ \frac{3-\nu}2}\Biggl|\frac{-16}{\kk^2}\right).
\label{eq:twocaps}
\end{equation}
The two terms have a simple meaning in the dual string theory. For the latitude the
string extends in the internal space $\mathbb{CP}^3$, where it lies on a two-sphere $\mathbb{CP}^1$ ending on a circle.
The circle divides the sphere into two caps, one that covers the fraction $\frac{1-\nu}2$ of the
sphere and the other that covers the remaining fraction $\frac{1+\nu}2$. The string can
cover either cap, so there are two classical worldsheets. Their weights follow from one rule: the string of the great circle $\nu=1$ has the weight $\kk\simeq e^{\pi\sqrt{2\lah}}$ and covering a fraction $f$ of the
sphere costs an extra factor $\kk^{-2f}$. Then the first cap is $G(\nu)$ that has weight $\kk^\nu$ and, as one-loop factor the ratio of Gamma
functions in \eqref{eq:twocaps}. This matches the computation made on the string side in \cite{Medina-Rincon:2019bcc,David:2019lhr}. The second cap is the first one with $\nu$ replaced by $-\nu$. The worldsheet can also wrap the whole sphere at the cost of a factor
$\kk^{-2}\simeq e^{-2\pi\sqrt{2\lah}}$ each time. These are the worldsheet instantons, collected by the series in $-16/\kk^2$.

\begin{figure}[t]
\centering
\includegraphics[width=\textwidth]{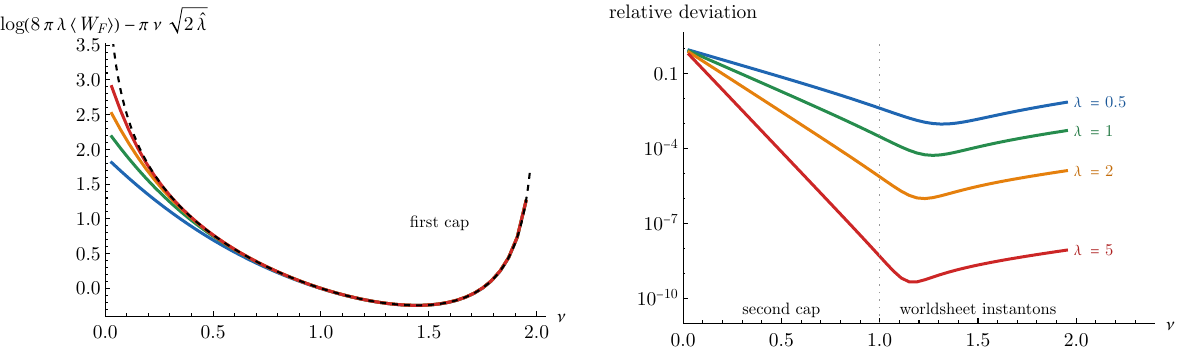}
\caption{The fermionic latitude at strong coupling and the two caps of the string, at $\la=0.5,1,2,5$ (same colours in
both panels). Left: $\log\big(8\pi\la\vev{W_F(\nu)}\big)$ minus the classical exponent $\pi\nu\sqrt{2\lah}$ of the first
cap. All couplings approach the same curve (dashed), the one-loop prefactor of the first cap in \eqref{eq:WFairy}, and
depart from it only at small $\nu$, where the second cap is not negligible. Right: the relative deviation of the latitude
from the first cap \eqref{eq:WFairy}. For $\nu<1$ it is the second cap, suppressed by $e^{-2\pi\nu\sqrt{2\lah}}$, and for
$\nu>1$ it is the worldsheet instantons, suppressed by $e^{-2\pi\sqrt{2\lah}}$.}
\label{fig:caps}
\end{figure}

Up to worldsheet instantons only the first cap survives (figure~\ref{fig:caps}) giving the following asymptotic form
\begin{equation}
\vev{W_F(\nu)}_\nu=-\frac{\Gamma(-\frac\nu2)}{8\pi^{3/2}\la\,2^\nu\,\Gamma(\frac{3-\nu}2)}\;e^{\pi\nu\sqrt{2\lah}} ,
\label{eq:WFairy}
\end{equation}
where we used $\kk=e^{\pi\sqrt{2\lah}}$. Inserted in \eqref{eq:PhiWF}, it
gives the phase of the bosonic latitude at strong coupling up to worldsheet instantons that reads
\begin{equation}
\tan\PhB(\nu)=\tan\frac{\pi\nu}{2}\,\Big[1+\frac{\Gamma(-\frac\nu2)}{\sqrt\pi\,2^\nu\,\Gamma(\frac{3-\nu}2)}\;
e^{\pi(\nu-1)\sqrt{2\lah}}\Big].
\label{eq:Phistrong}
\end{equation}
For fixed $\nu<1$ the exponential is negligible and $\PhB(\nu)\to\tfrac{\pi\nu}{2}$, while for $\nu>1$ it dominates and $\PhB(\nu)\to\tfrac\pi2$. At $\nu=1$ the
bracket vanishes while $\tan\frac{\pi\nu}2$ diverges, and the limit gives $\tan\PhB=2\sqrt{2\lah}-\frac2\pi$ that is the
strong-coupling $\Bhalf=\frac{\sqrt{2\lah}}{4\pi}-\frac1{4\pi^2}$ by \eqref{eq:framing}. The slope at $\nu=1$
interpolates between the two regimes.

These expansions can be compared with \cite{Bianchi:2018bke}, where the latitude matrix model was introduced and computed in
perturbation theory at weak coupling and through its Fermi gas at strong coupling. At weak coupling \eqref{eq:tanPhiweak}
reproduces their three-loop result. At strong coupling their Fermi gas gives the first cap alone. It is solved at $N\to\infty$
with $k$ fixed where the fermionic latitude is an Airy function \cite{Bianchi:2018bke} up to corrections exponentially
small in $\sqrt N$. The planar limit of the Airy form is \eqref{eq:WFairy}, where no worldsheet instantons appears. Indeed, they are negligible at
fixed $k$, but at fixed $\la$ they are the second cap and the spheres described above, so they are part of the planar answer.
This is why the large $N$ limit of the Airy form does not give \eqref{eq:latexact}. 
Equivalently, the planar limit cannot be taken directly in the matrix model
since for generic $\nu$ the double rescaling \eqref{eq:latmeasure} breaks down the usual eigenvalue methods (section~\ref{sec:twoloops}). Our route avoids
both obstacles: in the gas the latitude is the loop $W(1,\nu)$ of the undeformed theory \eqref{eq:latW}, whose planar limit is
a period of the curve (appendix~\ref{sec:gasperiods}) and contains all the worldsheet instantons at genus zero. Therefore, the two results
are complementary, the Airy form exact in $1/N$ but only at strong coupling, \eqref{eq:latexact} exact in the coupling
but only at genus zero. Beyond their overlap, \eqref{eq:latexact} gives the worldsheet instantons that the Fermi gas does not
yet provide. We leave the study of the latitude at any $N$ and $k$ for a future work.

By \eqref{eq:latwindexact} the first latitude derivative at the great circle is winding data in disguise, and the slope
\eqref{eq:dPhinumain} of the bosonic phase was obtained in this way. The closed form above
contains these data as its first Taylor coefficients. What the latitude adds here is a reading of the same data in terms of
the $\tfrac12$BPS line alone through the fermionic latitude. Indeed,
at $\nu=1$, where $\tan\tfrac{\pi\nu}2$ has a pole and the bracket of \eqref{eq:PhiWF} a zero, the pole shifts the orders by one: the phase on the
great circle fixes the first Taylor coefficient of the fermionic latitude, and its slope \eqref{eq:dPhinumain} fixes the
second,
\begin{equation}
\frac{\vev{W_F(\nu)}_\nu}{\vev{W^{1/2}}}=1+4\pi^2\Bhalf\,(\nu-1)+\pi^2\Lcl\,(\nu-1)^2+O\big((\nu-1)^3\big).
\label{eq:WFtaylor}
\end{equation}
The Bremsstrahlung function is the first latitude derivative of the fermionic line, as in its definition \eqref{eq:B12lat},
and the winding velocity is the second as follows
\begin{equation}
\Lcl=\frac1{2\pi^2}\,\frac{\partial^2_\nu\vev{W_F(\nu)}_\nu}{\vev{W_F(\nu)}_\nu}\,\Big|_{\nu=1},
\end{equation}
a second-order response of the $\tfrac12$BPS line, obtained from first-order data on the bosonic one. Inserted in
\eqref{eq:B16}, the two derivatives give the bosonic function in terms of the fermionic line alone
\begin{equation}
\Bsix=\frac{1}{\pi^2}\,\frac{\vev{W_F(\nu)}_\nu'\,\vev{W_F(\nu)}_\nu''}{\pi^2\vev{W_F(\nu)}_\nu^2+4\,\vev{W_F(\nu)}_\nu'^{\,2}}\bigg|_{\nu=1},
\label{eq:B16WF}
\end{equation}
with primes denoting derivatives in $\nu$. Like its ingredients, which are the relation \eqref{eq:BPVrel}, the
$\nu$-independence of the real part of the bosonic latitude and the cohomological relation \eqref{eq:cohomnu}, this holds
for equal ranks at any $N$ and $k$. The
radiation of the bosonic line, at the ultraviolet end of the flow of section \ref{sec:B16bounds}, is fixed by the response of the
line at the infrared end to the latitude deformation, through second order.

\subsection{The latitude and its framing}
\label{sec:latframing}

The latitude matrix model computes the latitudes at framing $f=\nu$, the framing at which the cohomological equivalence
\eqref{eq:cohomnu} is formulated \cite{Bianchi:2014laa,Bianchi:2018bke}. Although in its derivation from localisation the framing is an
input \cite{Griguolo:2021rke}, this identification is supported by perturbative computations at weak coupling  \cite{Bianchi:2018bke,Bianchi:2020cfn} and at strong coupling \cite{Medina-Rincon:2019bcc,David:2019lhr} that hold only at $f=\nu$. 
In the language of the loops with two charges, the framing is carried by the second charge: the loop $W(n_1,n_2)$ behaves as a
loop of winding $n_1$ at framing $f=n_2/n_1$, so that the latitude $(1,\nu)$ is the circle wound once at framing $\nu$. Two more facts
support this identification:
\begin{itemize}
\item \emph{One fermion.} At $N=1$ the theory is abelian but it still contains the matter fields. In
\eqref{eq:WtwoN1} they enter only through the real factor while the phase $e^{i\pi n_1n_2/k}$ comes only from the commutator $[\hat q,\hat p]=2\pi ik$. This is the framing factor of pure abelian Chern-Simons theory, $e^{i\pi n^2f/k}$ for a loop of charge $n$ at framing $f$
\cite{Witten:1988hf,Guadagnini:1989am}, with $n=n_1$ and $f=n_2/n_1$.
\item \emph{One loop.} In the planar limit \eqref{eq:Wtwoweak} gives $\im\vev{W(n_1,n_2)}=\pi\la\,n_1n_2+O(\la^3)$. The one-loop
phase of a loop of winding $n$ at framing $f$ is $\pi\la\,n^2f$ \cite{Bianchi:2018bke}. Then, the identification is the
same $n=n_1$ and $f=n_2/n_1$.
\end{itemize}
\begin{figure}[t]
\centering
\includegraphics[width=\textwidth]{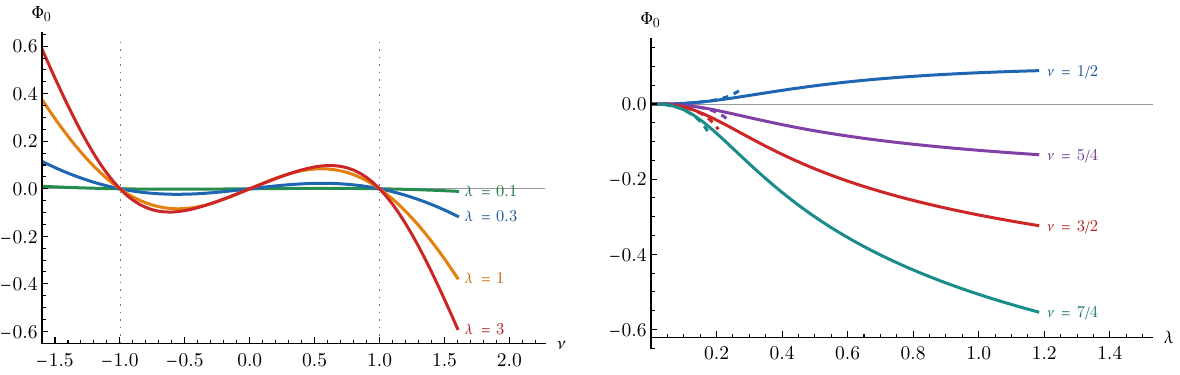}
\caption{The phase of the bosonic latitude that does not come from the framing $\Phi_0(\nu)=\PhB(\nu)-\nu\PhB$ of
\eqref{eq:Phizero}, which the latitude keeps at framing zero. Left: as a function of $\nu$ at four couplings. It is odd in
$\nu$, has the sign of $\nu(1-\nu^2)$, vanishes at $\nu=0$ and $\nu=\pm1$ (dotted), and grows with the coupling. Right:
as a function of $\la$ at four values of $\nu$, compared with the weak-coupling series \eqref{eq:Phizero}.}
\label{fig:Phi0}
\end{figure}
The second charge does more than set the framing. Indeed, it also changes the loop itself since the modulus depends on it. Moreover, beyond one loop the phase also contains a part that does not
come from the framing. If the whole phase of the latitude at framing $\nu$ came from the framing, it would be $\nu$ times the
phase per unit framing $\PhB(\nu)=\nu\PhB$. The difference
\begin{align}
\Phi_0(\nu)&:=\PhB(\nu)-\nu\PhB\label{eq:Phizero}\\
&\;=\nu(\nu^2-1)\left[-\frac{\pi^3\la^3}6+\frac{\pi^5\la^5}{60}\,\big(3\nu^2+13\big)
-\frac{\pi^7\la^7}{1680}\,\big(33\nu^4+173\nu^2+467\big)+O(\la^9)\right]\nonumber
\end{align}
is the part of the phase that the latitude has by itself. It is defined by the latitude and the great circle alone, and the
closed form gives it at every coupling (figure~\ref{fig:Phi0}). It vanishes in the two regimes above (at $N=1$ by \eqref{eq:WtwoN1} and at one loop) and also at $\nu=0$ and $\nu=\pm1$. In all the other cases, the latitude has a phase of its own which is still there at framing zero.

Feynman diagrams computed in the usual way, with dimensional regularisation and without point splitting, give the loops at
framing zero \cite{Guadagnini:1989am}. Defining the latitude at framing $f$ as \cite{Bianchi:2018bke}
\begin{equation}
\vev{W_B(\nu)}_f=e^{i(f-\nu)\PhB}\,\vev{W_B(\nu)}_\nu ,
\label{eq:linframing}
\end{equation}
the framing enters only through a phase at a rate $\PhB$ that does not depend on $\nu$. If \eqref{eq:linframing} holds at every order, setting $f=0$ removes
the framing phase $\nu\PhB$ from the matrix-model latitude, and the latitude at framing zero has the modulus
$\vev{W^{1/2}}/\cos\PhB(\nu)$ of \eqref{eq:latexact} and the phase $\Phi_0(\nu)$. Then, in the planar limit
\begin{equation}
\begin{split}
\vev{W_B(\nu)}_0={}&1+\frac{\pi^2\la^2}6\,(3\nu^2+2)-\frac{i\pi^3\la^3}6\,\nu(\nu^2-1)\\
&+\frac{\pi^4\la^4}{120}\,(5\nu^4-20\nu^2-14)-\frac{i\pi^5\la^5}{180}\,\nu(\nu^2-1)(6\nu^2-29)+O(\la^6),
\end{split}
\label{eq:Wframezero}
\end{equation}
where the first terms agree with the Feynman diagrams computed so far. At $\nu=1$ the great circle at framing zero is real and it matches the two loops of \cite{Drukker:2008zx,Chen:2008bp,Rey:2008bh}. For generic $\nu$ the result matches \cite{Bianchi:2018bke} up to three loops. The imaginary part is the first term of $\Phi_0$. It comes from a single diagram, with
three scalar bilinears on the contour, which does not depend on the framing and exists only for $\nu\neq1$, and it is the reason
why \eqref{eq:B16lat} uses the modulus of the latitude.

\subsection{The winding velocity as an integrated four-point function}
\label{sec:intcorr}

The latitude derivatives of section~\ref{sec:latclosed} are responses of the $\tfrac12$BPS line to a deformation, and in the
defect conformal field theory on the line such responses are integrated correlators. We denote the
logarithmic derivatives of the fermionic latitude by $\Ccor_n:=\partial_\nu^n\log\vev{W_F(\nu)}_\nu|_{\nu=1}$. The first derivative $\Ccor_1=4\pi^2\Bhalf$ is an integrated
two-point function in the displacement supermultiplet, which is how \eqref{eq:B12lat} was derived \cite{Bianchi:2017ozk}. 
The second derivative is given by
\begin{equation}
\Ccor_2=2\pi^2\Lcl-16\pi^4\Bhalf^2=\frac{\pi^2}4\,\big(1-\Upsilon\big),
\label{eq:WF2}
\end{equation}
where we used \eqref{eq:B16ratio}. The function $\Upsilon$ that orders the two lines is given by a second order response of the $\tfrac12$BPS line alone. In
this subsection we show that \eqref{eq:WF2}, the connected part of the second derivative, is an integrated four-point function of the
same multiplet. The same steps relate every $\Ccor_n$ to the integrated $2n$-point function.

The expectation values depend only on $\nu$, so we can set $\theta_0=0$. The latitude is then a great circle, and only its
coupling to the matter fields is deformed: it is tilted by the angle $\theta=\tfrac\pi2-2\alpha$ away from that of the
$\tfrac12$BPS circle, with the direction of the tilt turning once along the loop, and $\nu=\cos\theta$ by \eqref{eq:nudef}.
Expanding $\nu=1-\tfrac{\theta^2}2+\tfrac{\theta^4}{24}+\dots$, on functions of $\nu$ at the great circle
$\partial_\nu=-\partial_\theta^2$ and $\partial_\nu^2=\tfrac13(\partial_\theta^4+\partial_\theta^2)$, so that
\begin{equation}
\log\frac{\vev{W_F(\nu)}_\nu}{\vev{W^{1/2}}}=-2\pi^2\Bhalf\,\theta^2+\Big(\frac{\pi^2\Bhalf}6+\frac{\Ccor_2}8\Big)\theta^4+O(\theta^6).
\label{eq:WFtheta}
\end{equation}
The Bremsstrahlung function is the response at second order in the angle, and the winding velocity enters through $\Ccor_2$ at
fourth order, as the curvature of the latitude. Through the pole of \eqref{eq:PhiWF} the same curvature appears one order
earlier on the bosonic line where it is
two-point data while on the $\tfrac12$BPS line it is four-point data.

The deformation is generated by the tilt operators of the $\tfrac12$BPS line \cite{Bianchi:2017ozk} associated with the $R$-symmetry broken by the line $SU(4)\to SU(3)\times U(1)$. The latitude involves only one
tilt $O$ and its conjugate $\bar O$, with opposite charges under the $U(1)$ that rotates the plane of the latitude and
normalised as $\vev{O(t_1)\bar O(t_2)}=\Bhalf/t_{12}^2$ with $t_{ij}=t_i-t_j$ \footnote{In the notation of \cite{Bianchi:2020hsz}, $O=\tfrac12\,\mathbb O^1$ and $\bar O=\tfrac12\,\bar{\mathbb O}_1$ are the bosonic components of dimension one of the super-displacement multiplet, whose normalisation $C_\Phi=2\Bhalf$ gives $\vev{\mathbb O^a\bar{\mathbb O}_b}=4\Bhalf\,\delta^a_b/t_{12}^2$.}.

Because of charge conservation, the three-point functions of the tilts have to vanish and the four-point function have to contain two
$O$ and two $\bar O$. Then, the charges can then be alternate or adjacent, and this defines two independent connected correlators. Both are functions of the cross-ratio $\chi=t_{12}t_{34}/t_{13}t_{24}$ and they reads
\begin{equation}
\vev{O_1\bar O_2O_3\bar O_4}_c=\frac{k_A(\chi)}{t_{13}^2\,t_{24}^2},\qquad
\vev{O_1O_2\bar O_3\bar O_4}_c=\frac{k_D(\chi)}{t_{13}^2\,t_{24}^2}.
\label{eq:kAkD}
\end{equation}
with $t_1<t_2<t_3<t_4$.
Let's now consider the latitude as a deformation. We follow the strategy of \cite{Cavaglia:2022yvv}, where integrated four-point functions on the Wilson line of $\mathcal N=4$ SYM were obtained by deforming the line in conformal perturbation theory.
The latitude is a circle, and we label its points by their angle $\tau$. The map $t=\tan\frac\tau2$ sends the straight line, on
which \eqref{eq:kAkD} was written, to the circle. The
correlators on the circle are those on the line with every $t_{ij}$ replaced by the chord $s_{ij}=\sqrt{2(1-\cos\tau_{ij})}$. For
example $\vev{O(\tau_1)\bar O(\tau_2)}=\Bhalf/s_{12}^2$, and the cross-ratio becomes $\chi=s_{12}s_{34}/s_{13}s_{24}$.

The tilts come from the broken $R$-symmetry, so a small rotation of the coupling of the line to the matter fields inserts them
along the line. 
For the latitude, the coupling at the point $\tau$ of the loop is tilted by the angle $\theta$ in a direction rotated by $\tau$
under the $U(1)$ above, so that $O$ and $\bar O$ come with the phases $e^{-i\tau}$ and $e^{i\tau}$. Therefore the superconnection of the
latitude is that of the $\tfrac12$BPS circle in \eqref{eq:W12def} deformed by the tilts as follows
\begin{equation}
-i\mathcal L_\theta(\tau)=-i\mathcal L(\tau)+\theta\,\big(e^{-i\tau}O(\tau)+e^{i\tau}\bar O(\tau)\big)+\theta^2\mathcal L_2(\tau)
+O(\theta^3),
\label{eq:latdef}
\end{equation}
where $O$ and $\bar O$ are supermatrices, and $\theta^2\mathcal L_2$ collects the second-order changes of the couplings, to which we return below.
Inserted in the path-ordered exponential, \eqref{eq:latdef} turns the expansion of the latitude in $\theta$ into integrals of
correlators of the tilts on the $\tfrac12$BPS circle, in which each $O$ carries the phase $e^{-i\tau}$ of its position and each
$\bar O$ the phase $e^{i\tau}$. Along a fixed meridian, instead, the same deformation would rotate the whole line rigidly.

The integrals diverge where insertions collide, so the expansion needs a regulator. We use point
splitting, which keeps the insertions at angular distance larger than a cutoff. The divergences are then cancelled by counterterms, integrals along the loop of local functions of the coupling. Without
derivatives, such a function is invariant under the $U(1)$, so it depends on the size of the coupling but not on its direction,
and the latitude needs the same counterterms as the rigid $R$-symmetry rotation. The rigid rotation is a symmetry, so its
expectation value does not change: order by order its counterterms cancel its integrals. Subtracting it from the latitude
therefore removes the counterterms whatever the regulator, and in the difference each correlator is multiplied by its phase minus
one. The phase is just the product of the single phases of the inserted tilts.

It is instructive to compute the second derivative as an example. Expanding the path-ordered exponential of \eqref{eq:latdef} gives
\begin{equation}
\frac12\,\partial_\theta^2\log\vev{W_F}\Big|_{\theta=0}=\frac12\int_{|\tau_{12}|>\epsilon}d\tau_1\,d\tau_2\,
\Big(e^{-i\tau_{12}}\vev{O_1\bar O_2}+e^{i\tau_{12}}\vev{\bar O_1O_2}\Big)+2\pi c_\epsilon .
\end{equation}
Here $\tau_1$ and $\tau_2$ run over the circle at angular distance larger than the cutoff $\epsilon$ and the factor $1/2$ comes
from the path ordering. Each correlator carries the product of the phases of its insertions, while $c_\epsilon$ collects the local
terms of this order as the identity counterterm and the one-point function of $\mathcal L_2$. The rigid rotation gives the same
expression with both phases replaced by one and it vanishes as follows
\begin{equation}
0=\frac12\int_{|\tau_{12}|>\epsilon}d\tau_1\,d\tau_2\,\Big(\vev{O_1\bar O_2}+\vev{\bar O_1O_2}\Big)+2\pi c_\epsilon .
\end{equation}
because the rigid rotation is an $R$-symmetry transformation of the whole line, so that its expectation value is that of the
$\tfrac12$BPS circle for every $\theta$.
With $\vev{\bar O_1O_2}=\vev{O_1\bar O_2}=\Bhalf/s_{12}^2$, both integrals diverge as $4\pi\Bhalf/\epsilon$ fixing the counterterm
$c_\epsilon\simeq-2\Bhalf/\epsilon$ which is the identity counterterm of \cite{Cavaglia:2022yvv}. Subtracting the
second equation from the first removes $c_\epsilon$ and each of the two correlators is multiplied by its phase minus one
\begin{equation}\begin{split}
\frac12\,\partial_\theta^2\log\vev{W_F}\Big|_{\theta=0}&=\frac12\oint d\tau_1\,d\tau_2\,
\big(e^{-i\tau_{12}}-1+e^{i\tau_{12}}-1\big)\frac{\Bhalf}{s_{12}^2}\\
&=-\frac{\Bhalf}2\oint d\tau_1\,d\tau_2=-2\pi^2\Bhalf .
\label{eq:theta2}
\end{split}\end{equation}
Since $s_{12}^2=2-2\cos\tau_{12}$, the integrand is a constant and the cutoff is no longer needed! Moreover, 
$\partial_\theta^2=-\partial_\nu$ at $\theta=0$, then \eqref{eq:theta2} is $-\Ccor_1/2$, in agreement with \eqref{eq:B12lat} confirming the normalisation of the tilt.

A fourth order, the derivatives generate six terms that differs on the ordering of $O$ and $\bar O$. We
call such ordering a pattern $P$, labelled by the pair of points that carry $O$, and we write $e^{i\varphi_P}$ for its total
phase, the product of the phases of its four insertions, and $\vev P_c$ for its connected correlator. By \eqref{eq:kAkD}, in units
of $1/s_{13}^2s_{24}^2$, the alternating patterns $\{13\}$ and $\{24\}$ give $k_A(\chi)$ and $k_A(1-\chi)$, and the adjacent ones
give $k_D(\chi)$ for $\{12\}$ and $\{34\}$ and $k_D(1-\chi)$ for $\{23\}$ and $\{41\}$.

At this order the path-ordered exponential integrates over $\tau_1<\dots<\tau_4$. Subtracting the rigid rotation as at second order, each connected correlator is multiplied by its phase minus one and we obtain the following expansion
\begin{equation}
\log\frac{\vev{W_F}}{\vev{W^{1/2}}}=-2\pi^2\Bhalf\,r^2+r^4\int_{\tau_1<\dots<\tau_4}\sum_P\vev P_c\,\big(e^{i\varphi_P}-1\big)
+O(r^6).
\label{eq:logWr}
\end{equation}
where $r$ is the coupling of the tilt in conformal perturbation theory that coincides with $\theta$ only at leading order. 

The relation between $r$ and $\theta$ is fixed by the metric on the space of couplings.
Along a fixed direction, the line deformed by $r$ is conformal and the derivative of the line with respect to $r$ inserts $O+\bar O$. The inserted two-point function of it is
$g(r)/s_{ab}^2$, where $g(r)$ is the Zamolodchikov metric with $g(0)=2\Bhalf$. At order $r^2$, $g(r)$ is corrected by the connected four-point function of $O+\bar O$ with two points fixed and two integrated.
By conformal symmetry this correction does not depend on the two fixed points, so we can average them over the circle, and then
the four points enter in the same way. Since the four-point function of $O+\bar O$ is the sum over the patterns with all phases
equal to one, this gives
\begin{equation}
g(r)=2\Bhalf+\frac{r^2}{2\pi^2}\int_{\tau_1<\dots<\tau_4}\sum_P\vev P_c\sum_{i<j}s_{ij}^2+O(r^4).
\label{eq:metricr}
\end{equation}
The function $g(r)$ measures how much the line changes when the coupling changes: the lines with couplings $r$ and $r+dr$ are
at distance $\sqrt{g(r)}\,dr$, as measured by the two-point function of $O+\bar O$, the operator that takes one into the other.
The same two lines can be labelled by their angles, $\theta$ and $\theta+d\theta$. Since the rigid rotation is a symmetry, the line
looks the same at every angle, so a step $d\theta$ has the same length at every $\theta$ as at $\theta=0$, namely
$\sqrt{2\Bhalf}\,d\theta$. The angle is thus a uniform ruler, while the coupling is a ruler stretched by the deformation, and the
two measure the same step:
\begin{equation}
2\Bhalf\,d\theta^2=g(r)\,dr^2 .
\end{equation}
Integrating $d\theta=\sqrt{g(r)/2\Bhalf}\,dr$ at this order and inverting gives
\begin{equation}
r^2=\theta^2-\frac{\theta^4}{12\pi^2\Bhalf}\int_{\tau_1<\dots<\tau_4}\sum_P\vev P_c\sum_{i<j}s_{ij}^2+O(\theta^6),
\end{equation}
Inserting it in \eqref{eq:logWr}, the coefficient of $\theta^4$ in \eqref{eq:WFtheta} is
\begin{equation}
\frac{\pi^2\Bhalf}6+\frac{\Ccor_2}8=\int_{\tau_1<\dots<\tau_4}\sum_P\vev P_c\Big[e^{i\varphi_P}-1+\frac16\sum_{i<j}s_{ij}^2\Big],
\label{eq:c4int}
\end{equation}
where the first two terms of the bracket come from the latitude and the last one from the metric. The integrals in
\eqref{eq:logWr} and \eqref{eq:metricr} are taken with the cutoff, but their combination \eqref{eq:c4int} is finite without it.
Disconnected correlators do not enter, since we expand the logarithm. 

Other operators of the line appear as intermediate states in these four-point functions. Those that the cutoff or $\mathcal L_2$
add to the deformation, starting with the scalar bilinear along the direction of the line \cite{Cavaglia:2022yvv}, have
coefficients that vanish with the cutoff when $\Delta>1$. They contribute only where they meet a tilt, where the divergence of
their three-point function with the tilts compensates, and this renormalises $r$, as the sum over Wilson coefficients of
\cite{Cavaglia:2022yvv}. Since \eqref{eq:logWr} and \eqref{eq:metricr} are written in terms of the same $r$, this renormalisation
drops out of \eqref{eq:c4int}, which is the role played by the second invariance condition \cite{Cavaglia:2022yvv}.

The integrals in \eqref{eq:c4int} can be reduced to a single integral over the cross-ratio (details of this computation are in appendix \ref{app:kernels}). The resulting integrated correlator reads 
\begin{equation}
\begin{gathered}
\int_0^1d\chi\,\big[\mu_A(\chi)\,k_A(\chi)+\mu_D(\chi)\,k_D(\chi)\big]=\frac{\Bhalf}{12}+\frac{\Ccor_2}{16\pi^2},\\
\mu_A=\Big(\chi-\frac23\Big)\log\chi+\Big(\frac13-\chi\Big)\log(1-\chi),\qquad
\mu_D=\frac23\log\chi-\frac{2(3-\chi)}{3\chi}\log(1-\chi).
\end{gathered}
\label{eq:latII}
\end{equation}
The right-hand side of \eqref{eq:latII} contains $\Bhalf$, which is two-point data together with $\Ccor_2$ that is the genuine four-point data. 
Part of the identity is fixed by symmetry. A rigid rotation is an $R$-symmetry transformation, so the rotated $\tfrac12$BPS circles
are again conformal lines. They form a sphere: the $\tfrac12$BPS circle sits at the pole, and the rotation by $\theta$ in the
direction $\phi$ moves it to the point of polar angle $\theta$ and azimuth $\phi$. The two-point function of the tilts measures
distances on this sphere, and their integrated four-point function measures its curvature \cite{Drukker:2022pxk}. By symmetry, this sphere is round with metric
\begin{equation}
ds^2=2\Bhalf\big(d\theta^2+\sin^2\theta\,d\phi^2\big).
\end{equation}
The factor $2\Bhalf$ is the normalisation of the two-point function of $O+\bar O$, the tilt that moves the line along $\theta$. The
sphere therefore has radius $\sqrt{2\Bhalf}$ and curvature is the same at every point $K=1/2\Bhalf$. This
gives two identities. First, the coupling does not run. Then, the integral over $\chi$ of the four-point function of $O+\bar O$ which
would give its beta function vanishes. This four-point function is the sum over the six patterns: the two alternating ones give $k_A$ and the four adjacent ones give $k_D$. So we have
\begin{equation}
\int_0^1d\chi\,\big[k_A(\chi)+2k_D(\chi)\big]=0 .
\label{eq:margid}
\end{equation}
As a consequence, the kernels of \eqref{eq:latII} are defined only up to adding $c$ to $\mu_A$ and $2c$ to $\mu_D$. Second, the curvature
relation of \cite{Drukker:2022pxk}, written in our notation for the sphere of the latitude, is
\begin{equation}
\int_0^1d\chi\,\Big[-\log\chi(1-\chi)\;k_A+2\log\frac{\chi^2}{1-\chi}\;k_D\Big]=\Bhalf^2K=\frac{\Bhalf}2 .
\label{eq:curvid}
\end{equation}
Since both identities follow from symmetry, every consistent set of correlators satisfies them, and their right-hand sides know
only $\Bhalf$. In the same way, the displacement operator is associated with the conformal symmetry broken by the line, and the
integrated identities for its four-point function derived in \cite{Gabai:2025zcs} are fixed by its normalisation, which is
$C_D=12\Bhalf$ on this line.

The latitude case is different. Indeed, its direction turns once along the loop, so it is not a symmetry. Still, the term $\Bhalf/12$ of \eqref{eq:latII} is of the same kind as
\eqref{eq:curvid} because it is the first effect of the curvature of the sphere. Indeed, \eqref{eq:theta2} is $-\pi$ times the area of the
cap bounded by the latitude on the sphere of couplings. The curvature corrects this area by the factor $1-K\rho^2/12$, with
$\rho=\sqrt{2\Bhalf}\,\theta$ the radius of the cap, which gives $\Bhalf/12$.
Subtracting one sixth of \eqref{eq:curvid} removes it, and what remains is the latitude identity
\begin{equation}
\int_0^1\!d\chi\,\Big[\big(\chi-\tfrac12\big)\log\frac\chi{1-\chi}\;k_A(\chi)-\frac{2-\chi}\chi\log(1-\chi)\;k_D(\chi)\Big]
=\frac{\Ccor_2}{16\pi^2}=\frac{1-\Upsilon}{64},
\label{eq:intcorr}
\end{equation}
with $\Upsilon$ the function \eqref{eq:B16ratio} that orders the two lines. Its right-hand side is $\Ccor_2$ alone, which measures
how $\log\vev{W_F}$ deviates from $-\pi$ times the area of the cap. Unlike \eqref{eq:margid} and \eqref{eq:curvid}, \eqref{eq:intcorr} carries information beyond $\Bhalf$ since it contains the winding velocity \eqref{eq:WF2}.

Is it a new constraint? It depends on the theory. In $\mathcal N=4$ super Yang-Mills it is not: there the latitude is the
$\tfrac12$BPS circle at the rescaled coupling $\la\cos^2\theta$ \cite{Drukker:2006ga}, and both orderings of the tilts are fixed by
one reduced correlator and a topological constant \cite{Liendo:2018ukf,Cavaglia:2022qpg}. In terms of these, \eqref{eq:latII}
reduces to a combination of the two integrated constraints of \cite{Cavaglia:2022qpg}, with the same right-hand side, which checks
\eqref{eq:latII}, and \eqref{eq:intcorr} becomes equivalent to the curvature identity. We carry out this reduction explicitly in appendix~\ref{app:N4}. It uses only crossing symmetry and \eqref{eq:margid}, and it
reproduces the right-hand side of \eqref{eq:latII} exactly, through the recursion of the Bessel functions in \eqref{eq:BN4}. Since in
$\mathcal N=4$ every ingredient is known independently, this tests the derivation above. 
In ABJM, instead, the two orderings are independent. The tilts belong to the super-displacement multiplet, and supersymmetry fixes
the four-point functions of the whole multiplet in terms of two functions of the cross-ratio, $f$ for the alternating ordering and
$h$ for the adjacent one \cite{Bianchi:2020hsz}. They are normalised so that for a generalised free field $f=1-z$
and $h=1-\chi$, with $z=\chi/(\chi-1)$. The connected correlators of the tilts depend only on the deviations from these values,
$\delta f=f-1+z$ and $\delta h=h-1+\chi$, through derivatives fixed by supersymmetry as follows
\begin{equation}
\begin{gathered}
k_A(\chi)=\Bhalf^2\,\mathcal D\Big[\frac{\delta f(z)}\chi\Big],\qquad
k_D(1-\chi)=\Bhalf^2\Big(\mathcal D\Big[\frac{\delta h(\chi)}\chi\Big]-\partial_\chi\delta h(\chi)\Big),\\
\mathcal D=\partial_\chi\,\chi(1-\chi)\,\partial_\chi,
\end{gathered}
\label{eq:fhdict}
\end{equation}
where the factor $\Bhalf^2$ comes from the normalisation of the tilts. Since $f$ and $h$ are independent, \eqref{eq:intcorr} is a
new constraint on them. Since its right-hand side is known by \eqref{eq:B16}, the integrated four-point function of the $\tfrac12$BPS line is fixed by the Bremsstrahlung functions of the two lines.

At strong coupling the four-point functions of the tilts are computed from the string worldsheet in AdS$_2$, as a series in
$1/\sqrt\la$ \cite{Bianchi:2020hsz}. However, the right-hand side of \eqref{eq:intcorr} is a pure number. Up to worldsheet
instantons it is given by $\Ccor_2=1+\frac{\pi^2}{12}$, with no corrections in $1/\sqrt\la$. Then, the integral in \eqref{eq:intcorr} must vanish at any order (tree-level included), while at one-loop it has to be equal to $\frac1{16\pi^2}\big(1+\frac{\pi^2}{12}\big)$. 
At this order $f$ is given in \cite{Bianchi:2020hsz}, while $h$ is given by $f$ with $\log(-z)\to\log\chi$ and $z\to\chi$. Through \eqref{eq:fhdict}, we have
\begin{equation}
\begin{split}
k_A&=\frac{\Bhalf}{8\pi^2}\Big[-\frac{4\log\chi}{(1-\chi)^3}-\frac{4\log(1-\chi)}{\chi^3}-\frac{2(3\chi^2-3\chi+2)}{\chi^2(1-\chi)^2}\Big],\\
k_D&=\frac{\Bhalf}{8\pi^2}\Big[4\chi\log\frac\chi{1-\chi}-\frac{4\chi\log\chi}{(1-\chi)^3}-\frac{2(2\chi^2-3\chi+3)}{(1-\chi)^2}\Big],
\end{split}
\end{equation}
which are those of the $\tfrac12$BPS line of $\mathcal N=4$ SYM as expected from the universality of the
strong-coupling correlators of superconformal lines \cite{Pozzi:2025goj}. The integrals are elementary. Since
$\int_0^1k_A\,d\chi=6\Bhalf/8\pi^2$ and $\int_0^1k_D\,d\chi=-3\Bhalf/8\pi^2$, \eqref{eq:margid} holds while the kernels of
\eqref{eq:curvid} and \eqref{eq:intcorr} give $\Bhalf/2$ and $0$ as expected. The one-loop value $\frac1{16\pi^2}\big(1+\frac{\pi^2}{12}\big)$ comes from the one-loop determinant of the string that ends on
the latitude. The identity relates
it to the one-loop four-point functions of the fluctuations of the string of the $\tfrac12$BPS line in
AdS$_4\times\mathbb{CP}^3$, which are not known yet. 

At weak coupling the orders of the two sides of \eqref{eq:intcorr} need a comment. The right-hand side starts at one loop,
$\Ccor_2/16\pi^2=\la/16+O(\la^2)$, while the connected correlators carry the normalisation $\Bhalf^2=O(\la^2)$ of the tilts, so that
the integral of their leading weak-coupling form starts at order $\la^2$. The two are compatible because the expansion in the
coupling does not commute with the integral over $\chi$. The product of two tilts contains the scalar bilinear along the
direction of the line, whose dimension is $\Delta=1+O(\la)$. Near $\chi=0$ and $\chi=1$ its exchange gives terms that vanish at
fixed $\chi$ as $\la\to0$, but whose integrals against the logarithmic kernels are enhanced by inverse powers of $\Delta-1$ and
reach the order of the right-hand side. The same counting appears in \eqref{eq:curvid}, whose right-hand side $\Bhalf/2$ is also of
one-loop order although the identity follows from symmetry, so it is a feature of the weak-coupling expansion and not of the
latitude. In $\mathcal N=4$ the mechanism can be seen explicitly (appendix~\ref{app:N4}). A test of \eqref{eq:intcorr} at weak
coupling therefore needs the four-point functions of the tilts with the exchange of this operator resummed, which are not
available for the $\tfrac12$BPS line of ABJM.

The same steps work at every order in $\theta$. With $2n$ insertions, a pattern $P$ places $n$
insertions of $O$ and $n$ of $\bar O$ on the ordered points, and we define the integrated $2n$-point function of the tilts as
\begin{equation}
\mathcal J_{2n}=\int_{\tau_1<\dots<\tau_{2n}}\sum_P\vev P_c\,\big(e^{i\varphi_P}-1\big),
\label{eq:J2n}
\end{equation}
where each connected correlator is weighted by its phase minus one, as at second and fourth order. In terms of the coupling, the
logarithm of the latitude is $\sum_{n\ge1}r^{2n}\mathcal J_{2n}$, which generalises \eqref{eq:logWr}, and converting the coupling
into the angle with $g(r)(dr/d\theta)^2=2\Bhalf$ gives
\begin{equation}
\sum_{n\ge1}\frac{\Ccor_n}{n!}\,(\cos\theta-1)^n=\sum_{n\ge1}r(\theta)^{2n}\,\mathcal J_{2n} .
\label{eq:allorders}
\end{equation}
At order $\theta^{2n}$ this expresses $\Ccor_n$ through $\mathcal J_{2n}$, the metric at the same order and lower-point data. For
$n=1$ and $n=2$ it gives \eqref{eq:theta2} and \eqref{eq:c4int}. For $n=3$ it relates the third derivative $\Ccor_3$ to the
integrated six-point function $\mathcal J_6$. Unlike $\Ccor_1$ and $\Ccor_2$, $\Ccor_3$ is not winding data. By \eqref{eq:PhiWF}
\begin{equation}
\Ccor_3=3\pi^2\,T''(1)+2\pi^4\Bhalf-24\pi^4\Bhalf\Lcl+128\pi^6\Bhalf^3 ,
\label{eq:D3}
\end{equation}
where $T''(1)$ is the period with two logarithms of section~\ref{sec:latclosed}. The six-point identity is therefore not fixed by
the Bremsstrahlung functions of the two lines, and it is a constraint of a different kind. At strong coupling the exponent of \eqref{eq:WFairy} is linear in $\nu$, so every $\Ccor_n$ with $n\ge2$ comes from its
prefactor alone. They are the Taylor coefficients at $\nu=1$ of the one-loop factor of the first cap
\begin{equation}
\Ccor_n=\partial_\nu^n\log\frac{-\Gamma(-\frac\nu2)}{2^\nu\,\Gamma(\frac{3-\nu}2)}\bigg|_{\nu=1}
=(n-1)!\,\Big[\big(1-2^{1-n}\big)\,\zeta(n)+(-1)^n\Big],\qquad n\ge2,
\label{eq:Cnstrong}
\end{equation}
up to worldsheet instantons, with no correction in $1/\sqrt\la$. The first two are $\Ccor_2=1+\frac{\pi^2}{12}$ and
$\Ccor_3=\frac32\zeta(3)-2$, and every integrated $2n$-point identity has a pure number on its right-hand side.

\section{Discussion and outlook}
\label{sec:outlook}

In this manuscript we have computed the closed exact forms of the planar Bremsstrahlung functions of the two BPS Wilson lines of ABJM theory from the spectral curve of the
localisation matrix model. The fermionic function is a period of the second kind $\Bhalf=-\tfrac14(16+\kk^2)\partial_\kk^2\la$. Its
closed form, its Picard-Fuchs equation, its quasimodularity and the termination of its
strong-coupling expansion all follow from this one
statement. The bosonic function needs the derivative of the wound loop in its winding, whose imaginary part is the winding
velocity $\Lcl=\la-4\Bhalf$, a logarithmic period. It is the rational function \eqref{eq:B16} of the two, a projection rather
than a period. The derivation extends to ABJ theory.

The latitudes follow from the loops with two charges of the Fermi gas, whose planar limit is again a period of the curve. This
gives the bosonic and the fermionic latitudes in closed form at every $\nu$, with $\Bhalf$ and $\Lcl$ as their first two Taylor
coefficients at the great circle. In the defect conformal field theory on the
$\tfrac12$BPS line these coefficients are integrated correlators of the tilts, and the two Bremsstrahlung functions fix the integrated four-point function \eqref{eq:intcorr}. Some questions remain open. Between them two are the most urgent ones: what an integrability based computation should produce, and what survives at finite $N$ and $k$.

\paragraph{What integrability should compute.}
\label{sec:qscframe}
\label{sec:hring}

\begin{figure}[t]
\centering
\includegraphics[width=\textwidth]{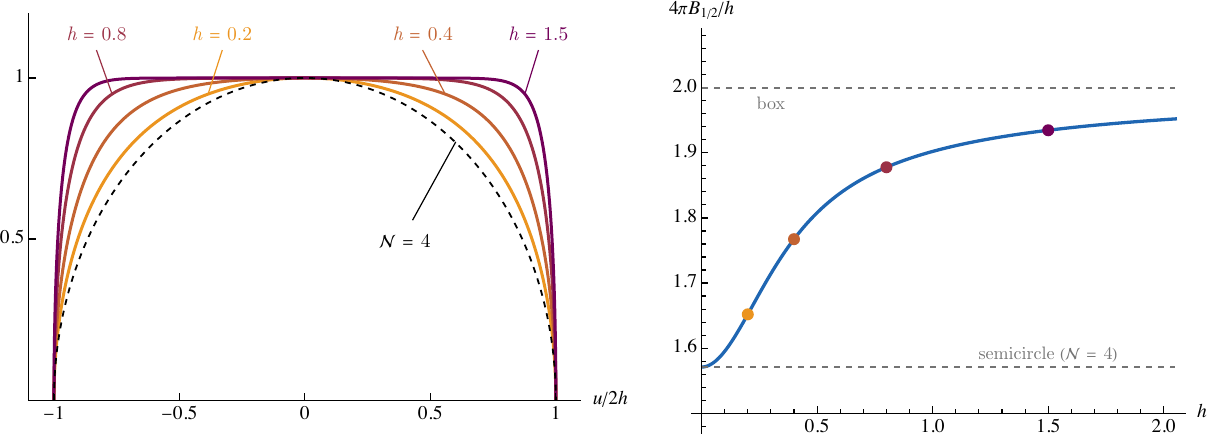}
\caption{The hyperbolic semicircle of \eqref{eq:cuttable}. Left: the profile $\sqrt{\Sh^2-\sinh^2\pi u}/\Sh$ against $u/2h$
for $h=0.2,0.4,0.8,1.5$ ($\la\simeq0.22,0.56,1.7,5.2$). At weak coupling it is the semicircle of $\mathcal N=4$ SYM (dashed), which in $\mathcal N=4$ keeps its shape at every coupling; in ABJM it flattens into a box. Right: the area
under the profile $4\pi\Bhalf/h$ grows from $\pi/2$ (semicircle) to $2$ (box). The dots mark the profiles on the left.}
\label{fig:semicircle}
\end{figure}

Integrability organises ABJM by the interpolating function $h(\la)$ which is conjectured to be related with $\kk$ as $\kk=4\sinh2\pi h$
\cite{Gromov:2014eha,Cavaglia:2016ide}. This dictionary was fixed by comparing the branch points of the
localisation curve with the cut of the QSC \cite{Gromov:2014eha}. A M\"obius map that sends branch points of $\Sigma_\kk$ to the ones of the QSC exists only if they have the same cross-ratio, which by \eqref{eq:crossratio} happens exactly for $\kk=4\sinh2\pi h$.
With $b$ the branch point of section~\ref{sec:curve} with $b+b^{-1}=2-i\kk$, the map and the curve are
\begin{equation}
\begin{gathered}
y=\frac{1+bZ}{Z+b}=e^{2\pi u},\\
\eta^2=y\,\big(y-e^{4\pi h}\big)\big(y-e^{-4\pi h}\big),\qquad\frac{dZ}{u}=-\frac12\,\frac{dy}{\eta}.
\end{gathered}
\label{eq:GScurve}
\end{equation}
Since the dictionary was fixed on localisation data, rewriting our results in $h$ tests nothing. What the map shows is the form
in which an integrability computation should return them. 
The cycle $A$ becomes a loop around the cut. Shrunk onto the cut, it runs along its upper edge and back along the lower one so that the $A$-period is twice the integral along the upper edge where
$\eta=2iy\sqrt{\Sh^2-\sinh^2\pi u}$ with $\Sh:=\sinh2\pi h=\kk/4$.
The integrability coupling is fixed by the
$\tfrac12$BPS circle $\sinh2\pi h=2\pi\la\vev{W^{1/2}}$. The Bremsstrahlung function and the winding velocity are the area
under this hyperbolic semicircle and its first moment in the rapidity, while the coupling is the same moment of the inverse
semicircle as follows
\begin{equation}
\begin{gathered}
\Bhalf=\frac1{8\pi\Sh}\int_{-2h}^{2h}\sqrt{\Sh^2-\sinh^2\pi u}\;du,\qquad
\Lcl=\frac1{2\Sh}\int_{-2h}^{2h}u\coth\pi u\,\sqrt{\Sh^2-\sinh^2\pi u}\;du,\\
\la=\frac{\Sh}{2}\int_{-2h}^{2h}\frac{u\coth\pi u}{\sqrt{\Sh^2-\sinh^2\pi u}}\;du .
\end{gathered}
\label{eq:cuttable}
\end{equation}
At weak coupling the semicircle becomes $\pi\sqrt{4h^2-u^2}$, the one of $\mathcal N=4$ super Yang-Mills on the cut
$[-2g,2g]$ and $\Bhalf\to h/8$. At strong coupling it is flat over most of the cut and $\Bhalf\to h/2\pi$ (figure~\ref{fig:semicircle}).

The comparison with $\mathcal N=4$ with $g=\sqrt\la/4\pi$, makes the targets precise. There the circle is a twisted moment of
the flat semicircle
\begin{equation}
\vev{W}_{\mathcal N=4}=\frac1{2\pi g^2}\int_{-2g}^{2g}e^{2\pi u}\sqrt{4g^2-u^2}\;du=\frac{2I_1(\sqrt\la)}{\sqrt\la},
\label{eq:N4cut}
\end{equation}
so the near-BPS limit of the cusp reduces the integrable description to Bessel functions, twisted moments of the same cut
\cite{Correa:2012hh,Drukker:2012de,Gromov:2012eu} and $B_{\mathcal N=4}$ \eqref{eq:BN4} is the ratio of two of them. In
ABJM the twisted moments of the hyperbolic semicircle are Legendre functions, and $\Bhalf$ is a single combination of two of
them with elementary coefficients as follows
\begin{equation}
\begin{gathered}
\Bhalf=\frac{\cosh(4\pi h)\,P_{-\frac12}(\cosh4\pi h)-P_{\frac12}(\cosh4\pi h)}{16\pi\sinh2\pi h} ,
\end{gathered}
\label{eq:BLegendre}
\end{equation}
where we used $\int_{-2h}^{2h}\frac{\cosh2\pi su}{\sqrt{\Sh^2-\sinh^2\pi u}}\;du=P_{s-\frac12}(\cosh4\pi h)$.

It is also curious to notice that the same coordinate simplifies the Picard-Fuchs equation for $B$ drastically. In $w=4\pi h$, with $\psi=\sqrt{\tanh(w/2)}\,\Bhalf$, the equation
becomes
\begin{equation}
-\psi''(w)+\frac{\ell(\ell+1)}{\sinh^2w}\,\psi(w)=0,\qquad \ell=\tfrac12 ,
\label{eq:coda-PT}
\end{equation}
the zero-energy Schr\"odinger equation of the hyperbolic P\"oschl-Teller potential, which is the Hamiltonian of the relative
motion of two particles in the hyperbolic Calogero-Sutherland model. The Bremsstrahlung function becomes a scattering problem.

A near-BPS computation of the cusp, in the QSC or in the Y-system of the cusped $\tfrac12$BPS line
\cite{Correa:2023lsm,Correa:2026akb}, which so far reproduces its one-loop anomalous dimension, should return these integrals,
and with them
\begin{equation}
\Bhalf=\frac h8+\frac{\pi^2h^3}{48}+O(h^5),\qquad
\Bhalf=\frac{h}{2\pi}-\frac{1-\log2}{4\pi^2}+O\big(e^{-4\pi h}\big),
\label{eq:Bhexp}
\end{equation}
with no correction in powers of $1/h$ at strong coupling. It would derive the dictionary $\kk=4\sinh2\pi h$ from observables. The one quantity that the QSC of ABJM has computed in this regime, the slope function
\cite{Gromov:2014eha}, is built from double integrals of the two powers of the same hyperbolic semicircle with a kernel.
Finally, in
$\mathcal N=4$ integrated constraints of the kind of \eqref{eq:intcorr}, together with the spectrum from integrability,
sharpened the numerical bootstrap of the $\tfrac12$BPS line \cite{Cavaglia:2021bnz,Cavaglia:2022yvv,Cavaglia:2022qpg,Cavaglia:2023mmu,Cavaglia:2024dkk}. We expect that, once the QSC approach to the spectrum of the local insertions on the Wilson line is developed, with \eqref{eq:intcorr} the same programme will be accessible for ABJM.

\paragraph{Beyond the planar limit.}
\label{sec:fNfuture}

At finite $N$ and $k$ the curve gives way to the Fermi gas of $N$ free fermions with $\hbar=2\pi k$, whose one-particle
Hamiltonian is the quantised mirror curve of local $\mathbb P^1\times\mathbb P^1$
\cite{Marino:2011eh,Kallen:2013qla,Grassi:2014zfa}. The two Bremsstrahlung functions should behave differently there. The
fermionic one is no longer fixed by a classical curve, and $\Bhalf(N,k)$ receives the $1/N$ corrections and the instantons of the
gas. The bosonic one remains the rational function \eqref{eq:B16} of $\Bhalf$ and of the winding velocity, since its derivation
uses only the polar decomposition of the bosonic loop and the reality of its winding derivative, two statements about averages in
the matrix model. The winding velocity, the one ingredient that needs the curve, should change by the single factor
\begin{equation}
\Lcl(N,k)=\frac Nk-4\,\xi\,\Bhalf(N,k),\qquad \xi=\frac{2\pi}k\cot\frac{2\pi}k ,
\label{eq:Lamfuture}
\end{equation}
the logarithmic derivative at unit winding of the factor $\csc\frac{2\pi n}k$ of the loop at large $N$
\cite{Klemm:2012ii}, which is the one-loop factor of the M2-brane that ends on the line \cite{Giombi:2023vzu}. As
$k\to\infty$ at fixed $\la$, $\xi\to1$ and \eqref{eq:Lamfuture} becomes \eqref{eq:Lamthm}. With \eqref{eq:B16} it gives the
bosonic function at every $N$ and $k$ in terms of the fermionic one alone as follows
\begin{equation}
\Bsix(N,k)=\frac{8\,\Bhalf\,\big(\frac{N}{k}-4\xi\Bhalf\big)}{1+64\pi^2\Bhalf^2} .
\end{equation}
where $\Bhalf=\Bhalf(N,k)$.
The natural tools are those of Boldis, Korchemsky and Testa: the operator representation of the loops in the Fermi gas, with the
exact relations of the resolvent of its one-particle operator that express the generating functions of the loops through the
partition function \cite{Boldis:2025yll}, and the functional relations that fix their instanton corrections
\cite{Boldis:2026zqu}. In this language \eqref{eq:Lamfuture} follows by differentiating the latitude at the great circle with all
instantons included.

Therefore, the central object is the latitude at finite $N$ and $k$. By \eqref{eq:latW} it is exactly the loop with charges
$(1,\nu)$ at every $N$ and $k$, but closed forms are known only at equal integer charges, for the wound loops
\cite{Klemm:2012ii,Boldis:2025yll}. What has to be verified is that the latitude is computed in the same way. At integer charges
the exact results are a sum of contributions from the four corners of the Fermi sea, up to exponentially small corrections, and
the hypothesis is that the same holds at the charges $(1,\nu)$. If it does, the tools of \cite{Boldis:2025yll,Boldis:2026zqu}
give the latitude in closed form at every $N$ and $k$, and with it both Bremsstrahlung functions, and they would provide an
analytic continuation of the loops to non-integer charges. Some checks already support the hypothesis: it reproduces the wound
loops at integer charges, the Airy form of the latitude found in \cite{Bianchi:2018bke} at large $k$, and the two caps
\eqref{eq:twocaps} in the planar limit. We leave this study for a future work.

\section*{Acknowledgements}
I am grateful to L.~Bianchi, M.~Bianchi, A.~Cavagli\`a, M.~Meineri, I.~Pesando, R.~Tateo and A.~Testa for useful
discussions, and especially to N.~Drukker, L.~Griguolo and D.~Seminara for inspiring exchanges and for their careful
reading of the manuscript.
The author work was supported by Marie Skłodowska-Curie Global Fellowship (HORIZON-MSCA-2022-PF-01) BOOTSTRABILITY-101109934 and by the INFN SFT specific initiative.
Claude Opus 5.5 (Anthropic) assisted with code development, editing and computational checks. The author takes full
responsibility for the content.
\appendix

\section{The Griffiths-Dwork reduction}
\label{sec:GM}

All the Gauss-Manin systems of this paper, for the periods \eqref{eq:GM0}-\eqref{eq:GM2}, for the logarithmic periods
\eqref{eq:logGM0}-\eqref{eq:logGM2} and for the ABJ curve \eqref{eq:ABJGM}, come from one algebraic step, the Griffiths-Dwork
reduction \cite{Griffiths:1969}. We describe it on the curve $u^2=P(Z;\kk)$ of \eqref{eq:curve}.

The cycle can be held fixed while $\kk$ varies, so the derivative of a period $J_n$ of \eqref{eq:Jdef} is the
period of the derivative of its differential. Since $\dk u^{-1}=-\tfrac12\,\dk P/u^3$,
\begin{equation}
\dk\Big(Z^n\frac{dZ}{u}\Big)=A_n\,\frac{dZ}{u^3},\qquad A_n=i\,Z^{n+1}\big(Z^2-i\kk Z+1\big).
\label{eq:GDAn}
\end{equation}
This differential has poles of order three at the branch points, so it is not in the basis $\{dZ/u,\,Z\,dZ/u,\,Z^2dZ/u\}$ of
section~\ref{sec:curve}.

We may add to it any exact differential, since exact differentials have no periods. With $\beta$ a polynomial
in $Z$ and the prime the derivative in $Z$, we have
\begin{equation}
d\Big(\frac{\beta}{u}\Big)=\Big(\beta'P-\tfrac12\,\beta P'\Big)\frac{dZ}{u^3} .
\label{eq:GDexact}
\end{equation}
Hence, if we find two polynomials $C_n=\sum_jc^{(n)}_jZ^j$ and $\beta_n$ such that
\begin{equation}
A_n=C_n\,P+\beta_n'P-\tfrac12\,\beta_nP' ,
\label{eq:GDcert}
\end{equation}
then $\dk(Z^n\,dZ/u)=C_n\,dZ/u+d(\beta_n/u)$, and $\dk J_n=\sum_jc^{(n)}_jJ_j$. Such polynomials exist because $P$ and $P'$ have no
common zero when the curve is smooth, and they are unique if $C_n$ has degree at most two and $\beta_n$ at most three. They are
found by matching the coefficients of \eqref{eq:GDcert}, and their only denominator is the resultant of $P$ and $P'$, that is the
discriminant $\Dsc$ of \eqref{eq:discdef}. This is how $\Dsc$ enters the connection.

For $n=0$ and $n=1$ the solution of \eqref{eq:GDcert} is
\begin{equation}
\begin{aligned}
C_0&=-\frac{\kk(\kk^2+4)}{\Dsc}-\frac{4i\kk^2}{\Dsc}\,Z+\frac{4\kk}{\Dsc}\,Z^2,&\qquad
\beta_0&=\frac{i\kk^2}{\Dsc}+\frac{\kk(2\kk^2+4)}{\Dsc}\,Z+\frac{7i\kk^2}{\Dsc}\,Z^2-\frac{4\kk}{\Dsc}\,Z^3,\\
C_1&=\frac{3i\kk^2}{\Dsc}+\frac{\kk^3}{\Dsc}\,Z+\frac{i\kk^2}{\Dsc}\,Z^2,&\qquad
\beta_1&=\frac{4\kk}{\Dsc}-\frac{7i\kk^2}{\Dsc}\,Z-\frac{\kk(2\kk^2+4)}{\Dsc}\,Z^2-\frac{i\kk^2}{\Dsc}\,Z^3,
\end{aligned}
\label{eq:GDbeta}
\end{equation}
and the coefficients of $C_0$ and $C_1$ are the rows \eqref{eq:GM0} and \eqref{eq:GM1}. For $n=2$ no new computation is needed,
since $A_2-A_0=-iZP+\tfrac i2Z^2P'$ is already of the form \eqref{eq:GDcert}, with $C=iZ$ and $\beta=-iZ^2$. Hence
\begin{equation}
C_2=C_0+iZ,\qquad \beta_2=\beta_0-iZ^2 ,
\label{eq:beta2}
\end{equation}
and the row \eqref{eq:GM2} is \eqref{eq:GM0} plus $iJ_1$, that is $\dk(J_2-J_0)=iJ_1$.

For the periods $\widehat J_m$ of \eqref{eq:Itilde} the differential is multiplied by $\log Z$, which
does not depend on $\kk$, so the same polynomials apply. However, in this case the exact term contributes since it is multiplied by the
logarithm. Integrating by parts along the cycle, on which $\log Z$ is single valued,
$\oint_A\log Z\,d(\beta_m/u)=-\oint_A\beta_m\,dZ/(Zu)$, which is the source $s_m$ of \eqref{eq:logGM}. It contains $J_{-1}$, which
equals $J_1$ by \eqref{eq:iotarel}, so it is a combination of the three basic periods. For $m=0$, with the row \eqref{eq:GM0}, we have
\begin{equation}
s_0=\dk J_0-\frac{4i\kk^2}{\Dsc}\Big(J_1-\frac{i\kk}4\,J_0\Big)=\dk J_0-\frac{2\pi\kk^2}{\Dsc},
\end{equation}
where the bracket is the flat invariant \eqref{eq:defQ}, with its value \eqref{eq:J1}. The source $s_1$ needs no simplification,
while $s_2=s_0+iJ_1$ by \eqref{eq:beta2}. These are the sources of \eqref{eq:logGM0}-\eqref{eq:logGM2}.

\section{The generating function of the wound loops}
\label{app:integertower}

To derive \eqref{eq:towerclosed} we carry out the reduction for all $n$ simultaneously. This is possible because \eqref{eq:contig}, whose coefficients are linear in $\ell$, becomes a first-order differential equation for the generating function of the moments. In the variable $z=i/Z$ we set
\begin{equation}
\mathcal F(z):=\sum_{m\ge0}J_m\Big(\frac zi\Big)^{m},\qquad
\mathcal P(z):=z^4P\big(\tfrac iz\big)=1-2\kk z+(\kk^2+2)z^2+2\kk z^3+z^4 .
\label{eq:Fmoments}
\end{equation}
We multiply \eqref{eq:contig} by $(z/i)^{\ell+3}$ and sum over $\ell\ge0$. In the $j$-th term we set $m=\ell-1+j$, so that
\begin{equation}
\sum_{j=0}^{4}p_j\Big(\frac zi\Big)^{4-j}\sum_{m\ge j-1}(2m+2-j)\,J_m\Big(\frac zi\Big)^{m}=0 .
\end{equation}
If every inner sum started at $m=0$, it would be $(2z\partial_z+2-j)\mathcal F$, and the sum over $j$ would produce the first-order operator $2z\mathcal P\partial_z-(2\mathcal P-z\mathcal P')$, because $\sum_jp_j(z/i)^{4-j}=\mathcal P$ and $\sum_j(4-j)\,p_j(z/i)^{4-j}=z\mathcal P'$. The inner sums start instead at $m=j-1$. For $j\le2$ this changes nothing, since the only terms that differ have the vanishing coefficient $2m+2-j$. For $j=3$ and $j=4$ the terms with $0\le m<j-1$ are missing, namely $p_3\frac zi\big(-J_0+J_1\frac zi\big)$ and $p_4\big(-2J_0+2J_2\frac{z^2}{i^2}\big)$. Moving them to the right-hand side, with $p_3=-2i\kk$ and $p_4=1$, we obtain
\begin{equation}
2z\mathcal P\,\mathcal F'-\big(2\mathcal P-z\mathcal P'\big)\mathcal F=-2J_0\,(1-\kk z)-2\big(J_2-i\kk J_1\big)z^2 .
\label{eq:Fode}
\end{equation}
The right-hand side thus contains exactly the moments that \eqref{eq:contig} leaves undetermined. Notice that $J_1$ and $J_2$ enter only through $J_2-i\kk J_1$, the period of the second-kind differential \eqref{eq:defsecondkind}.

Equation \eqref{eq:Fode} is linear and of first order, and it can be integrated explicitly. Its homogeneous solution is $z/\sqrt{\mathcal P}$. Then, one can write the inhomogeneous solution as
\begin{equation}
\mathcal F=c(z)\frac{z}{\sqrt{\mathcal P}},\qquad\text{with}\qquad
c'(z)=-J_0\,\frac{1-\kk z}{z^2\sqrt{\mathcal P}}-\frac{J_2-i\kk J_1}{\sqrt{\mathcal P}} .
\end{equation}
Integrating the equation with boundary conditions $\mathcal F(0)=J_0$ and $\mathcal F'(0)=-iJ_1$ we obtain 
\begin{equation}
\begin{gathered}
\mathcal F(z)=J_0-\frac{z}{\sqrt{\mathcal P(z)}}\Big[iJ_1+\big(J_2-i\kk J_1\big)\mathcal G(z)+J_0\,\mathcal H(z)\Big],\\
\mathcal G(z):=\int_0^z\frac{dt}{\sqrt{\mathcal P(t)}},\qquad
\mathcal H(z):=\int_0^z\frac{(\kk t+t^2)\,dt}{\sqrt{\mathcal P(t)}} .
\end{gathered}
\label{eq:Fsol}
\end{equation}
where $\mathcal{G}$ is the incomplete elliptic integral of the first kind while $\mathcal{H}$ is a combination of incomplete elliptic integrals of the first and second kind with an algebraic term.

The wound loops follow by summing \eqref{eq:Wn} with the same weight. Since $J_{n+1}$ and $J_{n-1}$ are the coefficients of $\mathcal F$ shifted by one step in either direction, the two shifts become a division and a multiplication by $z$
\begin{equation*}
\sum_{n\ge1}i^{1-n}\,4\pi^2n\la\vev{W^{1/6}_n}\,z^n=\frac{J_0-\mathcal F(z)}{z}-iJ_1-z\,\mathcal F(z) .
\end{equation*}
When \eqref{eq:Fsol} is inserted, $J_0$ cancels against $\mathcal F(0)$, and the two terms $-\mathcal F/z$ and $-z\mathcal F$ combine into $-(1+z^2)\mathcal F/z$, which turns the factor $z/\sqrt{\mathcal P}$ of \eqref{eq:Fsol} into $(1+z^2)/\sqrt{\mathcal P}$. This function is $\mathcal R(z)$ of \eqref{eq:dZrho}, the derivative of the resolvent along the curve. In terms of $\mathcal R$, all the wound loops are generated at once
\begin{equation}
\sum_{n\ge1}i^{1-n}\,4\pi^2n\la\vev{W^{1/6}_n}\,z^n
=iJ_1\big(\mathcal R-1\big)+\big(J_2-i\kk J_1\big)\,\mathcal R\,\mathcal G+J_0\big(\mathcal R\,\mathcal H-z\big).
\label{eq:WgenJ}
\end{equation}
This is the whole integer tower in terms of the three basic periods, each multiplying one explicit function of $z$: the third-kind period $J_1$ multiplies $\mathcal R-1$, the second-kind period $J_2-i\kk J_1$ multiplies $\mathcal R\mathcal G$, and the first-kind period $J_0$ multiplies $\mathcal R\mathcal H-z$. At $n=1$ it reduces to \eqref{eq:W1}, since $\mathcal R=1+\kk z+O(z^2)$, $\mathcal G=z+O(z^2)$ and $\mathcal H=O(z^2)$.

In order to compare with \eqref{eq:towerclosed}, it remains to insert the three basic periods of section~\ref{sec:threeperiods} in \eqref{eq:WgenJ}. From \eqref{eq:towerclosed}, we know that the right-hand side contains an elementary and a transcendental part, and that this split is unique. In \eqref{eq:WgenJ}, only the third-kind period $iJ_1$ has an elementary part, the constant $\frac\pi2$ fixed by the flat invariant \eqref{eq:J1}. Then we can deduce that $\frac\pi2(\mathcal R-1)$ is the elementary part of \eqref{eq:WgenJ} and that the polynomials $R_n$ of \eqref{eq:towerclosed} are the Taylor coefficients of $\mathcal R$. In the transcendental part, the coefficient of $\partial_\kk^2\la$ is $-i\pi^2\kk\,{(16+\kk^2)}\,\mathcal R\mathcal G$, and the coefficient of $\dk\la$ is $-i\pi^2\big[(4+\kk^2)\mathcal R\mathcal G+4(\mathcal R\mathcal H-z)-\kk(\mathcal R-1)\big]$, whose bracket can be rewritten as $\kk\,{(16+\kk^2)}\big[z^2/\mathcal P-\mathcal R\,\dk\mathcal G\big]$, in which $\mathcal H$ no longer appears. Altogether, the three polynomials of \eqref{eq:towerclosed} are generated by
\begin{equation}
\begin{gathered}
\sum_{n\ge0}R_nz^n=\mathcal R(z),\qquad
\sum_{n\ge1}T_n z^n=\kk\,\mathcal R(z)\,\mathcal G(z),\\
\sum_{n\ge1}S_n z^n=\kk\Big[\frac{z^2}{\mathcal P(z)}-\mathcal R(z)\,\dk\mathcal G(z)\Big],
\end{gathered}
\label{eq:STgen}
\end{equation}
with $R_0=1$. Since $\mathcal P$ has real coefficients on the physical slice and $\mathcal P(0)=1$, their Taylor coefficients are real polynomials in $\kk$, which is why the phase $i^{\,n-1}$ factors out of every loop.

Equations \eqref{eq:towerclosed} and \eqref{eq:STgen} solve the problem: every wound loop, at every coupling, is a combination of $1$, $\dk\la$ and $\partial^2_\kk\la$ whose coefficients are the Taylor coefficients of explicit functions of $z$. However, as it stands the result is not yet convenient. The function $\mathcal G$ is an incomplete elliptic integral $F$, and to obtain $S_n$ and $T_n$ from \eqref{eq:STgen} at a given winding one would have to differentiate it $n$ times, together with its $\kk$-derivative. Notably, this is never necessary! Indeed the Taylor coefficients of $\mathcal G$ are elementary, and the three polynomials follow from polynomial arithmetic alone.

Everything is built from the Taylor coefficients $q_n$ of $\mathcal P^{-1/2}$, defined in \eqref{eq:qn}, with $q_0=1$. They obey the following recursion with polynomial coefficients
\begin{equation}
n\,q_n=(2n-1)\kk\,q_{n-1}-(n-1)(\kk^2+2)\,q_{n-2}-(2n-3)\kk\,q_{n-3}-(n-2)\,q_{n-4},
\label{eq:u-recursion}
\end{equation}
obtained inserting \eqref{eq:qn} into $2\mathcal P\,\partial_z\mathcal P^{-1/2}+(\partial_z\mathcal P)\,\mathcal P^{-1/2}=0$. 
Exploiting \eqref{eq:STgen}, it is straightforward to write $R_n=q_n+q_{n-2}$. Since $\dk\mathcal R=(z\partial_z)^2\mathcal G$, also the incomplete elliptic integral $\mathcal G$ is reduced to the same data so that
\begin{equation}
\mathcal G(z)=\sum_{n\ge1}\frac{R_n'(\kk)}{n^2}\,z^n,\qquad R_n'=n\,q_{n-1},
\label{eq:Gseries}
\end{equation}
where the prime denotes $d/d\kk$. Thus, the $n$-th Taylor coefficient is the $\kk$-derivative of the polynomial $R_n$ of the fermionic loop of winding $n$ up to the factor $1/n^2$. Inserting \eqref{eq:Gseries} into \eqref{eq:STgen}, together with $z^2/\mathcal P=(z\partial_z\mathcal G)^2$, the two transcendental generating functions become the finite convolutions \eqref{eq:ST-convolutions}, with $R_0=1$. These relations determine $R_n$, $S_n$ and $T_n$ at any winding by polynomial arithmetic alone, and they make the leading terms evident. Since $R_j$ is a polynomial of degree $j$ with leading coefficient 1, every term of the sum for $T_n$ contributes $\kk^n/j$, so that $T_n=H_n\kk^n+\dots$, while the two sums of $S_n$ contribute $(n-1)\kk^{n-2}$ and $(n-H_n)\kk^{n-2}$, so that $S_n=(H_n-1)\kk^{n-1}+\dots$.

\section{The periods as the planar limit of the gas}
\label{sec:gasperiods}

The aim of this appendix is to justify \eqref{eq:Wtwoplanar}. We call $\vev{W_{\rm gas}(n_1,n_2)}$ the planar limit of the loop with two charges
\eqref{eq:Wtwodef}, and $\vev{W_{\rm per}(n_1,n_2)}$ the periods \eqref{eq:Wtwoplanar}. What
has to be shown is that
\begin{equation}
\vev{W_{\rm gas}(n_1,n_2)}=\vev{W_{\rm per}(n_1,n_2)}
\label{eq:gasper}
\end{equation}
for all charges $n_1$ and $n_2$. This is equivalent to say that in the planar limit the fermions fill the region enclosed by the cycle $A$ uniformly, so that the
average of $X^{n_1}Y^{n_2}$ in the gas is its average over that region. On the diagonal \eqref{eq:gasper} holds since both
sides are the wound loops by \eqref{eq:Wtwodef} and \eqref{eq:phasespace}. Off the diagonal it is not automatic for two
reasons. First, the matrix model fixes only the wound loops, which depend on the gas through $XY=e^{(\hat q+\hat p)/k}$ alone:
they tell how the fermions are distributed in $\hat q+\hat p$, but not in $\hat q$ and $\hat p$ separately, which is what the
loops with two charges measure. Second, the gas is not classical in the 't~Hooft limit. Although $\hat q/k$ and $\hat p/k$
commute at large $k$, the density $\hat\rho=\chi(\hat q)\chi(\hat p)$ varies on the scale $1/k$ in these variables. Treated
classically, the gas would fill a region of the real phase space, over which every loop is real, whereas the planar loops are
complex already at order $\la$.

We therefore proceed as for a differential equation with initial data on a line. We find an equation that relates the loops at
neighbouring charges and that $\vev{W_{\rm gas}}$ satisfies, we show that $\vev{W_{\rm per}}$ satisfies it too and that its solution
is fixed by its values on the diagonal. The equation comes from a second operator $\widetilde{\mathcal H}$ with two properties: it
commutes with $\hat\rho$, and it is a smooth function of $\hat q/k$ and $\hat p/k$ whose classical value is
$(X-X^{-1})(Y-Y^{-1})$, the left-hand side of the curve \eqref{eq:XYcurve}. We do not use it to describe the states of the gas
one by one, but through the Ward identity that it implies. The argument is divided in four steps and none of them requires a semiclassical description of the states of the gas.

\emph{(i) The dual Hamiltonian.} The operator we need is
\begin{equation}
\widetilde{\mathcal H}:=2\cosh\frac{\hat q+\hat p}k-2\cosh\frac{\hat q-\hat p}k=Z+Z^{-1}-\frac XY-\frac YX ,
\label{eq:Odual}
\end{equation}
built from the operator $Z=XY$ of the wound loop and from the second coordinate $X/Y$ of the curve with the exponentials
ordered symmetrically. To see that it commutes with $\hat\rho$ we work in the position representation identifying $\hat q=x$ and
$\hat p=-2\pi ik\,\partial_x$.  Thus, the momentum generates translations in the imaginary direction, so a symmetrically ordered
exponential acts as $e^{(a\hat q+b\hat p)/k}f(x)=e^{a(x-i\pi b)/k}f(x-2\pi ib)$, where we used the commutation relation \eqref{eq:rhogas}. The four exponentials appearing in \eqref{eq:Odual} have
$b=\pm1$, so $\widetilde{\mathcal H}$ acts on a function as the following difference operator
\begin{equation}
\big(\widetilde{\mathcal H}f\big)(x)=s_-(x)\,f(x-2\pi i)-s_+(x)\,f(x+2\pi i),\qquad s_\pm(x)=2\sinh\frac{x\pm i\pi}k .
\end{equation}
Acting with $\widetilde{\mathcal H}$ on the left translates the first variable of the kernel $\rho(x,y)$ of $\hat\rho$. Acting on
the right, a shift of the integration variable moves the translations onto the second variable, in the opposite direction.
The kernels of the two products are therefore
\begin{equation}
\begin{aligned}
\big(\widetilde{\mathcal H}\hat\rho\big)(x,y)&=s_-(x)\,\rho(x-2\pi i,y)-s_+(x)\,\rho(x+2\pi i,y),\\
\big(\hat\rho\,\widetilde{\mathcal H}\big)(x,y)&=s_+(y)\,\rho(x,y+2\pi i)-s_-(y)\,\rho(x,y-2\pi i).
\end{aligned}
\end{equation}
We can recall now the form \eqref{eq:rhogas} of the kernel, $\rho(x,y)=\chi(x)/\big(4\pi k\cosh\frac{x-y}{2k}\big)$. Since
$\chi(x\pm2\pi i)=-\chi(x)$ and the rest depends only on $x-y$, translating $x$ is, up to a sign, the same as translating $y$ in the opposite
direction $\rho(x\mp2\pi i,y)=-\rho(x,y\pm2\pi i)$. The two kernels then coincide if
\begin{equation}
\big[s_-(x)+s_+(y)\big]\rho(x,y+2\pi i)=\big[s_+(x)+s_-(y)\big]\rho(x,y-2\pi i) ,
\label{eq:Kident}
\end{equation}
and this holds because, by $\sinh A+\sinh B=2\sinh\frac{A+B}2\cosh\frac{A-B}2$,
\begin{equation*}
s_\mp(x)+s_\pm(y)=4\sinh\frac{x+y}{2k}\,\cosh\frac{x-y\mp2\pi i}{2k},
\end{equation*}
which contains exactly the $\cosh$ in the denominator of $\rho(x,y\pm2\pi i)$, so both sides of \eqref{eq:Kident} are equal to
$\frac{\chi(x)}{\pi k}\sinh\frac{x+y}{2k}$. Hence, we conclude that
\begin{equation}
\big[\hat\rho,\widetilde{\mathcal H}\,\big]=0 .
\label{eq:rhoOcomm}
\end{equation}
The four exponentials do not commute with $\hat\rho$ separately, only the combination \eqref{eq:Odual} does. Since $\widetilde{\mathcal H}$ commutes with $\hat\rho=e^{-\hat H}$, it has the same eigenstates as the Hamiltonian $\hat H$ of the gas, so we call it the dual Hamiltonian. The name refers to the quantised mirror curve:
$-i\widetilde{\mathcal H}$ is the modular dual \cite{Faddeev:1995nb,Faddeev:2000if} of the quantum curve $\hat\rho^{-1}=e^{\hat H}$
of local $\mathbb P^1\times\mathbb P^1$ \cite{Marino:2011eh,Grassi:2014zfa,Kashaev:2015kha}, whose Planck constant $4\pi/k$
vanishes in the 't~Hooft limit.

\emph{(ii) The Ward identity.} The one-body density matrix $\gamma$ of the $N$ fermions tells how the levels of $\hat\rho$ are
occupied. It is a function of $\hat\rho$, so it commutes with $\widetilde{\mathcal H}$. By the cyclicity of the trace, the average of the
commutator of $\widetilde{\mathcal H}$ with any one-body operator $O$ vanishes as follows
\begin{equation*}
\Big\langle\sum_i\,[\widetilde{\mathcal H},O]_i\Big\rangle=\mathrm{Tr}\big(\gamma\,[\widetilde{\mathcal H},O]\big)
=\mathrm{Tr}\big([\gamma,\widetilde{\mathcal H}]\,O\big)=0 .
\end{equation*}
We can now apply it to the loops. We write $E(u)=e^{(u_1\hat q+u_2\hat p)/k}$ so that the loop with charges $(n_1,n_2)$ is the average of
$E(n_1,n_2)$ and $\widetilde{\mathcal H}=E(1,1)+E(-1,-1)-E(1,-1)-E(-1,1)$. Given the commutation relation \eqref{eq:rhogas}, we have
\begin{equation}
\big[E(u),E(v)\big]=2i\sin\frac{\pi(u_1v_2-u_2v_1)}k\;E(u+v),
\label{eq:commE}
\end{equation}
so that for $O=E(n_1,n_2)$ each term of $\widetilde{\mathcal H}$ shifts the charges by $(\pm1,\pm1)$ with a sine
as coefficient
\begin{equation}
\begin{aligned}
\big[\widetilde{\mathcal H},E(n_1,n_2)\big]={}&2i\sin\frac{\pi(n_2-n_1)}k\,\big[E(n_1+1,n_2+1)-E(n_1-1,n_2-1)\big]\\
&\quad-2i\sin\frac{\pi(n_1+n_2)}k\,\big[E(n_1+1,n_2-1)-E(n_1-1,n_2+1)\big].
\end{aligned}
\end{equation}
Its average vanishes. In the planar limit $k\to\infty$ at fixed charges, the sine become linear and the planar loops $W=W_{\rm gas}$ satisfy
\begin{equation}
\begin{aligned}
(n_2-n_1)&\Big[\vev{W(n_1+1,n_2+1)}-\vev{W(n_1-1,n_2-1)}\Big]\\
&\qquad-(n_1+n_2)\Big[\vev{W(n_1+1,n_2-1)}-\vev{W(n_1-1,n_2+1)}\Big]=0.
\end{aligned}
\label{eq:ward}
\end{equation}
This is the relation between loops at neighbouring charges that we were looking for.

\emph{(iii) The periods satisfy the Ward identity.} We now show that $W=W_{\rm per}$ satisfies \eqref{eq:ward} too.
For this we write the periods as averages over the region $S$ enclosed by the cycle $A$ as for \eqref{eq:phasespace}. In the
coordinates $Q=\log X$ and $P=\log Y$, applying the Stokes' theorem to the period \eqref{eq:Wtwoplanar} we have
\begin{equation}
\la\,{\vev{W_{\rm per}(n_1,n_2)}}=\frac1{4\pi^2n_2}\oint_A e^{n_1Q+n_2P}\,dQ=
\frac1{4\pi^2}\int_S e^{n_1Q+n_2P}\,dQ\wedge dP .
\end{equation}
For them the Ward identity is the classical version of step (ii): the commutator is replaced by the Poisson bracket with
$\{Q,P\}=1$, and $\widetilde{\mathcal H}$ by its classical value $H_{\rm cl}=(X-X^{-1})(Y-Y^{-1})$ obtained by replacing the operators $\hat{q}/k$ and $\hat{p}/k$ with the numbers $Q$ and $P$. The bracket of two exponentials
\begin{equation}
\big\{e^{u_1Q+u_2P},e^{v_1Q+v_2P}\big\}=(u_1v_2-u_2v_1)\,e^{(u_1+v_1)Q+(u_2+v_2)P},
\end{equation}
is the planar limit of \eqref{eq:commE}. Applied to $H_{\rm cl}=2\cosh(Q+P)-2\cosh(Q-P)$ and $f=e^{n_1Q+n_2P}$, it gives
the four shifted exponentials with the coefficients of \eqref{eq:ward},
\begin{equation*}
\begin{aligned}
\{H_{\rm cl},f\}={}&(n_2-n_1)\big[e^{(n_1+1)Q+(n_2+1)P}-e^{(n_1-1)Q+(n_2-1)P}\big]\\
&\quad-(n_1+n_2)\big[e^{(n_1+1)Q+(n_2-1)P}-e^{(n_1-1)Q+(n_2+1)P}\big],
\end{aligned}
\end{equation*}
so that its integral over $S$ is $4\pi^2\la$ times the difference of the two sides of \eqref{eq:ward} for $\vev{W_{\rm per}}$. Then, the Ward
identity for the periods is 
\begin{equation}
\int_S\{H_{\rm cl},f\}\,dQ\wedge dP=\int_S dH_{\rm cl}\wedge df=-\int_Sd\big(f\,dH_{\rm cl}\big)=\oint_A f\,dH_{\rm cl}=0 .
\end{equation}
that holds because the integral of a Poisson bracket
over a region is a boundary term, and on the boundary $A$ the function $H_{\rm cl}$ is constant and equal to $i\kk$.

\begin{figure}[t]
\centering
\includegraphics[width=0.5\textwidth]{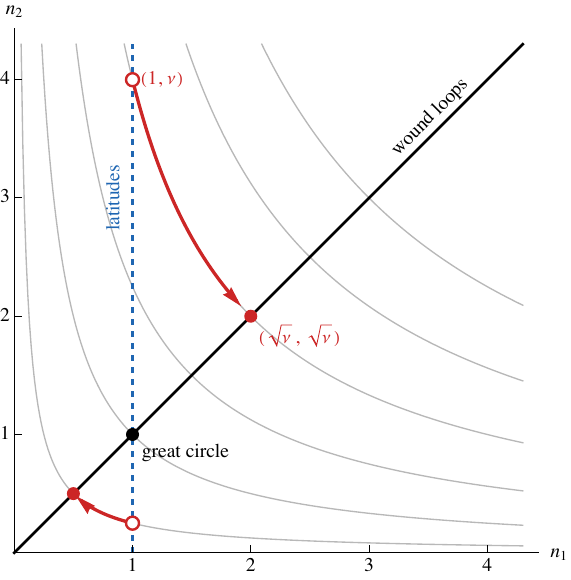}
\caption{The plane of the charges. On the diagonal the loops of the gas and the periods are
both the wound loops, so $\delta=0$. The latitudes lie on the line $n_1=1$ (dashed), which meets the diagonal at the great
circle. At the highest degree in the charges the Ward identity \eqref{eq:ward} becomes \eqref{eq:topward}, whose
characteristics are the hyperbolas $n_1n_2=$ const (grey), the orbits of the squeeze $(n_1,n_2)\to(e^sn_1,e^{-s}n_2)$. Along
each of them $\delta_D$ is constant, and each crosses the diagonal, so $\delta_D=0$. The arrows carry the latitudes with $\nu=4$
and $\nu=\tfrac14$ to the diagonal at $(\sqrt\nu,\sqrt\nu)$. In the double-scaling limit, $\la\to0$ at fixed $\la\nu$, only the
highest degree survives and the latitude $\nu$ is the loop wound $\sqrt\nu$ times.}
\label{fig:hyperbolas}
\end{figure}

\emph{(iv) The Ward identity and the diagonal fix the loops.} It remains to show that \eqref{eq:ward} and the values on the
diagonal determine its solution. Since \eqref{eq:ward} is linear, the difference
\begin{equation}
\delta(n_1,n_2):=\vev{W_{\rm gas}(n_1,n_2)}-\vev{W_{\rm per}(n_1,n_2)}
\end{equation}
satisfies it too, and it vanishes on the diagonal, where both functions are the wound loops. We show that this forces $\delta=0$,
order by order at weak coupling.

First, at each order in $\la$ the loops are polynomials in the charges. Expanding the exponential $e^{(n_1\hat q+n_2\hat p)/k}$,
the term of degree $m$ in the charges has as coefficients the moments of order $m$ of $\hat q/k$ and $\hat p/k$. At weak coupling
these operators are of order $\sqrt\la$, like the eigenvalues of the two nodes, whose wound loops are the averages of
$e^{n(\hat q\pm\hat p)/k}$. The term of degree $m$ is then of order $\la^{m/2}$, and the coefficient of $\la^j$ in $\vev{W_{\rm gas}}$ is a
polynomial of degree at most $2j$. For $\vev{W_{\rm per}}$ this is explicit in \eqref{eq:Jtwoseries}, where the coefficient of $\kk^j$ is
a product of two binomial coefficients, polynomials of degree $j$ in $n_1$ and in $n_2$. Since the coefficients of
\eqref{eq:ward} do not depend on $\la$, it holds order by order, and at a given order $\delta$ is a polynomial that satisfies
\eqref{eq:ward} and vanishes on the diagonal. The first property is essential: $\sin\big(\pi(n_2-n_1)/2\big)$ also satisfies
\eqref{eq:ward} and vanishes on the diagonal.

Second, we look at the terms of $\delta$ of highest degree $D$, which we call $\delta_D$, because on them \eqref{eq:ward} becomes
a differential equation. For a polynomial a symmetric difference is twice the derivative plus terms of lower degree, so we have
\begin{equation}
\delta(n_1+1,n_2\pm1)-\delta(n_1-1,n_2\mp1)=2\,(\partial_{n_1}\pm\partial_{n_2})\,\delta_D+\text{terms of degree}\le D-2 .
\end{equation}
Since the coefficients appearing in \eqref{eq:ward} have degree one, the terms of degree $D$ give
$(n_2-n_1)(\partial_{n_1}+\partial_{n_2})\delta_D=(n_1+n_2)(\partial_{n_1}-\partial_{n_2})\delta_D$. The mixed derivatives cancel
between the two sides, and what remains is
\begin{equation}
\big(n_1\partial_{n_1}-n_2\partial_{n_2}\big)\,\delta_D=0 .
\label{eq:topward}
\end{equation}
This equation has a simple meaning. The operator $n_1\partial_{n_1}-n_2\partial_{n_2}$ generates the squeeze
$(n_1,n_2)\to(e^sn_1,e^{-s}n_2)$, which preserves $n_1n_2$, so $\delta_D$ is constant along the hyperbolas $n_1n_2=$ const of
figure~\ref{fig:hyperbolas}. Explicitly, it multiplies the monomial $n_1^an_2^b$ by $a-b$, so that only the powers of $n_1n_2$
survive: $\delta_D=c\,(n_1n_2)^{D/2}$ for even $D$, and $\delta_D=0$ for odd $D$.

Third, the diagonal fixes $c$. On it each homogeneous part of $\delta$ gives a single power of $n$,
$\delta(n,n)=\sum_d\delta_d(1,1)\,n^d$, and since $\delta(n,n)=0$ all the coefficients vanish, in particular $c=\delta_D(1,1)=0$.
In figure~\ref{fig:hyperbolas}: $\delta_D$ is constant along each hyperbola and vanishes where the hyperbola crosses the
diagonal. Hence $\delta$ has degree at most $D-1$, its new top part satisfies \eqref{eq:topward} in the same way, and repeating
the argument down to the constant term gives $\delta=0$. This proves \eqref{eq:gasper}: \eqref{eq:Wtwoplanar} is the planar
limit of the loop with two charges for all charges and, at the charges $(1,\nu)$, the planar limit of the latitude matrix model
by \eqref{eq:latW}. The results at $\nu=1$, the phase and the slope \eqref{eq:latwindexact}, do not depend on it.

The same steps show how \eqref{eq:ward} carries the wound loops off the diagonal. Its polynomial solutions of degree at most four
are the combinations of $1$, $n_1n_2$ and $(n_1n_2)^2-n_1^2-n_2^2$, and the wound loop
$\vev{W^{1/6}_n}=1+i\pi n^2\la+\frac{\pi^2}3(2n^2-n^4)\la^2+O(\la^3)$ fixes the combination at each order: the result is the expansion \eqref{eq:Wtwoweak} of the periods through order
$\la^2$, obtained here from the Ward identity and the wound loops alone. At large charges only the terms of highest degree matter, $(n_1n_2)^j$ at order $\la^j$, so the loop depends on the charges only
through $n_1n_2$: this is the double-scaling limit of section~\ref{sec:twoperiods}.

\section{The kernels of the latitude identity}
\label{app:kernels}

In this appendix we reduce the integral over the positions of four points in \eqref{eq:c4int} to the integral over the
cross-ratio in \eqref{eq:latII}, and compute the kernels $\mu_A$ and $\mu_D$. The connected correlators depend on the positions
only through $\chi$ and $\vev P_c$ so the task is to integrate the bracket of \eqref{eq:c4int} at fixed
$\chi$ over the conformal transformations of the circle.

Since the integrand $F$ of \eqref{eq:c4int}, summed over the patterns, does not change under a cyclic
relabelling of the four points, we can let the first point run over the whole circle and divide by the four choices of which
point comes first as follows
\begin{equation}
\int_{0\le\tau_1<\dots<\tau_4<2\pi}\prod_{i=1}^4d\tau_i\;F=\frac14\int_0^{2\pi}d\tau_1\int_{\tau_1<\tau_2<\tau_3<\tau_4<\tau_1+2\pi}
\prod_{i=2}^4d\tau_i\;F .
\label{eq:appquarter}
\end{equation}
At fixed $\tau_1$, $\tau_3$ and $\tau_4$, the cross-ratio $\chi=s_{12}s_{34}/s_{13}s_{24}$ grows from $0$ to $1$
as $\tau_2$ moves from $\tau_1$ to $\tau_3$, with $\partial\chi/\partial\tau_2=s_{14}s_{34}/s_{13}s_{24}^2$. Trading $\tau_2$ for
$\chi$ we can write
\begin{equation}
\frac{d\tau_1\,d\tau_2\,d\tau_3\,d\tau_4}{s_{13}^2\,s_{24}^2}=d\chi\,\frac{d\tau_1\,d\tau_3\,d\tau_4}{s_{13}\,s_{14}\,s_{34}} .
\label{eq:appmeasure}
\end{equation}
At fixed $\chi$ any two configurations are related by a conformal transformation of the circle, since the cross-ratio is the
only invariant of four points, so we can use the transformation itself as coordinates for the remaining three angles. We take as
reference the configuration in which $\chi$ is manifest, the points $t_1=0$, $t_2=\chi$, $t_3=1$ and $t_4=\infty$ of the real
line, and send it to the circle with the M\"obius map
\begin{equation}
z=e^{i\psi}\,\frac{t-w}{t-\bar w},\qquad 0\le\psi<2\pi,\qquad w=x+iy,\quad y>0 .
\label{eq:appmap}
\end{equation}
Every configuration on the right-hand side of \eqref{eq:appquarter} with cross-ratio $\chi$ is obtained in this way for exactly
one choice of $\psi$ and $w$, with angles $\tau_a=\psi+2\arg(t_a-w)$ for $a=1,2,3$ and $\tau_4=\psi$: the angle $\psi$ rotates the
circle, while $w$ moves the points along it at fixed $\chi$. In these variables the measure that multiplies $d\chi$ in
\eqref{eq:appmeasure} is the product of the rotations and of the invariant area of the upper half-plane
\begin{equation}
\frac{d\tau_1\,d\tau_3\,d\tau_4}{s_{13}\,s_{14}\,s_{34}}=d\psi\,\frac{dx\,dy}{2y^2} ,
\label{eq:apphaar}
\end{equation}
as follows from the chords $s_{a4}=2y/|t_a-w|$ and $s_{13}=2y/|w|\,|1-w|$, and from the Jacobian $4y/|w|^2|1-w|^2$ of
$(\tau_1,\tau_3,\tau_4)$ with respect to $(\psi,x,y)$.
With \eqref{eq:appquarter}, the right-hand side
of \eqref{eq:c4int} becomes
\begin{equation}
\int_0^1d\chi\sum_Pk_P(\chi)\,m_P(\chi),\qquad
m_P(\chi)=\frac14\int_0^{2\pi}d\psi\int_{y>0}\frac{dx\,dy}{2y^2}\,\Big[\cos\varphi_P-1+\frac16\sum_{i<j}s_{ij}^2\Big],
\label{eq:appmP}
\end{equation}
where we kept the real part of the bracket, since its imaginary parts cancel between conjugate patterns.

In these coordinates the chords are
\begin{equation}
s_{a4}=\frac{2y}{|t_a-w|},\qquad s_{ab}=\frac{2y\,|t_a-t_b|}{|t_a-w|\,|t_b-w|},\qquad a,b=1,2,3,
\end{equation}
and the phase of a pattern is $e^{i\varphi_P}=N_P/\bar N_P$, where $N_P$ is the product of the factors $t_a-w$ and $t_a-\bar{w}$ corresponding to the insertions of $\bar O$ and $O$ respectively. For example,
$N_{13}=-\bar w\,(\chi-w)(1-\bar w)$. The chords do not depend on $\psi$, and neither does the phase, since the
charges add up to zero. Conjugate patterns give the same bracket, so that $m_{24}=m_{13}$, $m_{34}=m_{12}$ and $m_{41}=m_{23}$, and
the integral over $\psi$ gives
\begin{equation}
m_P(\chi)=\frac\pi4\int_0^\infty dy\int_{-\infty}^\infty\frac{dx}{y^2}\,\Big[\cos\varphi_P-1+\frac16\sum_{i<j}s_{ij}^2\Big].
\end{equation}

The bracket is a rational function of $x$, and its integral over $x$ follows from the residues at
$x=t_a+iy$. The result depends only on how the three distances between $t_1$, $t_2$ and $t_3$ are shared between the charges. We
call $d_0$ the distance between the two points with equal charge and $d_1$, $d_2$ the other two, so that $(d_1,d_2,d_0)$ is
$(\chi,1-\chi,1)$ for $P=13,24$, $(1,1-\chi,\chi)$ for $P=12,34$ and $(\chi,1,1-\chi)$ for $P=23,41$. With $u=4y^2$ we have
\begin{equation}
\int_{-\infty}^\infty\frac{dx}{y^2}\,\Big[\dots\Big]=\frac{\pi}{3y}\bigg[\frac{6\big((d_1^2+d_2^2-d_0^2)\,u-2d_1^2d_2^2\big)}
{(d_1^2+u)(d_2^2+u)}+\sum_{i=0}^2\frac{4d_i^2}{d_i^2+u}\bigg].
\end{equation}
At $u=0$, where three points collide, this vanishes thanks to the term from the metric, without which the integral over $y$ would
diverge logarithmically. The integral over $y$ is then elementary and gives
\begin{equation}
m_P=-\frac{\pi^2}{12}\left[\left(6\frac{d_1^2+3d_2^2-d_0^2}{d_2^2-d_1^2}-4\right)\!\log d_1\!+\!\left(6\frac{d_0^2-3d_1^2-d_2^2}{d_2^2-d_1^2}-4\right)\!\log d_2-4\log d_0\right],
\end{equation}
that explicitly is
\begin{equation}
\begin{aligned}
m_{13}=m_{24}&=\frac{\pi^2}3\Big((3\chi-2)\log\chi+(1-3\chi)\log(1-\chi)\Big),\\
m_{12}=m_{34}&=\frac{\pi^2}3\Big(\log\chi+\log(1-\chi)-\frac3\chi\,\log(1-\chi)\Big),\\
m_{23}=m_{41}&=\frac{\pi^2}3\Big(\log(1-\chi)-\frac{2+\chi}{1-\chi}\,\log\chi\Big).
\end{aligned}
\label{eq:appm}
\end{equation}

The patterns $\{24\}$, $\{23\}$ and $\{41\}$ multiply $k_A(1-\chi)$ and $k_D(1-\chi)$. Changing
$\chi\to1-\chi$ in their terms, the right-hand side of \eqref{eq:c4int} becomes
\begin{equation*}
\int_0^1d\chi\,\Big[2\,m_{13}(\chi)\,k_A(\chi)+2\big(m_{12}(\chi)+m_{23}(1-\chi)\big)\,k_D(\chi)\Big],
\end{equation*}
which by \eqref{eq:appm} is $2\pi^2$ times the left-hand side of \eqref{eq:latII} with $2m_{13}=2\pi^2\mu_A$ and $2\big(m_{12}(\chi)+m_{23}(1-\chi)\big)=2\pi^2\mu_D$.

\section{The latitude identity in \texorpdfstring{$\mathcal N=4$}{N=4} SYM}
\label{app:N4}

In this appendix we write the identities of section~\ref{sec:intcorr} for the $\tfrac12$BPS line of $\mathcal N=4$ SYM where the four-point functions of the tilts are known in terms of one function and one constant
\cite{Liendo:2018ukf}, and we show that \eqref{eq:latII} is a combination of the two integrated constraints of
\cite{Cavaglia:2022qpg}. Since every ingredient is known independently, this is a non trivial test of the derivation of
section~\ref{sec:intcorr}.

The latitude is the $\tfrac12$BPS circle at the coupling $\la\cos^2\theta$ \cite{Drukker:2006ga}. With $\nu=\cos\theta$ as in
section~\ref{sec:intcorr}, its first two logarithmic derivatives at the great circle are
\begin{equation}
\Ccor_1=4\pi^2B,\qquad \Ccor_2=8\pi^2\la\,\partial_\la B-4\pi^2B,
\label{eq:N4C}
\end{equation}
where $B=B_{\mathcal N=4}$ is given in \eqref{eq:BN4}. The tilts in the plane of the latitude are $O=\tfrac12(\Phi^1+i\Phi^2)$ and its
conjugate, which are normalised as in section~\ref{sec:intcorr} since $\vev{\Phi^a(t_1)\Phi^b(t_2)}=2B\,\delta^{ab}/t_{12}^2$
\cite{Correa:2012at}. Superconformal symmetry writes their four-point functions in terms of a reduced correlator $f(\chi)$ and a
constant $\mathbb F$ \cite{Liendo:2018ukf}. In the conventions of \cite{Cavaglia:2022qpg}, the four-point function of four
identical tilts $\Phi^1=O+\bar O$ is given by
\begin{equation}
\vev{\Phi^1(t_1)\Phi^1(t_2)\Phi^1(t_3)\Phi^1(t_4)}=\frac{4B^2\,G(\chi)}{t_{12}^2\,t_{34}^2},\qquad
G=\mathbb F\chi^2+\Big(\frac2\chi-1\Big)f-\big(\chi^2-\chi+1\big)\,\partial_\chi f .
\label{eq:N4G}
\end{equation}
Its disconnected part is the generalised free field, namely the sum over the three pairings of the insertions of the products of
two-point functions that read
\begin{equation}
G_{\rm GFF}=1+\frac{\chi^2}{(1-\chi)^2}+\chi^2,\qquad
f_{\rm GFF}=\frac{\chi(2\chi-1)}{\chi-1},\qquad \mathbb F_{\rm GFF}=3,
\label{eq:N4GFF}
\end{equation}
where the three terms of $G_{\rm GFF}$ come from the pairings $(12)(34)$, $(14)(23)$ and $(13)(24)$. The connected correlators are
what is left after subtracting it, so we define the deviations from the generalised free field
\begin{equation}
\begin{gathered}
\delta f=f-f_{\rm GFF},\qquad \delta\mathbb F=\mathbb F-3,\\
\delta G=G-G_{\rm GFF}=\delta\mathbb F\,\chi^2+\Big(\frac2\chi-1\Big)\delta f-\big(\chi^2-\chi+1\big)\,\partial_\chi\delta f ,
\end{gathered}
\label{eq:N4delta}
\end{equation}
so that $4B^2\,\delta G/t_{12}^2t_{34}^2$ is the connected four-point function of $\Phi^1$. In terms of these deviations, the
connected correlators of \eqref{eq:kAkD} are
\begin{equation}
k_A=-B^2\,\partial_\chi\Big(\frac{\delta f}{\chi^2}\Big)\qquad\text{and}\qquad
k_D=B^2\big(\delta\mathbb F-\partial_\chi\delta f\big).
\label{eq:N4kAkD}
\end{equation}
Since $\Phi^1=O+\bar O$, the connected function of $\Phi^1$ is the sum over the six patterns of section~\ref{sec:intcorr}, and
indeed $4B^2\,\delta G/\chi^2=k_A(\chi)+k_A(1-\chi)+2k_D(\chi)+2k_D(1-\chi)$. Crossing symmetry requires
$(1-\chi)^2f(\chi)+\chi^2f(1-\chi)=0$. The generalised free field satisfies it, and therefore so does $\delta f$.

Note that the generalised free field is not the correlator at zero coupling! Indeed, the tree-level of the four-point function is given by Wick contractions, but in the planar limit only the two pairings that do not
cross survive, while the crossed pairing $(13)(24)$ is suppressed by $1/N^2$. Hence, at zero coupling we have
\begin{equation}
G^{(0)}=1+\frac{\chi^2}{(1-\chi)^2}=G_{\rm GFF}-\chi^2,\qquad f^{(0)}=f_{\rm GFF},\qquad \mathbb F^{(0)}=2 .
\label{eq:N4free}
\end{equation}
The reduced correlator is the same, and only the constant changes. The free correlator already has a connected
part $\delta G=-\chi^2$, which is minus the missing crossed pairing. In \eqref{eq:N4kAkD} it corresponds to $\delta f=0$ and
$\delta\mathbb F=-1$ that give $k_A=0$ and $k_D=-B^2$. This is what one expects. In the adjacent ordering $OO\bar O\bar O$ the
crossed pairing joins $O_1$ with $\bar O_3$ and $O_2$ with $\bar O_4$, and its absence gives $k_D=-B^2$. In the alternating
ordering $O\bar OO\bar O$ the crossed pairing would join two equal charges, so it never contributes and $k_A=0$.

With \eqref{eq:N4kAkD}, integrating by parts and using crossing, the left-hand side of \eqref{eq:latII} becomes
\begin{equation}\begin{split}
\int_0^1d\chi\,\big[\mu_Ak_A+\mu_Dk_D\big]=&\;4B^2\,\int_0^1d\chi\,\Big[\frac{\delta f}{\chi}-\frac23\,\delta G\,\frac{1+\log\chi}{\chi^2}\Big]+B^2\Big(\frac{\pi^2}3-4\Big)\delta\mathbb F\\
&+\frac43\int_0^1d\chi\,\big[k_A+2k_D\big]\,.
\label{eq:N4latII}
\end{split}\end{equation}
The last term of \eqref{eq:N4latII} is the integral of \eqref{eq:margid}. Its integrand is a total derivative, and crossing
relates its boundary value at $\chi=1$ to the one at $\chi=0$, so that
\begin{equation}
\int_0^1d\chi\,\big[k_A+2k_D\big]=2B^2\Big(\delta\mathbb F+2\lim_{\chi\to0}\frac{\delta f}{\chi^2}\Big).
\label{eq:N4marg}
\end{equation}
The right-hand side vanishes at every finite coupling. In the OPE of the reduced correlator
\cite{Liendo:2018ukf,Cavaglia:2021bnz}
\begin{equation}
f=\chi+C_{\rm BPS}^2\Big(-\frac{\chi^2}2+O(\chi^3)\Big)+\sum_n\frac{C_n^2}{1-\Delta_n}\,\chi^{\Delta_n+1}\big(1+O(\chi)\big),
\qquad C_{\rm BPS}^2=\mathbb F-1,
\label{eq:N4OPE}
\end{equation}
the identity operator gives $\chi$, the protected multiplet gives a term $\chi^2$ whose coefficient is fixed by $\mathbb F$, and a
long multiplet of dimension $\Delta_n$ starts at $\chi^{\Delta_n+1}$. At finite coupling all $\Delta_n>1$, so the term $\chi^2$ comes
only from the protected multiplet. Since $f_{\rm GFF}=\chi-\chi^2+O(\chi^3)$, this gives \cite{Cavaglia:2022qpg}
\begin{equation}
\delta f=\frac{3-\mathbb F}2\,\chi^2+O\big(\chi^{\Delta_1+1}\big)\qquad\text{for}\quad\chi\to0,
\label{eq:N4small}
\end{equation}
with $\Delta_1$ the lowest long dimension. Hence $\lim_{\chi\to0}\delta f/\chi^2=-\delta\mathbb F/2$, and the last term of
\eqref{eq:N4latII} vanishes. Thus, in $\mathcal N=4$ the identity \eqref{eq:margid} follows from crossing symmetry and the OPE alone.

To compute the other integral appearing in \eqref{eq:N4latII}, we need to recall the integrated constraints found in \cite{Cavaglia:2022qpg}. In terms of the quantities defined in \eqref{eq:N4delta}, they are given by
\begin{equation}
\int_0^1d\chi\,\delta G\,\frac{1+\log\chi}{\chi^2}=\frac{3\mathbb C-B}{8B^2},\qquad
\int_0^1d\chi\,\frac{\delta f}{\chi}=\frac{\mathbb C}{4B^2}+\mathbb F-3,
\label{eq:CGJPconstr}
\end{equation}
with $\mathbb C$ the curvature function of the near-BPS cusp \cite{Gromov:2015dfa}. In \cite{Cavaglia:2022qpg} the
regularised functions $\delta G$ and $\delta f$ are measured from the correlator at zero coupling. This gives the same $\delta f$, and a $\delta G$ larger than
ours by $\chi^2$, whose integral against $(1+\log\chi)/\chi^2$ vanishes since $\int_0^1(1+\log\chi)\,d\chi=0$. As the right-hand side
of the second constraint contains $\mathbb F-3$ explicitly, \eqref{eq:CGJPconstr} holds unchanged with our subtraction. Then, one can recognise that the
combination appearing in the first line of \eqref{eq:N4latII} is the second constraint minus two thirds of the first, in which $\mathbb C$ cancels, so that
\begin{equation}
\int_0^1d\chi\,\Big[\frac{\delta f}{\chi}-\frac23\,\delta G\,\frac{1+\log\chi}{\chi^2}\Big]=\delta\mathbb F+\frac{1}{12B}
\end{equation}
Then, using $\mathbb F=3(g^2-B)/\pi^2B^2$ and $g^2=\la/16\pi^2$ \cite{Cavaglia:2022qpg}, the left-hand side of \eqref{eq:latII} is
\begin{equation}
B^2\Big[\frac{\pi^2}3\,\delta\mathbb F+\frac1{3B}\Big]=\frac{\la}{16\pi^2}-\frac23\,B-\pi^2B^2 .
\end{equation}
By \eqref{eq:N4C} its right-hand side is $\frac B{12}+\frac{\Ccor_2}{16\pi^2}=\frac12\la\,\partial_\la B-\frac B6$, and the two agree if
and only if
\begin{equation}
\la\,\partial_\la B=\frac{\la}{8\pi^2}-B-2\pi^2B^2 ,
\label{eq:N4Bessel}
\end{equation}
which follows from the recursion relations $I_1'=I_2+I_1/x$ and $I_2'=I_1-2I_2/x$ of the Bessel functions in \eqref{eq:BN4}. In
$\mathcal N=4$ the latitude identity is therefore the combination of the two constraints of \cite{Cavaglia:2022qpg} from which the
curvature function drops out, and its right-hand side is reproduced exactly.

In the same way, we have
\begin{equation}
\int_0^1d\chi\,\Big[-\log\chi(1-\chi)\,k_A+2\log\frac{\chi^2}{1-\chi}\,k_D\Big]
=\frac32\int_0^1d\chi\,\big[\mu_Ak_A+\mu_Dk_D\big]-\frac{\pi^2}2\,B^2\,\delta\mathbb F ,
\end{equation}
which with the values above equals $B/2$, as \eqref{eq:curvid} requires. Since $\mathbb F$ is a known function of $B$, in
$\mathcal N=4$ SYM the identities \eqref{eq:curvid}, \eqref{eq:latII} and \eqref{eq:intcorr} are equivalent. In ABJM the two orderings
of the tilts are independent functions and there is no topological constant, and \eqref{eq:intcorr} is not implied by
\eqref{eq:curvid}.

The last remaining thing to address is the fact that the vanishing of \eqref{eq:N4marg} requires all long dimensions to be larger than one. This fails only at zero coupling,
where the lowest long operator (the scalar $\Phi^6$ along the line) has $\Delta_1\to1$. Indeed, the correlator at zero coupling has
$\delta f=0$ and $\delta\mathbb F=-1$, and \eqref{eq:N4marg} gives $-2B^2$ instead of zero. This mismatch is the weak-coupling
``anomaly'' found in \cite{Cavaglia:2022qpg}, where it appears in a different form. By the sum over the six patterns,
$\int_0^1d\chi\,[k_A+2k_D]=2B^2\int_0^1d\chi\,\delta G/\chi^2$. Since the $\delta G$ of \cite{Cavaglia:2022qpg} is measured from
the correlator at zero coupling, its integral equals $1$ at finite coupling but it is $0$ at zero coupling.

The missing $2B^2$ comes from $\Phi^6$. Let $\Delta_1=1+\gamma$, with $\gamma=4g^2+O(g^4)$ \cite{Cavaglia:2021bnz}. At zero coupling
the protected multiplet gives only half of the term $-\chi^2$ in $f_{\rm GFF}$, and the other half comes from $\Phi^6$. This fixes
$C_1^2\simeq\gamma/2$ \cite{Cavaglia:2022qpg}, and near $\chi=0$ the block of $\Phi^6$ in \eqref{eq:N4OPE} gives
\begin{equation}
k_A\simeq\frac\gamma2\,B^2\,\chi^{\gamma-1}.
\label{eq:N4contact}
\end{equation}
This term vanishes at every fixed $\chi$ as $\gamma\to0$, but its integral over $(0,1)$ is $\tfrac12B^2$ for every $\gamma$. As
$\gamma\to0$ it therefore tends to $\tfrac12B^2\,\delta(\chi)$, a contact term at $\chi=0$. Similar contact terms in $k_A$ and $k_D$
appear at $\chi=1$, and together they give the missing $2B^2$. The weak-coupling expansion of the correlators misses them, and
expanding \eqref{eq:N4contact} as $\chi^{\gamma-1}=\chi^{-1}(1+\gamma\log\chi+\dots)$ gives integrals that diverge at $\chi=0$ at
every order. In \cite{Cavaglia:2022qpg} this was solved by keeping the block of $\Phi^6$ exact. In \cite{Drukker:2025dfm} the
anomaly was explained in general: a nearly marginal operator gives a contact term in the OPE of two tilts, which must be kept for
the integrated identities to hold. The first constraint in \eqref{eq:CGJPconstr} has the same anomaly at weak coupling
\cite{Cavaglia:2022qpg}.

\bibliographystyle{JHEP.bst}
\bibliography{refs}
\end{document}